\documentclass[useAMS,usenatbib,babel]{mn2e}

\usepackage[english]{babel}
\usepackage{amsmath}
\usepackage{amssymb,textcomp}
\usepackage{array}
\usepackage{supertabular}
\usepackage{hhline}
\usepackage[draft]{hyperref}
\usepackage[usenames]{color}
\hypersetup{dvips, colorlinks=true, linkcolor=black, citecolor=black, filecolor=black, urlcolor=black}
\usepackage{graphicx}

\def\gtrsim{\lower.5ex\hbox{$\; \buildrel > \over \sim \;$}}

\newcommand{\hagn}{\mbox{{\sc \small Horizon-AGN}}}
\newcommand{\hnoagn}{\mbox{{\sc \small Horizon-noAGN}}}

\definecolor{grey}{rgb}{0.75,0.75,0.75}
\definecolor{Orange}{rgb}{1.0,0.5,0.15}
\definecolor{brown}{rgb}{0.7,0.25,0.0}
\definecolor{pink}{rgb}{1.0,0.5,0.5}
\definecolor{darkerred}{rgb}{0.8,0,0}
\definecolor{darkerblue}{rgb}{0,0,0.8}
\definecolor{Blue}{rgb}{0,0.08,0.65}
\definecolor{Red}{rgb}{0.65,0.08,0.05}
\definecolor{Green}{rgb}{0.15,0.45,0.25}

\begin{document}

\author[Dubois et al.]{
\parbox[t]{\textwidth}
{Yohan Dubois$^1$\thanks{E-mail: dubois@iap.fr}, S\'ebastien Peirani$^{1,2}$, Christophe Pichon$^{1,3}$, Julien Devriendt$^{4,5}$,\\ Rapha\"el Gavazzi$^1$, Charlotte Welker$^{1,6}$ and Marta Volonteri$^1$}
\vspace*{6pt} \\ 
$^{1}$ CNRS and UPMC Universit\'e Paris 06, UMR 7095, Institut d'Astrophysique de Paris, 98 bis Boulevard Arago, Paris 75014, France\\
$^{2}$ Kavli IPMU (WPI), UTIAS, The University of Tokyo, Kashiwa, Chiba 277-8583, Japan\\
$^{3}$ Korea Institute of Advanced Studies (KIAS) 85 Hoegiro, Dongdaemun-gu, Seoul, 02455, Republic of Korea\\
$^{4}$ Sub-department of Astrophysics, University of Oxford, Keble Road, Oxford OX1 3RH\\
$^{5}$ Observatoire de Lyon, UMR 5574, 9 avenue Charles Andr\'e, Saint Genis Laval 69561, France\\
$^{6}$ ICRAR, University of Western Australia, 35 Stirling Highway, Crawley, Western Australia 6009, Australia
}
\date{Accepted 2016 September 06. Received 2016 September 02; in original form 2016 June 09}
\title[Morphology of galaxies in Horizon-AGN]{The Horizon-AGN simulation: 
morphological diversity of galaxies 
promoted by AGN feedback
}

\maketitle

\begin{abstract}
{The interplay between cosmic gas accretion on to galaxies and galaxy mergers drives the observed morphological diversity of galaxies.
By comparing the state-of-the-art hydrodynamical cosmological simulations \hagn\, and \hnoagn, we unambiguously identify the critical role of active galactic nuclei (AGN) in setting up the correct galaxy morphology for the massive end of the population. 
With AGN feedback, typical kinematic and morpho-metric properties of galaxy populations as well as the  galaxy-halo mass relation are in much better agreement with observations.
Only AGN feedback allows massive galaxies at the centre of groups and clusters to become ellipticals, while without AGN feedback those galaxies reform discs.
It is the merger-enhanced AGN activity that is able to freeze the morphological type of the post-merger remnant by durably quenching its quiescent star formation.
Hence morphology is shown to be driven not only by mass but also by the nature of cosmic accretion: at constant galaxy mass, ellipticals are galaxies that are mainly assembled through mergers, while discs are preferentially built from the in situ star formation fed by smooth cosmic gas infall.
}
\end{abstract}


\section{Introduction}

Observations of galaxies establish that a wide variety of morphological types are witnessed at all times in the Universe, from discs to ellipticals and irregular galaxies (the so-called Hubble sequence), although morphologies strongly depend on the galaxy mass~\citep[e.g.][]{conselice06,ilbertetal10,bundyetal10}.
In the $\Lambda$ cold dark matter (CDM) paradigm, late-type (disc) galaxies are supposed to form from the smooth gas accretion of circum-galactic gas in order to form a rotating disc of gas, which will lead, through the \emph{in situ} star formation, to the build up of a stellar rotating disc.
In contrast, early-type (ellipticals) galaxies are thought to be the product of numerous galaxy mergers that lead to dispersion-dominated remnants~\citep{toomre&toomre72}, hereby dominated by an \emph{ex situ} component of stars.

In this cosmological scenario, gas accretion is a very efficient process, in particular at high redshift, where gas is funnelled in cold cosmic streams of gas towards the centre of dark matter (DM) haloes
\citep{binney77, keresetal05, dekel&birnboim06, ocvirketal08, dekeletal09}, and contribute to the rapid disc built-up at high redshift~\citep{pichonetal11,kimmetal11,stewartetal13,danovichetal15}.
As haloes increase their mass, the gas temperature increases and the cosmic gas, which is not able to efficiently cool, enters the haloes in a hot and diffuse fashion due to shocks~\citep[e.g.][]{birnboim&dekel03}.
Nonetheless, this hot circum-galactic gas trapped at the core of DM haloes, has short cooling times compared to a typical Hubble time, and a cooling catastrophe would occur if it were not for any feedback process.
Therefore, two main mechanisms have been proposed in order to account for the inefficient star formation witnessed for both the low-mass dwarf galaxies and for the massive ellipticals at the centre of groups and clusters, feedback from Supernovae (SNe)~\citep{dekel&silk86} and from Active Galactic Nuclei (AGN)~\citep{silk&rees98} respectively.
It has been shown that the sharp decrease at the high-mass end of the galaxy stellar mass function is not achievable with SN feedback alone without destroying the less massive galaxies, leading the predicted faint-to-intermediate galaxy stellar mass function to be in stark disagreement with observations~\citep{oppenheimeretal10, daveetal11}. Therefore, it motivates the use of a completely different feedback process amongst which AGN feedback is the most plausible~\citep[e.g.][]{silk&rees98, crotonetal06}.

Mergers between galaxies explain the rapid morphological transformation from disc-like to elliptical-like galaxies, and this is being possible only if very little gas is still present after the interaction, otherwise, the remnant will rebuild its stellar disc.
Several cosmological simulations have established that massive galaxies are preferentially built through ex situ mass acquisition (i.e. mergers) instead of in situ star formation~\citep{oseretal10, lackneretal12, duboisetal13, lee&yi13, rodriguez-gomezetal16}, even though there is still little consensus on the exact amount of ex situ mass contribution.
In particular, those of~\cite{oseretal10} lead to similarly large levels of ex situ fractions for massive galaxies as in~\cite{duboisetal13} (or~\citealp{rodriguez-gomezetal16}) despite the absence of AGN feedback in the former.

Another strong evidence of the important role of mergers for massive ellipticals is that
galaxies appear compact at high redshift ($z>2$), and are more extended in the local Universe by a factor 3 to 5 at constant stellar mass~\citep{daddietal05, trujilloetal06, cimattietal08, franxetal08, vanderweletal08, vandokkumetal10}.
This evolution in size can be explained by the succession of mergers~\citep{khochfar&silk06, bournaudetal07,nipotietal09,naabetal09,oseretal12,duboisetal13,shankaretal13,welkeretal16discs}.
Since massive ellipticals are the product of mergers, satellites (and in particular low-mass satellites) are torn apart by tidal forces, and preferentially deposit their stars in the outskirts of the remnant~\citep[e.g.][]{hilzetal13}.
However, for the mergers to be efficient enough to turn the remnant into an elliptical galaxy and increase its size, it needs to be dissipation-less, i.e. contain a low amount of cold star-forming gas, or, otherwise will decrease the galaxy size~\citep{khochfar&silk06,wellonsetal15,welkeretal16discs}.

This view is supported by direct observational evidence, where disturbed features are commonly witnessed for all morphological types~\citep{schweizer&seitze88,vandokkum05,ducetal15}.
Also the fraction of galaxies in interaction is higher for more massive galaxies and the accompanying burst of star formation is lower with decreasing redshift~\citep{bundyetal09}, even though it remains moderate even at the peak epoch of star formation ($z\sim2$) with a factor of 2 enhancement compared to non-interacting galaxies~\citep{kavirajetal13,kavirajetal15}.
Therefore, it indicates that mergers are an important ingredient of the growth of massive galaxies and that their role increases with time.

With the advent of large-sized ($\sim 100\,\rm Mpc$) cosmological hydrodynamical simulations including the necessary physics to produce a morphological mix of galaxies such as \hagn~\citep{duboisetal14},  {\sc Illustris}~\citep{vogelsbergeretal14}, or {\sc Eagle}~\citep{schayeetal15}, it now becomes feasible to investigate in detail, and with statistical significance, the reasons of the morphological transformations of galaxies.

The aim of this paper is to highlight the role of AGN feedback in shaping the transformation of massive galaxies into ellipticals by tuning the amount of in situ versus ex situ origin of stars. 
For that purpose we analyse the state-of-art hydrodynamical cosmological simulations \hagn\,~\citep{duboisetal14} and its companion \hnoagn, with strictly identical initial conditions and physics but for a prescription for AGN feedback which is included in the former one but not in the latter.
This joint publication is one part in four papers which investigates the properties of \hagn, focusing respectively on the 
redshift evolution of the mass and luminosity functions \citep{kavirajetal16}, the AGN population \citep{volonterietal16} and the inner density profiles of both galaxies and DM haloes (Peirani et al., in preparation).
The other papers can be seen as complementary validation for the core properties of its synthetic galaxies.
In brief, it was shown that \hagn\, matches well the observed cosmic evolution of all these properties, with exception at the low-mass end where resolution effects are the strongest.
Here we will focus on {\sl understanding} how the 
 interplay between smooth gas infall on to galaxies and galaxy mergers is regulated by AGN feedback,
which produces the observed morphological diversity of galaxies by preventing in situ star formation 
from washing out the effect of mergers.

The outline of this paper is as follows.
Section~\ref{section:simulation} briefly presents the simulation, and the cross-identification of galaxies in the two simulations.
Section~\ref{section:morpho} 
shows the impact of AGN feedback on morphology.
Section~\ref{section:star-formation} investigates the changes on 
the in situ star formation and the ex situ mass acquisition.
In Section~\ref{section:morphotoinsitu}, we make a direct connection between the morphometric changes of galaxies due to AGN feedback and the in situ versus ex situ mass budget.
Section~\ref{section:mergers} illustrates the morphological transformation of galaxies during mergers and how the AGN feedback freezes the post-merger morphology.
Section~\ref{section:conclusion} provides the final conclusions.

\begin{figure*}
 \includegraphics[width=0.95\hsize]{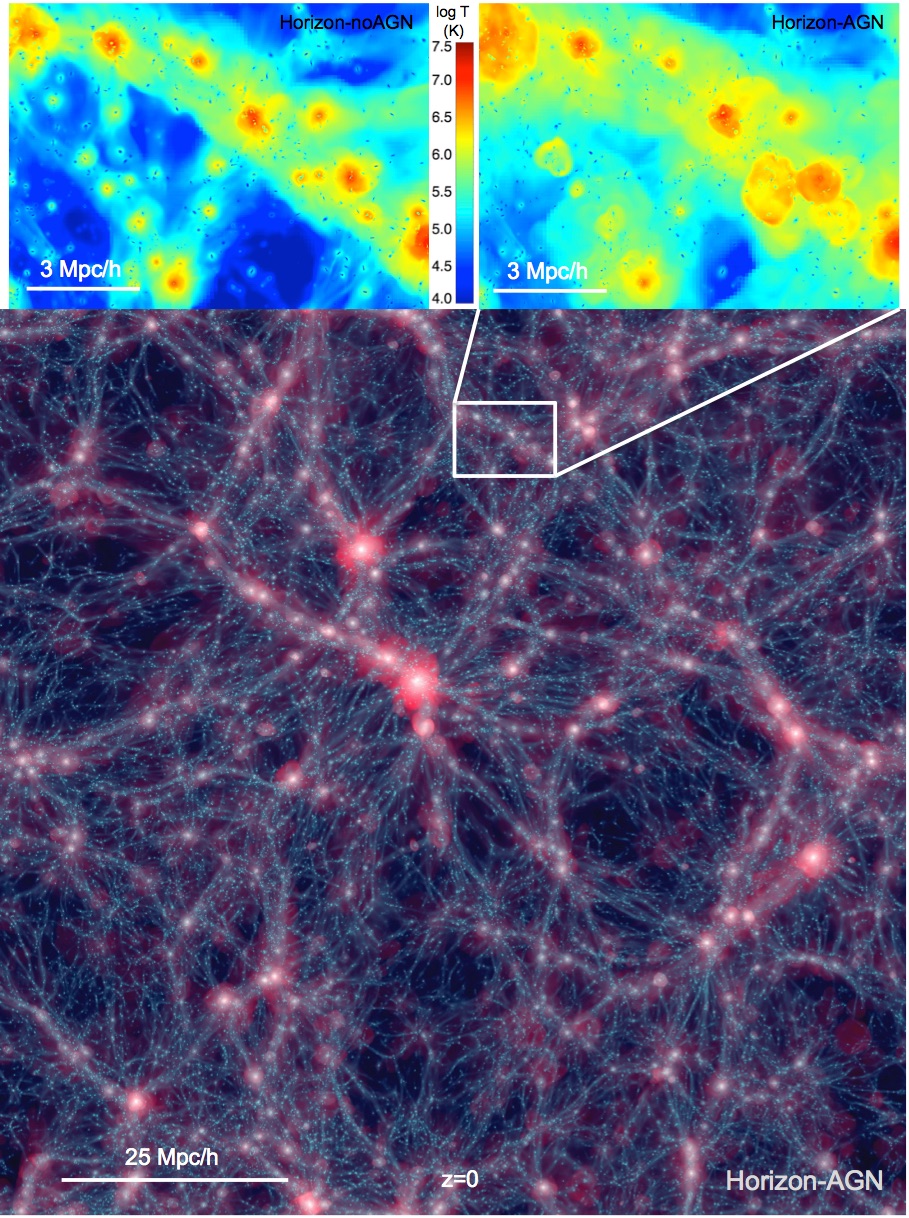}
 \caption{Main large panel: projection of \hagn\, at $z=0$ with the gas density in silver blue and the gas temperature in red. The depth of the projection is $25 \, h^{-1}\,\rm Mpc$ and the box extent on the side is $100\,h^{-1}\,\rm Mpc$. Top two panels: temperature maps of \hnoagn\, (left) and \hagn\, (right) zoomed into a small region of the box. }
\label{fig:nice}
\end{figure*}

\section{Numerical Methods}
\label{section:simulation}

We first briefly introduce the two simulations employed in this work \hagn\, and \hnoagn, and the procedure we use to identify haloes and galaxies.

\subsection{Horizon-AGN simulations}

The details of \hagn\,, which we only briefly describe here, can be found in~\cite{duboisetal14}. The \hagn\, simulation is run in a $L_{\rm box} = 100\, h^{-1}\,\rm Mpc$ cube with a $\Lambda$CDM cosmology with total matter density $\Omega_{\rm m}=0.272$, dark energy density $\Omega_\Lambda=0.728$, amplitude of the matter power spectrum $\sigma_8=0.81$, baryon density $\Omega_{\rm b}=0.045$, Hubble constant $H_0=70.4 \, \rm km\,s^{-1}\,Mpc^{-1}$, and $n_s=0.967$ compatible with the 7-year Wilkinson Microwave Anisotropy Probe data~\citep{komatsuetal11}. The total volume contains $1024^3$ DM particles, corresponding to a DM mass resolution of $M_{\rm DM, res}=8\times10^7 \, \rm M_\odot$, and initial gas resolution of $M_{\rm gas,res}=1\times 10^7 \, \rm M_\odot$. It is run with the {\sc ramses} code~\citep{teyssier02}, and the initially coarse $1024^3$ grid is adaptively refined down to $\Delta x=1$ proper kpc, with refinement triggered in a quasi-Lagrangian manner: if the number of DM particles becomes greater than 8, or the total baryonic mass reaches eight times the initial DM mass resolution in a cell.

Heating of the gas from a uniform UV background takes place after redshift $z_{\rm reion} = 10$ following~\cite{haardt&madau96}. Gas can cool down to $10^4\, \rm K$ through H and He collisions with a contribution from metals using rates tabulated by~\cite{sutherland&dopita93}. 
Star formation occurs in regions of gas number density above $n_0=0.1\, \rm H\, cm^{-3}$ following a Schmidt law: $\dot \rho_*= \epsilon_* {\rho_{\rm g} / t_{\rm ff}}$, where $\dot \rho_*$ is the star formation rate mass density, $\rho_{\rm g}$ the gas mass density, $\epsilon_*=0.02$ the constant star formation efficiency, and $t_{\rm ff}$ the local free-fall time of the gas. The stellar mass resolution is $2\times 10^6 \,\rm M_\odot$. Feedback from stellar winds, supernovae Type Ia and Type II are included into the simulation with mass, energy and metal release.

The formation of black holes (BHs) is implemented in \hagn. They can grow by gas accretion at a Bondi-capped-at-Eddington rate and coalesce when they form a tight enough binary. BHs release energy in a heating or jet mode (respectively ``quasar'' and ``radio'' mode) when the accretion rate is respectively above and below one per cent of Eddington, with efficiencies tuned to match the BH-galaxy scaling relations at $z=0$~\citep[see][for details]{duboisetal12}.

Finally, a simulation called \hnoagn\, was performed with no BH formation and, therefore, no AGN feedback, but with the exactly same set of initial conditions and sub-grid modelling.
This reference simulation is used to assess the importance of AGN feedback on galaxy properties.

\hagn\, and \hnoagn\, have both $6.6\times 10^9$ leaf cells (i.e. gas resolution elements) at $z=0$ (for comparison, {\sc Illustris}, \citealp{vogelsbergeretal14}, {\sc Eagle}, \citealp{schayeetal15}, and {\sc MassiveBlack-II}, \citealp{khandaietal15} have respectively $5.3\times 10^9$ gas cells, $3.4\times 10^9$, $5.8\times 10^9$ gas particles), and $\sim 10$ and $4$ million CPU hours, respectively, have been used to reach the end of each simulation.

Fig.~\ref{fig:nice} shows a large-scale projection of the gas distribution with density and temperature in the \hagn\, simulation at $z=0$, with two insets of zoomed regions with the gas temperature distribution in \hagn\, and \hnoagn. 
From this simple visual inspection, it is quite clear that AGN feedback manifests on the scale of dark matter haloes by driving hot gas out of DM haloes.

\subsection{Identification of halos and galaxies}
\label{section:catalogs}

In order to extract DM haloes from the simulation, we use the AdaptaHOP halo finder from~\cite{aubertetal04}.
For the positioning of the centre of the DM halo, we start from the first AdaptaHOP guess of the centre (densest particle in the halo) and from a sphere of size the virial radius of the halo, we use a shrinking sphere~\citep{poweretal03} by recursively finding the centre of mass of the halo within a sphere 10\% smaller than the previous iteration. 
We stop the search once the sphere has a size smaller than $3 \Delta x$ and take the densest particle in the final region.
20 neighbours are used to compute the local density. 
Only structures with density larger than 80 times the average total matter density and with more than 100 particles are taken into account.
The galaxy stellar mass $M_{\rm s}$ used throughout the paper is the total stellar mass of one galaxy given by the galaxy finder, while the halo mass $M_{\rm h}$ is the DM halo virial mass that satisfies the virial equilibrium.

The original AdaptaHOP finder is applied to the stellar distribution in order to identify galaxies with more than 50 particles, e.g. with a minimum galaxy stellar mass of $10^8 \,\rm M_\odot$.
Note that this identification efficiently removes the substructures in DM haloes and in galaxies.
Therefore, for galaxies, satellites which are embedded in the intra-cluster light of a massive galaxy are naturally taken out of the mass of the galaxy.
Merger trees of galaxies are build with merger tree construction from~\cite{tweedetal09} from the catalogue of galaxies built on 56 outputs at different times equally spaced in time ($\sim 250\, \rm Myr$).

In order to quantify the role of mergers in shaping elliptical galaxies, we compute the fraction of in situ star formation in galaxies.
We walk back the merger tree to follow the main progenitor branch and compute the amount of new stars formed in situ between two time steps of the simulation. 
Thus, the in situ mass of stars formed at $z=0$ is the cumulative amount of the mass of stars formed in the main branch of the merger tree, and, consequently, the ex situ component is obtained from the remaining stars not formed in the main branch (i.e. stars acquired by mergers). 
We can define the mass fraction of in situ formed stars $f_{\rm in situ}$ in a galaxy as the ratio of mass of in situ-formed stars above a given redshift over the stellar mass of the galaxy at this redshift.
With this definition of in situ versus ex situ stars, their amount does \emph{not} depend on the count of the various branches in the tree since it only depends on how the main progenitor is followed over time.
Note that we have varied the number of outputs used to generate the merger trees of galaxies by a factor of 2, and also used a lower number of star particles to detect galaxies, and this does not affect the measurement of the in situ star formation that will be shown later.

\begin{figure*}
\center \includegraphics[width=2.1\columnwidth]{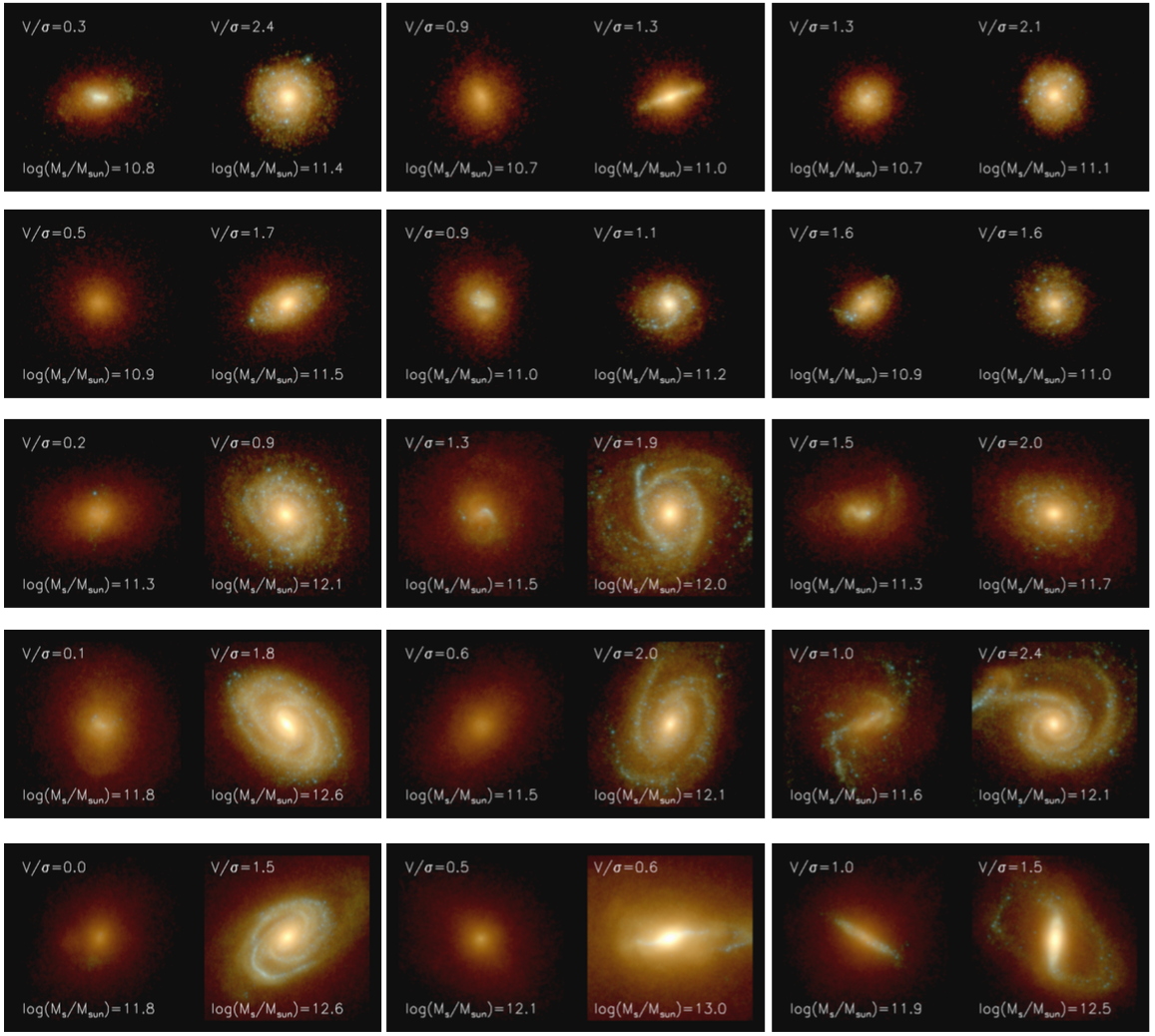}
 \caption{Composite images of the stellar emission in $u-g-i$ bands of a few selected massive galaxies in \hagn\, (first, third and fifth columns) matched with their corresponding galaxy in \hnoagn\, (second, fourth and sixth columns) at $z=0$. The galaxy mass is indicated in the panels and increases in \hagn\, from top to bottom. The ratio of rotation to dispersion of stars is also indicated in each panel, and increases in \hagn\, from left to right. Each thumbnail extends to 100 kpc in the vertical direction. It is clear from these images that the absence of AGN feedback in \hnoagn\, strongly favours the formation of disc-like blue galaxies with spiral arms for massive galaxies, while in the AGN feedback in \hagn\, red ellipticals are favoured for massive galaxies.}
\label{fig:composite}
\end{figure*}

At a few places (Figs.~\ref{fig:composite}, \ref{fig:individuals}, and  \ref{fig:individualsmaps}), we had to match the galaxies between the two simulations.
For this matching procedure, we first associate the DM haloes between \hagn\, and \hnoagn\, with haloes having more than 75\% of their particles with the same identity.
Note that in doing so, there is a small fraction of objects that cannot
be matched, in particular, sub-structures that got their particles stripped in the main halo host at a different time and with a different intensity in the two simulations (see Peirani et al., in preparation, for more details).
Once haloes are identified, we associate galaxies to DM haloes as the most massive galaxy within $0.1r_{\rm vir}$ of the halo starting from main haloes with decreasing mass, and then using sub-haloes with decreasing substructure hierarchy once the previous halo (sub-halo) hierarchy is completed.
Finally, the galaxies are matched through their halo hosts.

\section{Galaxy Morphologies}
\label{section:morpho}

\subsection{Visual inspection}

Fig.~\ref{fig:composite} shows a few examples of galaxies randomly selected in the stellar mass ($M_{\rm s}$ is the total galaxy stellar mass obtained by running the galaxy finder) range $5\times 10^{10}\le M_{\rm s}\le10^{12}\,\rm M_\odot$ in \hagn\, and compared with their matched galaxy in \hnoagn.
These images are composite images in $u$, $g$ and $i$ filter bands of the light of stars simulated from their mass, age and metallicity using single stellar population models from~\cite{bruzual&charlot03} assuming a Salpeter initial mass function.
For simplicity, the absorption of light by dust is not taken into account.
From this figure, it is clear that galaxies have endured a morphological transformation from a preferential disc morphology in \hnoagn\, to a preferentially elliptical morphology with AGN feedback in \hagn.
This sample of galaxies suggests that massive galaxies in \hagn\, appear rounder, redder and smoother than they are in \hnoagn.
In the following sections, we will quantify this morphological transformation and show how this is caused by the action of AGN activity and driven by mergers of galaxies.

\subsection{Galaxy kinematics at $z=0$}

\begin{figure}
\center \includegraphics[width=0.995\columnwidth]{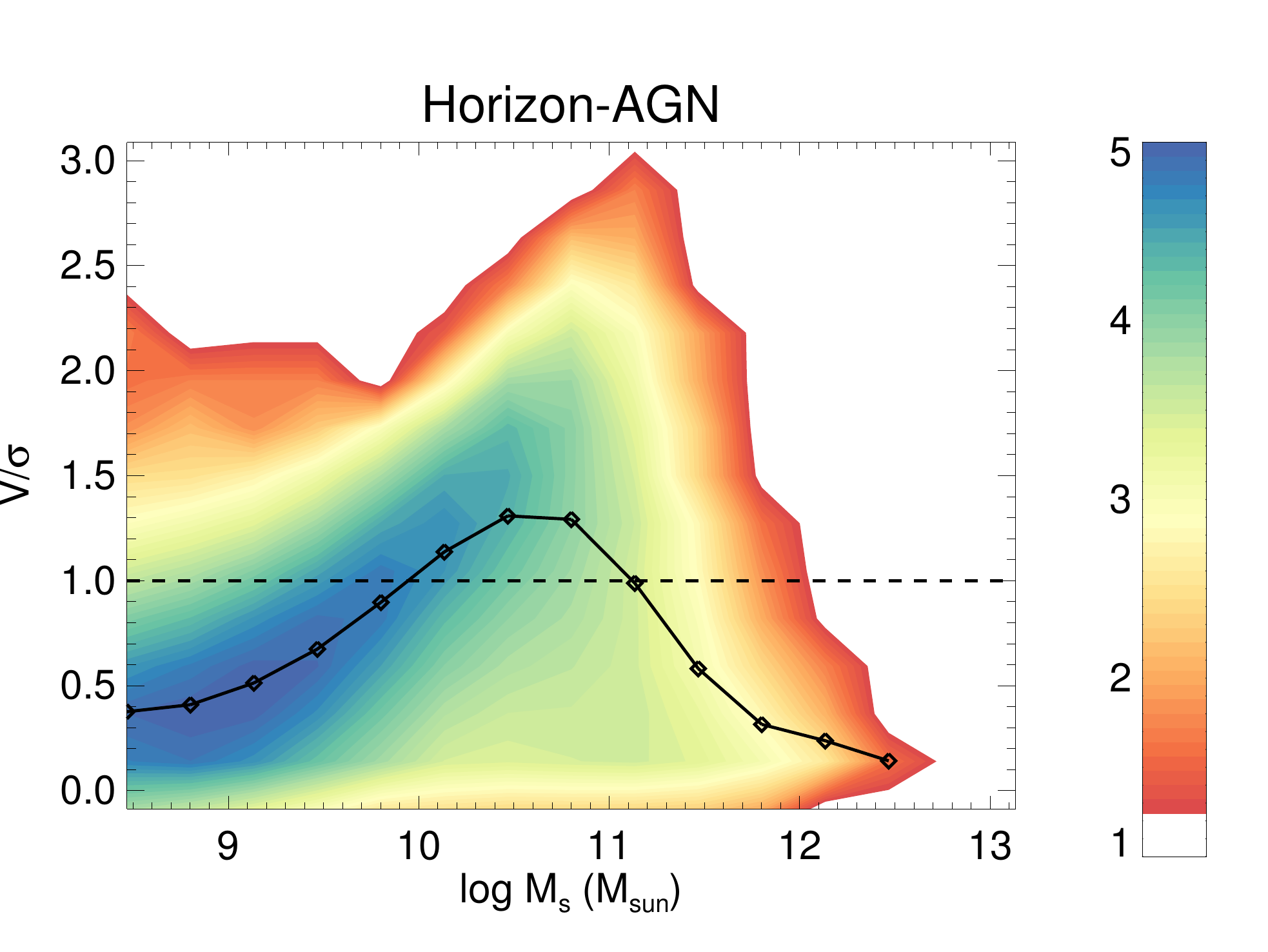}\vspace{-0.6cm}
\center \includegraphics[width=0.995\columnwidth]{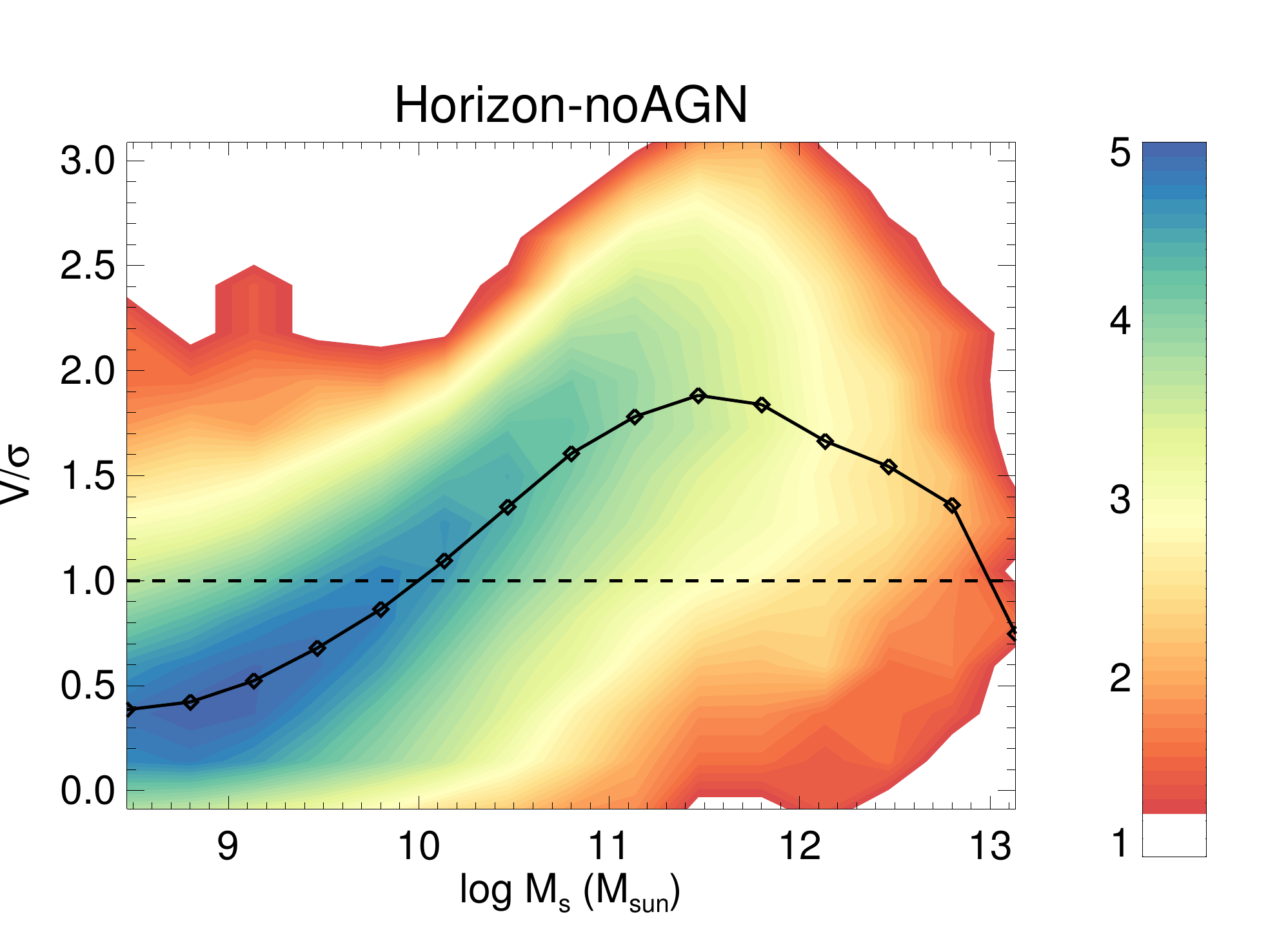}
 \caption{Kinematics-mass distribution of galaxies in the \hagn\, and \hnoagn\, at $z=0$, respectively top and bottom panels. The black solid lines correspond to the average $V/\sigma$ as a function of galaxy stellar mass. The colour scale represents $\log [N_{\rm gal}/\rm d\log M_{\rm s}/d(V/\sigma)]$. Galaxies with stellar mass larger than $M_{\rm s}>5\times 10{11}\,\rm M_\odot$ have significantly smaller values of $V/\sigma$ in \hagn\, than in \hnoagn. With AGN feedback massive galaxies are dispersion-dominated.}
\label{fig:vsig}
\end{figure}

\begin{figure}
\center \includegraphics[width=0.995\columnwidth]{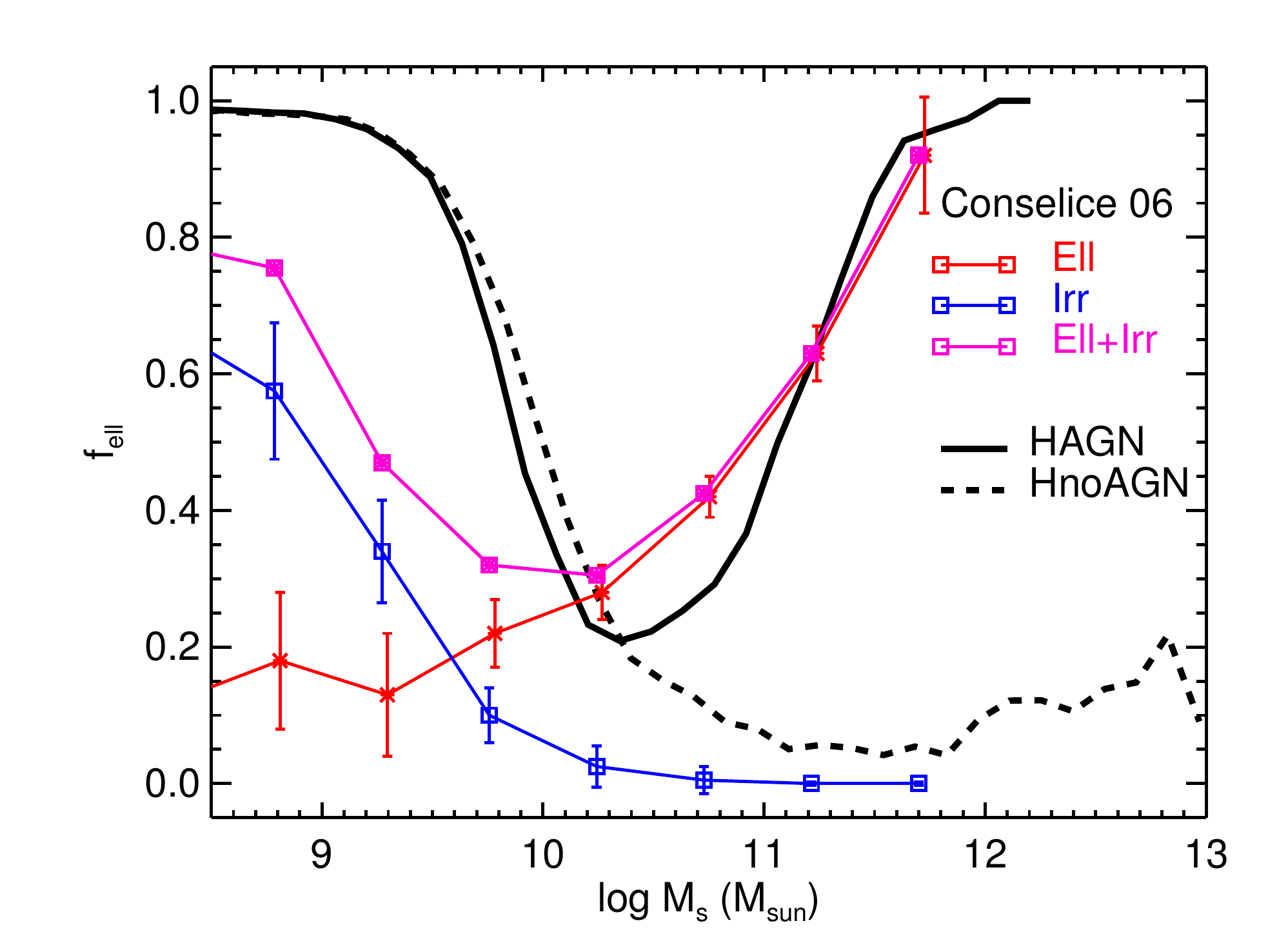}
 \caption{Fraction $f_{\rm ell}$ of elliptical galaxies (with $V/\sigma<1$) as a function of galaxy stellar mass at $z=0$ in the \hagn\, and \hnoagn\, simulations, respectively the black solid lines and the dashed lines. For comparison, the data points from~\citet{conselice06} of the fraction of ellipticals (red), irregulars (blue), and elliptical plus irregulars (magenta) are reported in the top panel. We see that the population of the most massive galaxies embedded in the most massive haloes is clearly dominated by elliptical galaxies in \hagn, whereas it is dominated by disc galaxies in \hnoagn.}
\label{fig:ellfraction}
\end{figure}

\begin{figure}
\center \includegraphics[width=0.995\columnwidth]{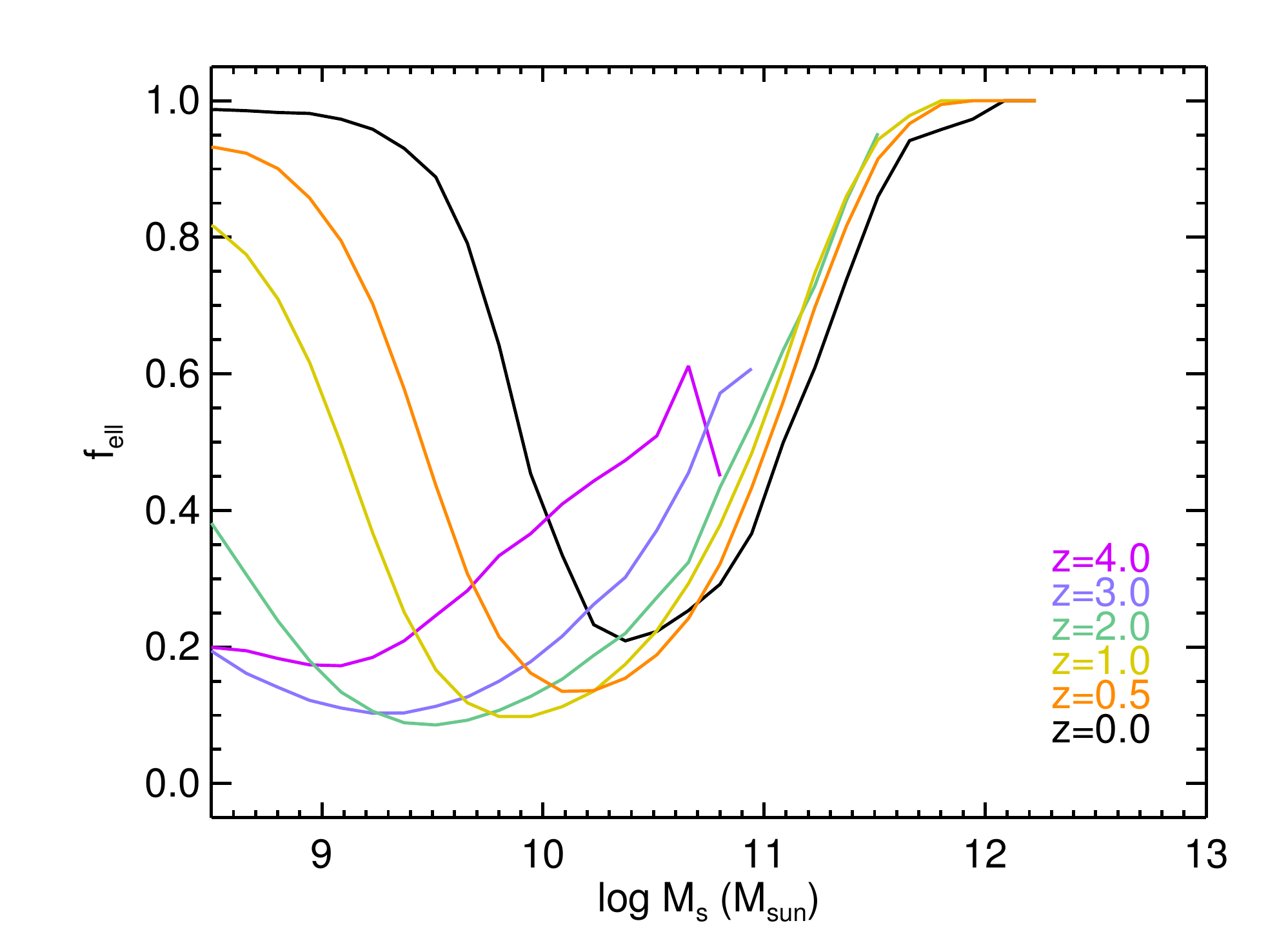}
 \caption{Same as Fig.~\ref{fig:ellfraction} for the \hagn\ only and measured at different redshifts as indicated by the different colours. The minimum of the fraction of ellipticals $f_{\rm ell}$ (maximum of fraction of discs) is found at lower stellar mass with increasing redshift.}
\label{fig:ellvsz}
\end{figure}

In order to infer galaxy morphology, we use their stellar kinematics as a proxy. 
We first find the galaxy spin by measuring its angular momentum vector from stars.
This spin vector defines the orientation of the z-axis cylindrical coordinate system within which we compute the radial, tangential and vertical velocity components of each stellar particle.
The rotational velocity $V$ of the galaxy is the average of the tangential velocity component.
The velocity dispersion is obtained from the dispersion of the radial $\sigma_{\rm r}$, the tangential $\sigma_{\rm \theta}$ and the vertical velocity $\sigma_{\rm z}$ components around their averaged values, i.e. $\sigma=(\sigma_{\rm r}^2+\sigma_{\rm \theta}^2+\sigma_{\rm z}^2)^{1/2}/3$.
In Fig.~\ref{fig:composite}, the ratio of $V/\sigma$, rotation over dispersion, is indicated in the top left of each vignette, and it is clear that the \hagn\, galaxies in the high-mass end have on average lower $V/\sigma$ ratios than those of \hnoagn\, galaxies, and with different amplitude variations.

Fig.~\ref{fig:vsig} shows the distribution of the ratio of $V/\sigma$ as a function of stellar mass for galaxies in \hagn\, and \hnoagn\, simulations at redshift $z=0$.
This ratio allows us to put a separation between rotation-dominated $V/\sigma >1$ and dispersion-dominated $V/\sigma <1$ galaxies, i.e. what we define discs and ellipticals respectively from now on.
We see that AGN feedback reduces significantly the formation of galaxies with stellar mass above $M_{\rm s}>10^{12}\,\rm M_\odot$.
In both cases, lowest-mass galaxies $M_{\rm s}\lesssim5\times 10^9\,\rm M_\odot$ are dispersion-dominated with no significant difference implied by the presence of AGN feedback.
Galaxies above this mass start to be rotation-dominated with a swift turn towards dispersion-dominated above $M_{\rm s}\gtrsim 10^{11}\,\rm M_\odot$ for \hagn\, and which appears, marginally, at much higher masses $M_{\rm s}\gtrsim 10^{13}\,\rm M_\odot$ for \hnoagn.
In the simulation without AGN feedback, massive galaxies are rotationally-supported disc galaxies, while AGN feedback allows for the formation of massive dispersion-dominated elliptical galaxies.

We measure the fraction of elliptical galaxies $f_{\rm ell}$ as a function of stellar mass in Fig.~\ref{fig:ellfraction}.
Above $2\times 10^{10}\,\rm M_\odot$, there is a clear change in the fraction of elliptical galaxies between the two simulations.
The absence of AGN feedback creates a population of massive galaxies dominated by discs with a disc fraction $f_{\rm d}=1-f_{\rm ell}$ of more than $90\%$ at all masses above $M_{\rm s}\gtrsim2\times 10^{10}\,\rm M_\odot$.
In contrast, the action of AGN feedback allows for a mixed population of discs and ellipticals in the stellar mass-range $2\times 10^{10} \lesssim M_{\rm s} \lesssim2\times 10^{11}\,\rm M_\odot$, where the fraction of elliptical galaxies increases with mass.
Above $M_{\rm s} \gtrsim4\times 10^{11}\,\rm M_\odot$, all galaxies are ellipticals in the \hagn\, simulation:  $f_{\rm ell}>95\%$.
The same trend is obtained as a function of halo mass (see Appendix~\ref{appendix}) since galaxy mass correlates with haloes mass.
The minimum of the probability of finding an elliptical (peak fraction of finding a disc) is obtained at a galaxy mass of $M_{\rm s}\simeq 2\times 10^{10}\, \rm M_\odot$ in \hagn\, (or a halo mass of $M_{\rm h}\simeq4\times 10^{11}\,\rm M_\odot$).
The fraction of ellipticals is also measured as a function of the halo mass in Appendix~\ref{appendix}, and similar trends are found: AGN feedback is responsible for the formation of a population of elliptical-like massive galaxies above the group-mass haloes $M_{\rm h}>5\times 10^{12}\,\rm M_\odot$ .

Compared to observations from~\cite{conselice06}, the fraction of ellipticals in \hagn\, as a function of stellar mass is in fairly good agreement with the data, especially for the most massive galaxies, which are better resolved.
The most noticeable difference with observations corresponds to the position of the minimum probability of observing ellipticals (or maximum probability of observing discs), which is found at higher galaxy stellar mass in \hagn.
Although we also find an increase in the fraction of ellipticals towards low-mass galaxies (below $2\times 10^{10}\,\rm M_\odot$) as in the observations (irregulars plus ellipticals), these galaxies are also those most affected by resolution; it is possible that the kinematics of the low-mass simulated galaxies is mostly driven by lack of resolution.
Note that this increase of ellipticals at the low-mass end also does not depend on AGN feedback since it is seen for both \hagn\, and \hnoagn.
On the one hand, it is expected that the effect of AGN feedback in this mass range is irrelevant thanks to supernova-quenching of BH activity~\citep{duboisetal15snbh, habouzitetal16}.
On the other hand, lack of resolution most certainly affects the possible growth of BHs and the impact of AGN feedback for this class of galaxies.
Despite those potential numerical shortcomings at the low-mass end, it is extremely encouraging that comparable trends are obtained in \hagn, especially in the intermediate- and high-mass range, and this reinforces the case that AGN feedback drives the morphological properties of this population.

\subsection{Evolution with redshift}

We now investigate the evolution in galaxy morphology as a function of redshift by repeating the measurement of the fraction of elliptical galaxies at higher redshifts in \hagn\, (Fig.~\ref{fig:ellvsz}).
At high redshift, massive galaxies are more likely to be elliptical of lower masses, because of a progenitor bias: massive ellipticals at low redshift are more likely to have progenitors that are ellipticals (and, conversely, disc galaxies at low redshift are more likely to have disc progenitors), and since their progenitors  at high redshift have lower masses, massive ellipticals have lower masses.
The detailed inspection of the progenitor bias of low-redshift galaxies will be covered in a forthcoming paper.
The U-shaped distribution at $z=0$ (with a minimum fraction of ellipticals found at $M_{\rm s}\simeq 2\times 10^{10}\, \rm M_\odot$; see Fig.~\ref{fig:ellfraction}) is also not fully recovered at the highest redshift ($z=3$ and 4).
At $z\gtrsim3$, the fraction of ellipticals flattens (or does not significantly increase) towards lower masses, and, hence, low-mass galaxies are more likely to be discs at high redshifts than at low redshifts.

Note that this is a possible indication that the tendency for low-mass galaxies to be ellipticals at low redshift is not entirely due to resolution  since galaxies of the  same mass  are more likely to be discs at high redshift.
As galaxies are more gas-rich at high redshift~\citep{santinietal14}, they are also more likely to have a disc of stars built from the net rotation of the gas, rather than to have a dispersion-dominated stellar disc.
Low-mass galaxies are also more likely to be part of a sub-halo at low redshift since structure formation is more advanced at later times. 
Thus, gas accretion rate on to low-mass galaxies should be relatively reduced at low redshift compared to its  high-redshift counterpart, hence the lower gas content and higher  dispersion in the stellar kinematics.
However, to clearly assess the role of resolution  versus evolutionary and environmental effects on low-mass galaxies, we would need a dedicated large-scale simulation (with smaller volume, improved spatial and mass resolution), which is clearly beyond the scope of this paper.
Nevertheless, the most massive galaxies already show a significant fraction of ellipticals at early times, e.g. 50 \% of $M_{\rm s}=5\times 10^{10}\,\rm M_\odot$ are ellipticals at $z=4$ and 3.

\begin{figure}
\center \includegraphics[width=0.995\columnwidth]{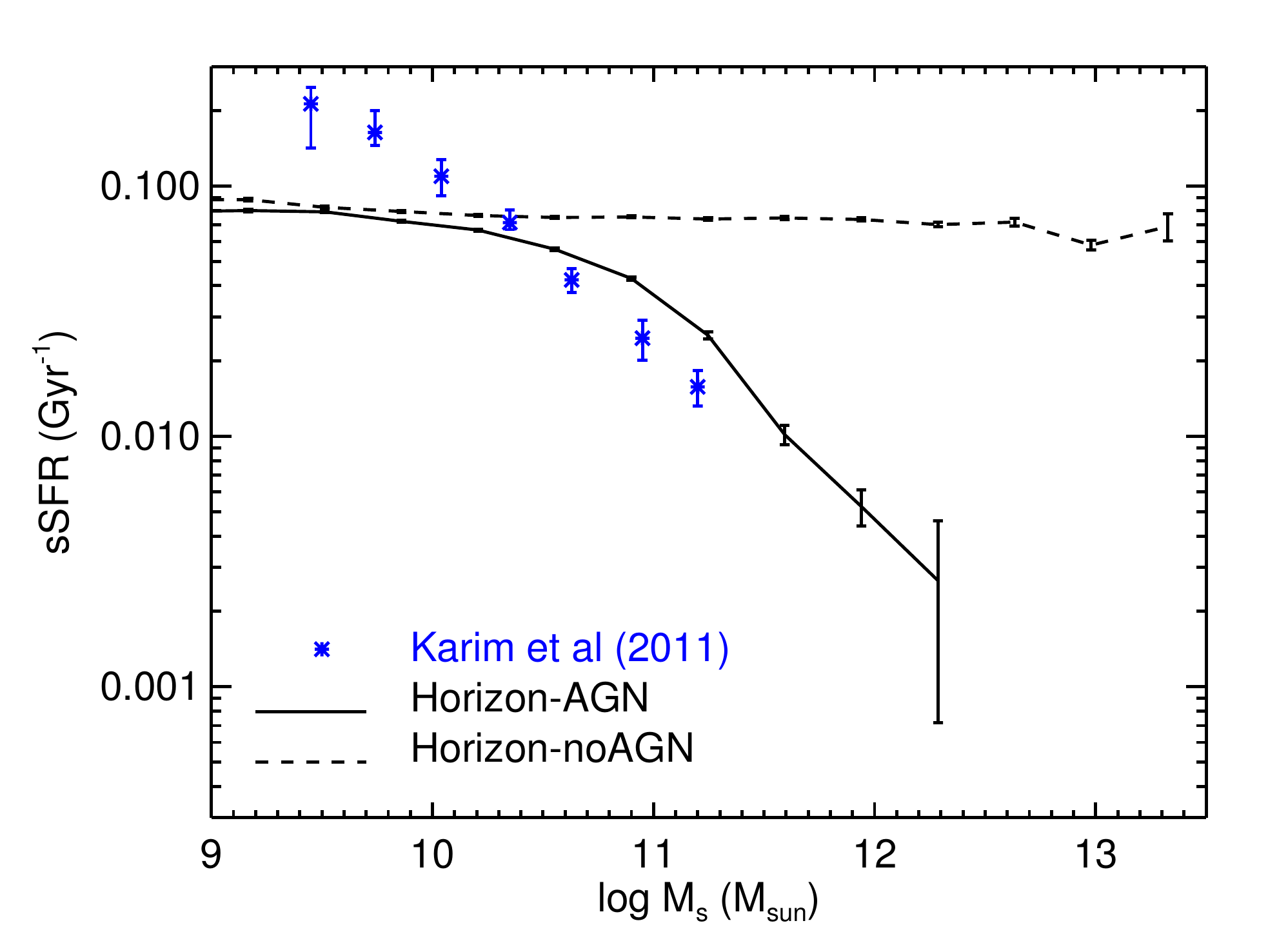}
 \caption{Mean value of the $\rm sSFR$ as a function of galaxy stellar mass for \hagn\, (solid) and \hnoagn\, (dashed) at $z=0.3$ with observations from~\citet{karimetal11} in blue at the same redshift. The error bars stand for the error around the mean. The presence of AGN feedback creates a population of passive galaxies at high stellar masses.}
\label{fig:ssfrvsmgal}
\end{figure}

\section{Star formation signatures}
\label{section:star-formation}

\subsection{Impact of AGN feedback on star formation}

In order to quantify the impact of AGN feedback on star formation activity, we measure the specific star formation rate $\rm sSFR=SFR/M_{\rm s}$ as a function of stellar mass.
In Fig.~\ref{fig:ssfrvsmgal}, we measure the $\rm sSFR$ with the amount of stars formed over the last $100\,\rm Myr$ at $z=0.3$.
Without AGN feedback, galaxies are, on average, actively forming stars along the main sequence with a sSFR$\simeq 0.08 \,\rm Gyr^{-1}$.
AGN feedback, in \hagn, allows for the formation of passive galaxies at the high-mass end with more than an order of magnitude decrease in $\rm sSFR$ for galaxies with $M_{\rm s}\gtrsim 5\times 10^{11} \,\rm M_\odot$ compared to \hnoagn.
Since AGN feedback can reduce the star formation in massive galaxies, we expect that the amount of stars formed in the main progenitor, i.e. formed in situ, will be decreased, and that the stellar mass budget will lean more towards merger-acquired stars (ex situ acquisition) than in the absence of AGN activity.

Compared to observations at $z=0.3$ from~\cite{karimetal11}, \hagn\, is in much better agreement than \hnoagn.
However, at the low-mass end (below $10^{10}\, \rm M_\odot$), simulated galaxies have up to a factor of 2 lower sSFR than in observations (similar in \hagn\ and \hnoagn), and have a constant sSFR with stellar mass, instead of a continuous decrease.
The agreement between \hagn\, and observations is much better above $10^{10}\, \rm M_\odot$: 
it shows a similar decrease with stellar mass, and the magnitude of the sSFRs is close, though galaxies have a larger residual star formation in the simulation.
On the contrary, in this high-mass range, \hnoagn\, fails at producing a population of passive galaxies.
The dispersion in the relation (not represented here) is large, in particular at the low-mass end (1 dex) due to the variety of galaxies being either galaxies of the field or satellites in groups or clusters of galaxies.
For a more complete overview of the redshift evolution of star formation activity of galaxies in \hagn\, and its comparison with observations, we refer the reader to~\cite{kavirajetal16}.

\subsection{In situ star formation vs accreted stars}

\begin{figure}
\center \includegraphics[width=0.995\columnwidth]{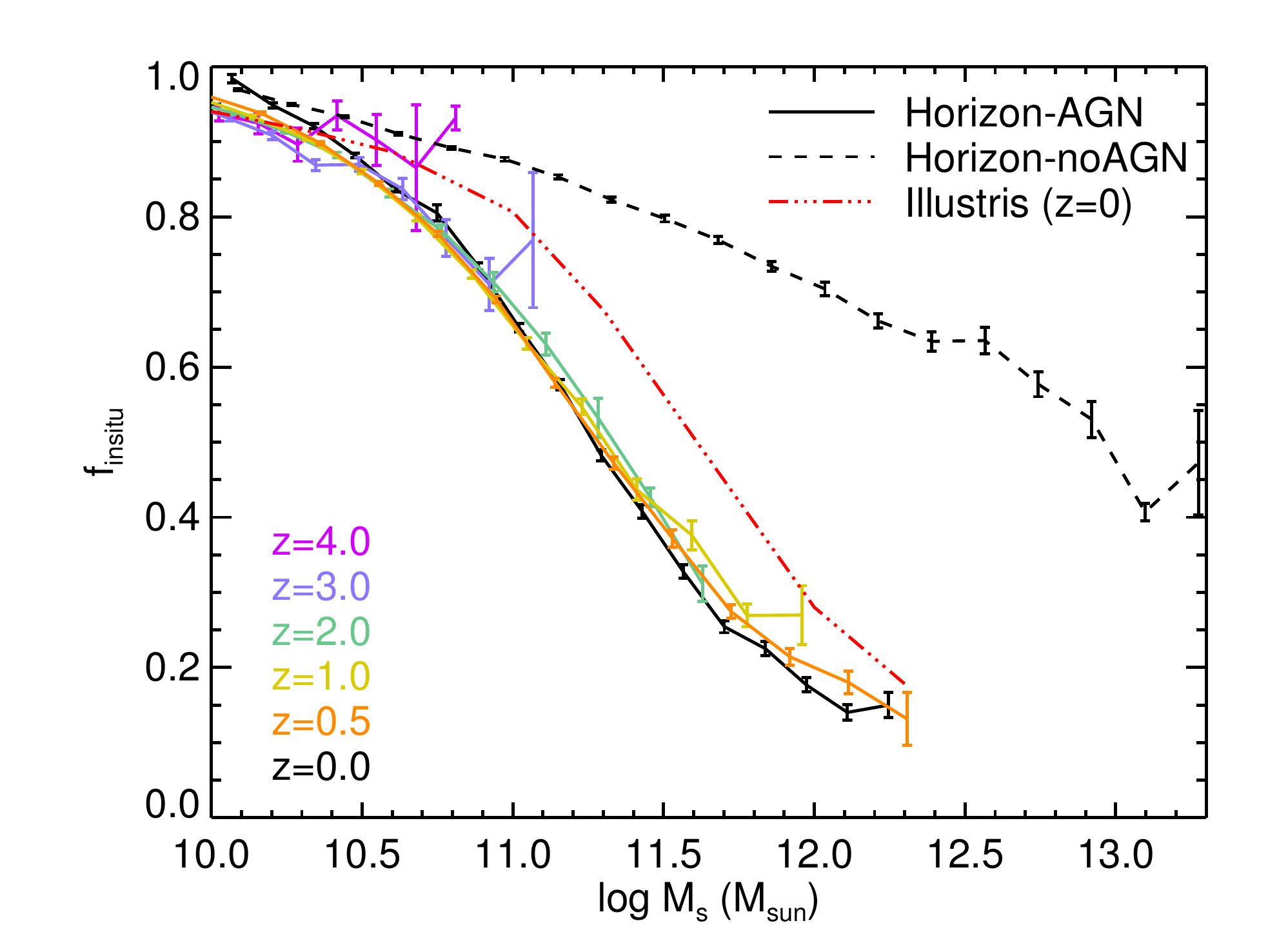}
 \caption{Mean fraction of in situ $f_{\rm insitu}$ formed stars as a function of stellar mass at $z=0$ for \hagn\, (solid) and \hnoagn\, (dashed) with errors around the mean indicated by error bars. The result from the Illustris simulation~\citet{rodriguez-gomezetal16} at $z=0$ is also shown in triple-dotted-dashed red for comparison. Both simulations show a decrease in $f_{\rm insitu}$ with $M_{\rm s}$, however, only in \hagn\, this fraction becomes negligible. This relation shows no significant evolution with redshift.}
\label{fig:finsitu}
\end{figure}

\begin{figure}
\center \includegraphics[width=0.995\columnwidth]{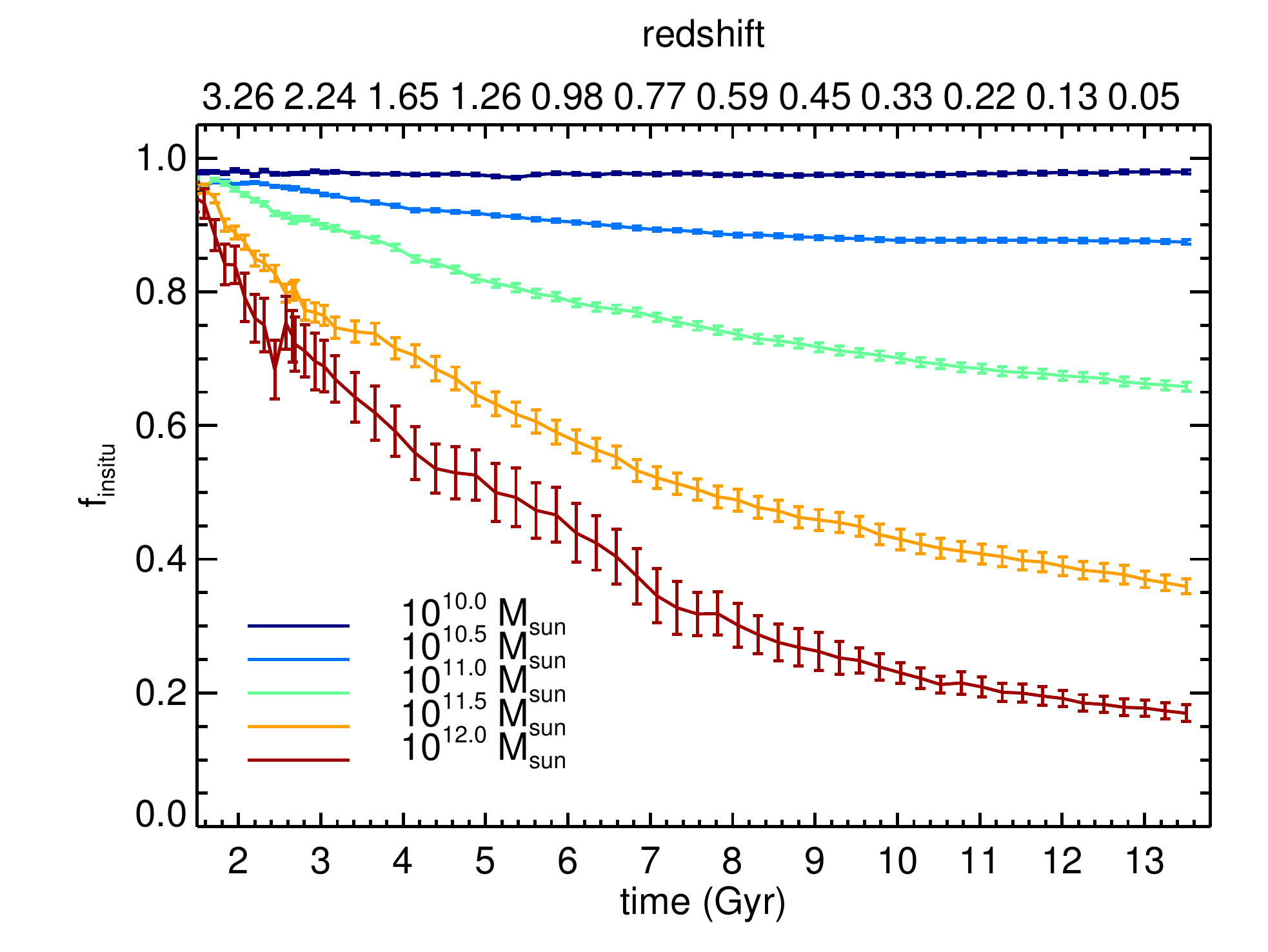}
 \caption{Mean in situ fraction as a function of time and redshift for various galaxy mass bins in the \hagn\, simulation. The galaxy stellar mass is determined at $z=0$ and we trace back the progenitors of these objects. The mass bin width is $\pm10\%$ of the indicated mass on the plot.}
\label{fig:finsituvsz}
\end{figure}

As the accreted cosmic gas settles into a rotating disc, the induced in situ star formation naturally leads to the formation of a stellar disc.
Conversely, mergers induce  violent relaxation of the stellar component -- even though accretion shares a preferential filamentary  direction when it penetrates the halo of the central galaxy, and a coplanar distribution as it gets closer to the central galaxy~\citep{welkeretal16plane} -- they increase significantly the velocity dispersion of their stars and turn discs into ellipticals.
This competition between in situ star formation and the (ex situ) stellar accretion  is what ultimately defines the final morphological fate of the galaxy.
Note that in the absence of feedback, the number of stars acquired by mergers can remain very large\footnote{At fixed halo mass, the ex situ \emph{mass} acquired by the central galaxy is larger in the absence of AGN feedback. However, the ex situ \emph{fraction} is lower than in the AGN feedback case.}, yet the unimpeded in situ star formation forces the galaxy to get back to disc-like configurations, leading to unrealistic disc-like morphologies for massive galaxies~\citep{duboisetal13}.

\begin{figure*}
\includegraphics[width=\columnwidth]{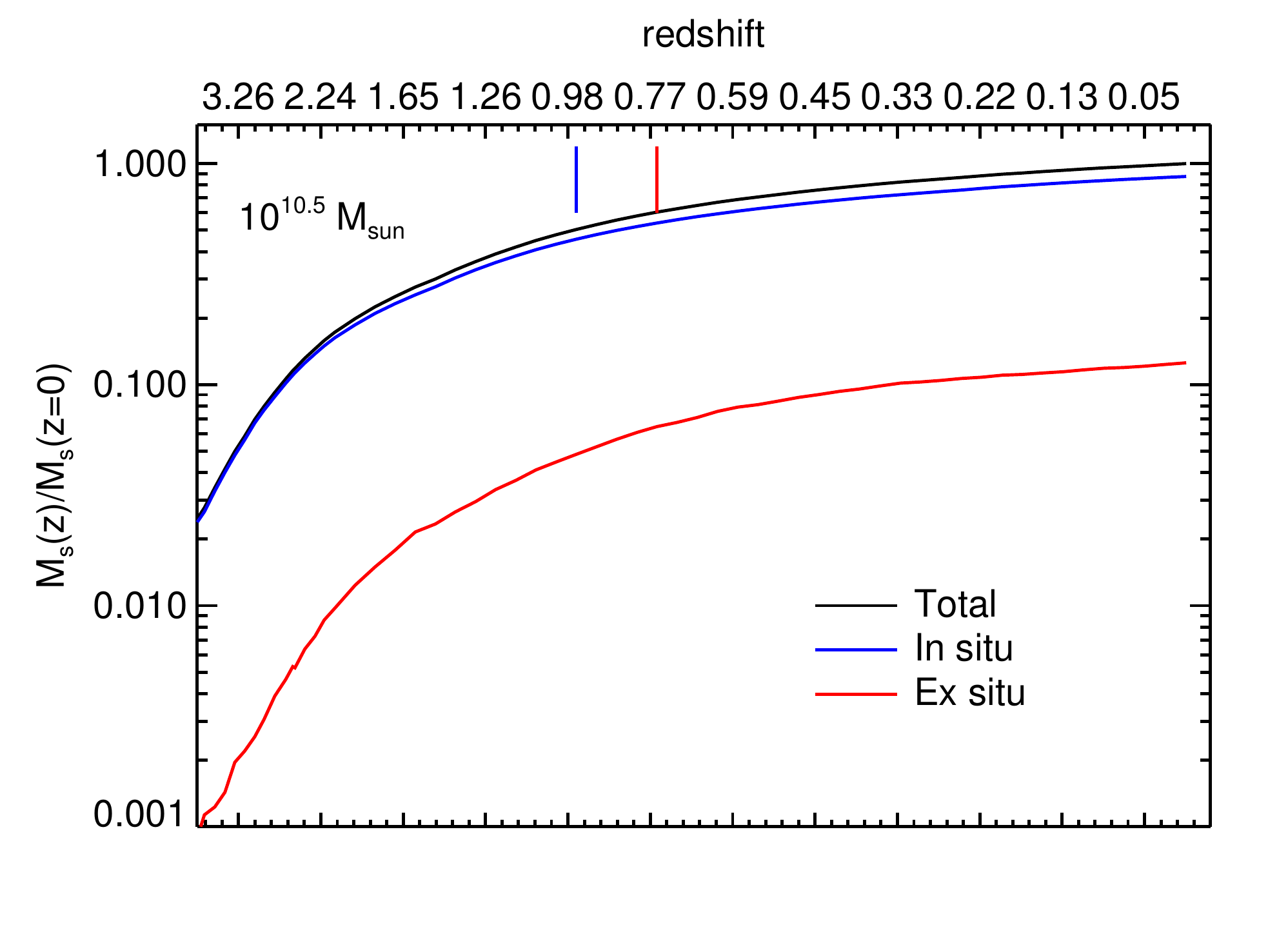}\hspace{-1.6cm}
\includegraphics[width=\columnwidth]{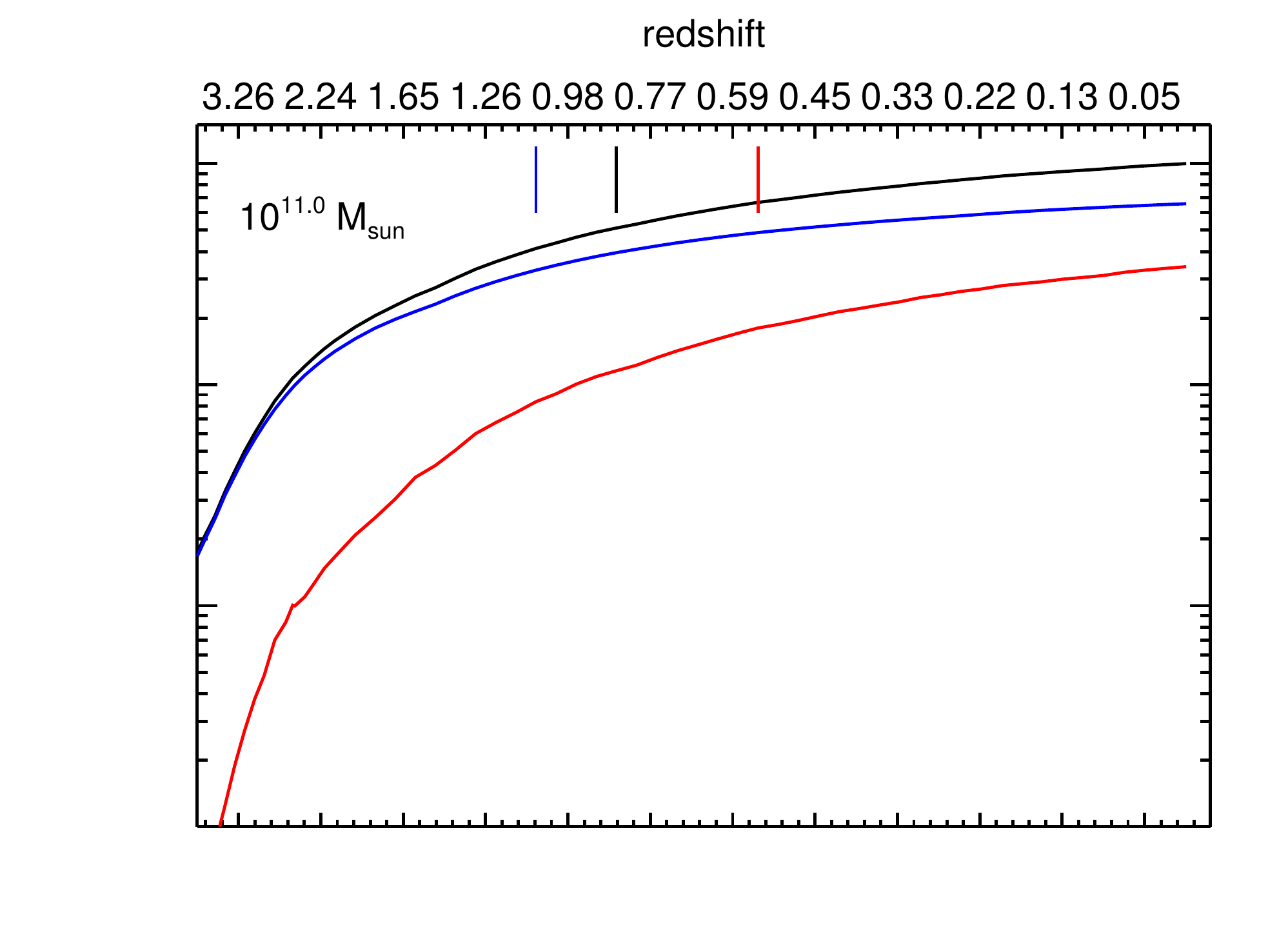}\vspace{-1.5cm} \\ 
\includegraphics[width=\columnwidth]{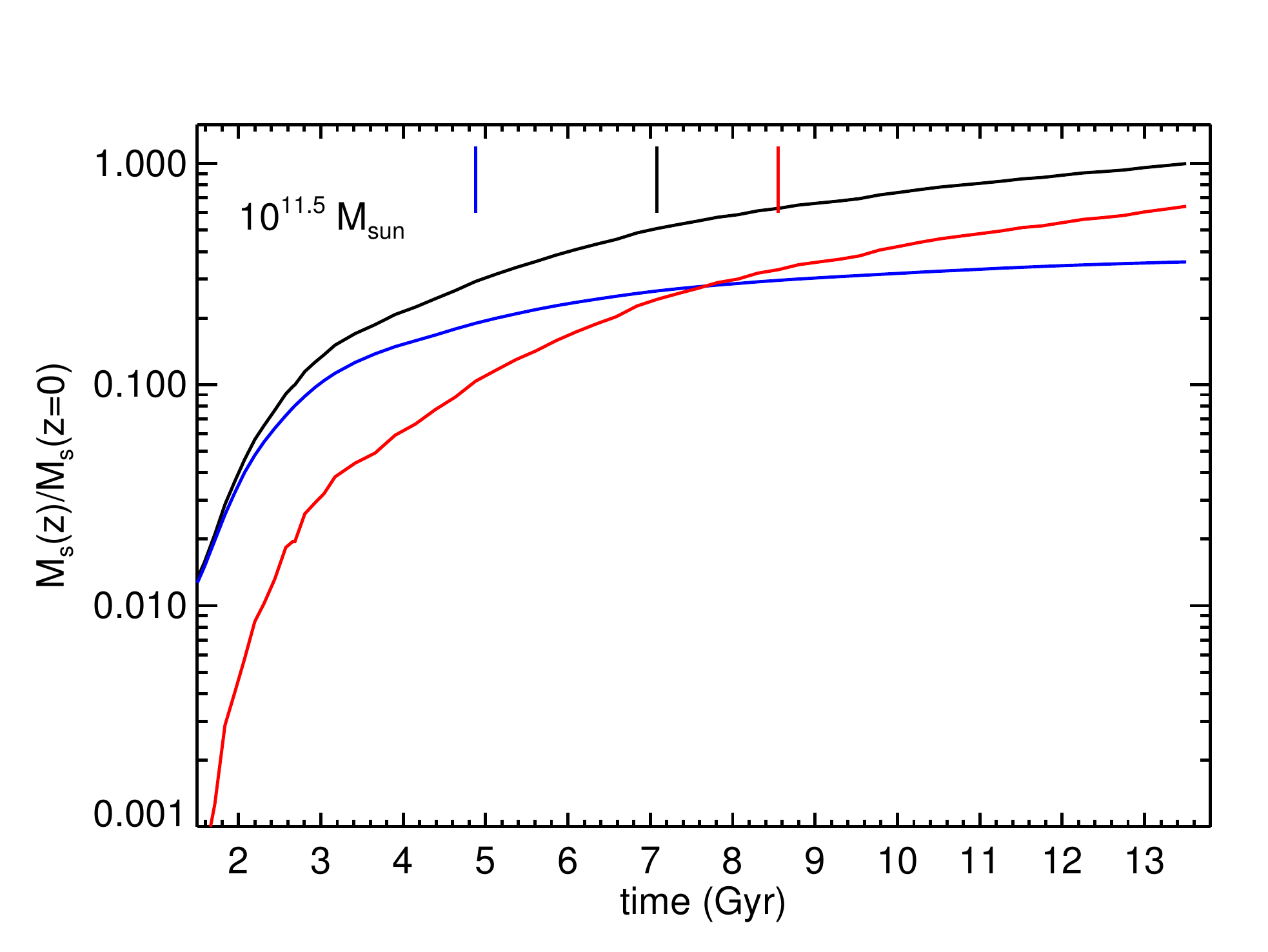}\hspace{-1.6cm}
\includegraphics[width=\columnwidth]{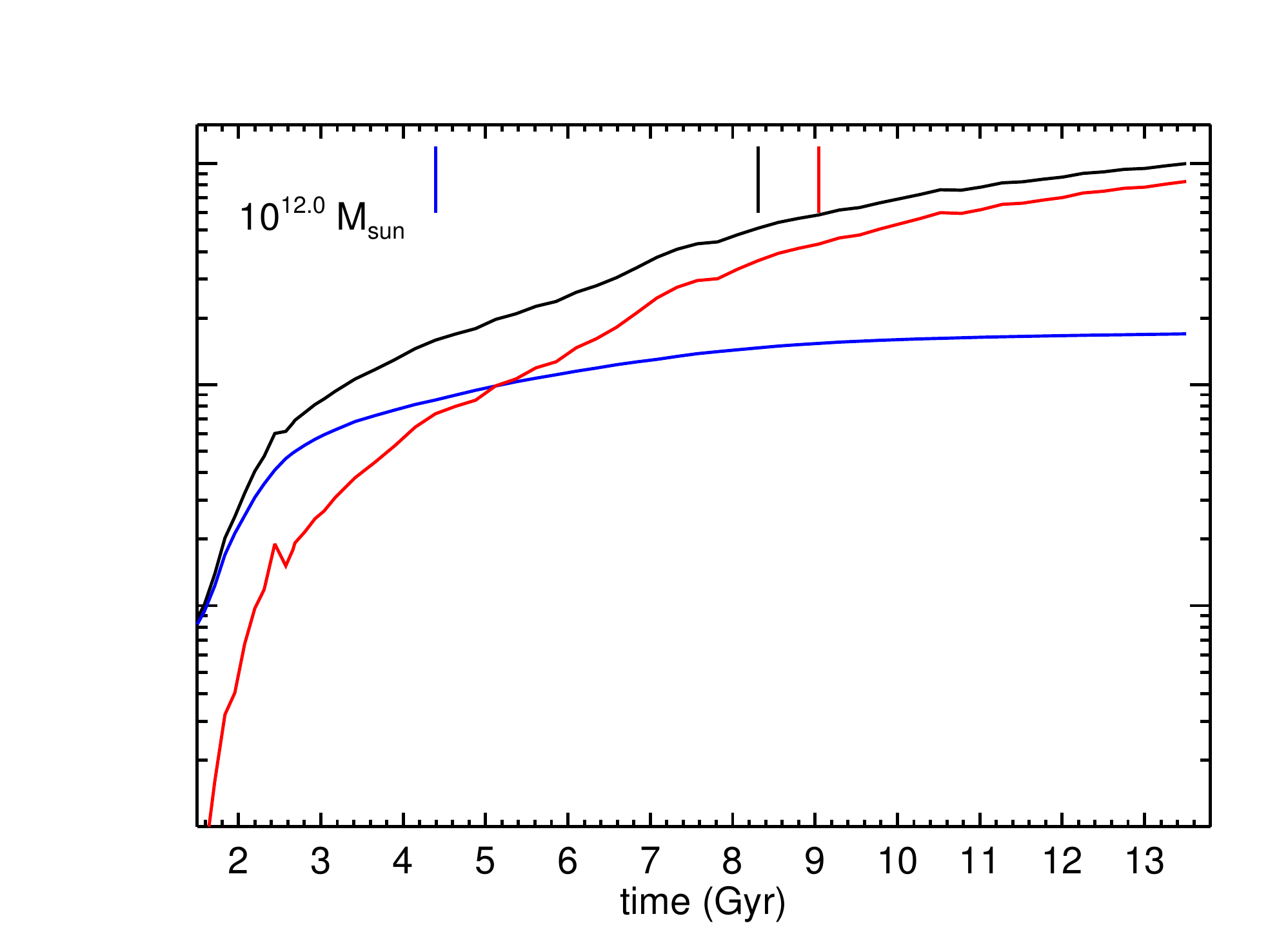}
 \caption{Mean stellar mass evolution in \hagn\, as a function of time and redshift normalised to its value at $z=0$ for the total stellar mass (black), in situ component (blue) or ex situ component (red) for different galaxy stellar masses at $z=0$ as indicated in the top left corner of each panel ($\pm 10\%$ of the indicated mass). Vertical bars indicate the time at which half of the corresponding mass component is acquired by the galaxy. We see that galaxies first form their stellar component from in situ star formation, and for the most massive ones assemble the ex situ component at late times. }
\label{fig:minsituvsz}
\end{figure*}

Fig.~\ref{fig:finsitu} shows the mean fraction of in situ formed stars $f_{\rm insitu}$ (see Section~\ref{section:catalogs} for definition) as a function of stellar mass at $z=0$ for the two simulations.
In both simulations, a decreasing trend of $f_{\rm insitu}$ as a function of $M_{\rm s}$ is observed.
However, the average value of $f_{\rm insitu}$ is always above 0.5 in the \hnoagn\, simulation at any mass. 
Therefore, on average, in the absence of AGN feedback, galaxies are dominated by in situ star formation,  become dominated by rotation, hence  look like disc galaxies whatever their mass.
Galaxies in the \hagn\, simulation show a stronger decrease for the in situ fraction as a function of stellar mass, reaching an average value of 0.1 for the most massive galaxies $M_{\rm s}\gtrsim 10^{12}\,\rm M_\odot$.
Above a galaxy mass of $M_{\rm s} >2\times 10^{11}\,\rm M_\odot$, those galaxies including AGN feedback end up with an average $f_{\rm insitu}$ below 0.5, which corresponds to the mass transition from disc-like galaxies to elliptical-like galaxies (see Fig.~\ref{fig:ellfraction}).
The in situ fraction as a function of stellar mass shows very little evolution with redshift (represented in colour in Fig.~\ref{fig:finsitu} only for \hagn\, for the sake of clarity, though \hnoagn\, shows also no redshift evolution).
Compared to previous numerical work~\citep{oseretal10, lackneretal12, duboisetal13, lee&yi13}, we find comparable trends: more massive galaxies form less  in situ stars, while  the transition from in situ-dominated to ex situ-dominated stellar content happens earlier on with more massive galaxies.
However, it is worth noting that~\cite{oseretal10} obtained similar levels of ex situ fraction {\it without} AGN feedback to our \hagn\, simulation, and this fundamental difference with our work (and that of \citealp{lackneretal12, duboisetal13, lee&yi13}) is probably due to a stronger prescription for SN feedback or a more fundamental difference between smoothed particle hydrodynamics and grid-based codes leading to different galaxy morphologies in absence of AGN as highlighted by~\cite{scannapiecoetal12}~\citep[see][for a discussion on the results from different simulations in terms of in situ versus ex situ mass content]{duboisetal13}.
We also show in Fig.~\ref{fig:finsitu} the result from the Illustris simulation~\citep{rodriguez-gomezetal16}, where the in situ fraction of formed stars is larger than in \hagn, though the trend with stellar mass is extremely similar.

We recall that the minimum galaxy stellar mass extracted by our galaxy finder is $10^8\,\rm M_\odot$ (50 star particles).
With that minimum mass threshold, mergers of galaxies $M_{\rm s}=10^{10}\,\rm M_\odot$ are complete down to merger mass ratios of $1:100$ at $z=0$.
However, galaxies of such a mass at $z=0$, have indeed lower progenitor stellar mass at higher redshift.
For example, galaxies with $M_{\rm s}=10^{10}\,\rm M_\odot$ have a mass of $5\times 10^8\,\rm M_\odot$ at $z=4$ (not shown here but see Fig.~\ref{fig:minsituvsz} for the behaviour at higher stellar mass).
Therefore, galaxies of $M_{\rm s}=10^{10}\,\rm M_\odot$ at $z=0$ are complete down to merger mass ratios of $1:5$ up to $z=4$.

Fig.~\ref{fig:finsituvsz} shows the evolution with time and redshift of the fraction of stars formed in situ for different galaxy masses selected at $z=0$.
The mass content of the progenitors of galaxies at $z=0$ is dominated by the in situ star formation at high redshift, and, as the most massive galaxies become passive (see Fig.~\ref{fig:ssfrvsmgal}), the fraction of stars formed in situ decreases in favour of the ex situ component.
For the least massive galaxies selected at $z=0$ ($M_{\rm s}=10^{10}\, \rm M_\odot$), the in situ fraction remains constant over time and close to unity.
We show in Fig.~\ref{fig:minsituvsz} what fraction of the $z=0$ in situ and ex situ mass is acquired by galaxies as a function of time and redshift for four different mass samples selected at $z=0$.
We see the same behaviour as in Fig.~\ref{fig:finsituvsz}: low-mass galaxies are dominated by the in situ component at all redshifts, while high-mass galaxies are dominated by the in situ component at high redshift and by the ex situ component at low redshift.
The time at which galaxies have formed half of their in situ mass of stars (indicated by the vertical blue bar) happens earlier for more massive galaxies since the AGN activity suppresses the SFR more efficiently in those galaxies.
The time at which half of the ex situ mass is assembled shows very little variation with selected galaxy mass at $z=0$, though it is slightly later for more massive galaxies.
Finally, the time at which half of the total stellar mass is in the galaxy occurs later with the selected galaxy mass at $z=0$, as this is the result of massive galaxies having more stars from the ex situ component (late assembly phase) than from the in situ component (early assembly phase). 
Massive galaxies form their stars early in situ and assemble late~\citep{delucia&blaizot07}.

\begin{figure}
\center \includegraphics[width=0.995\columnwidth]{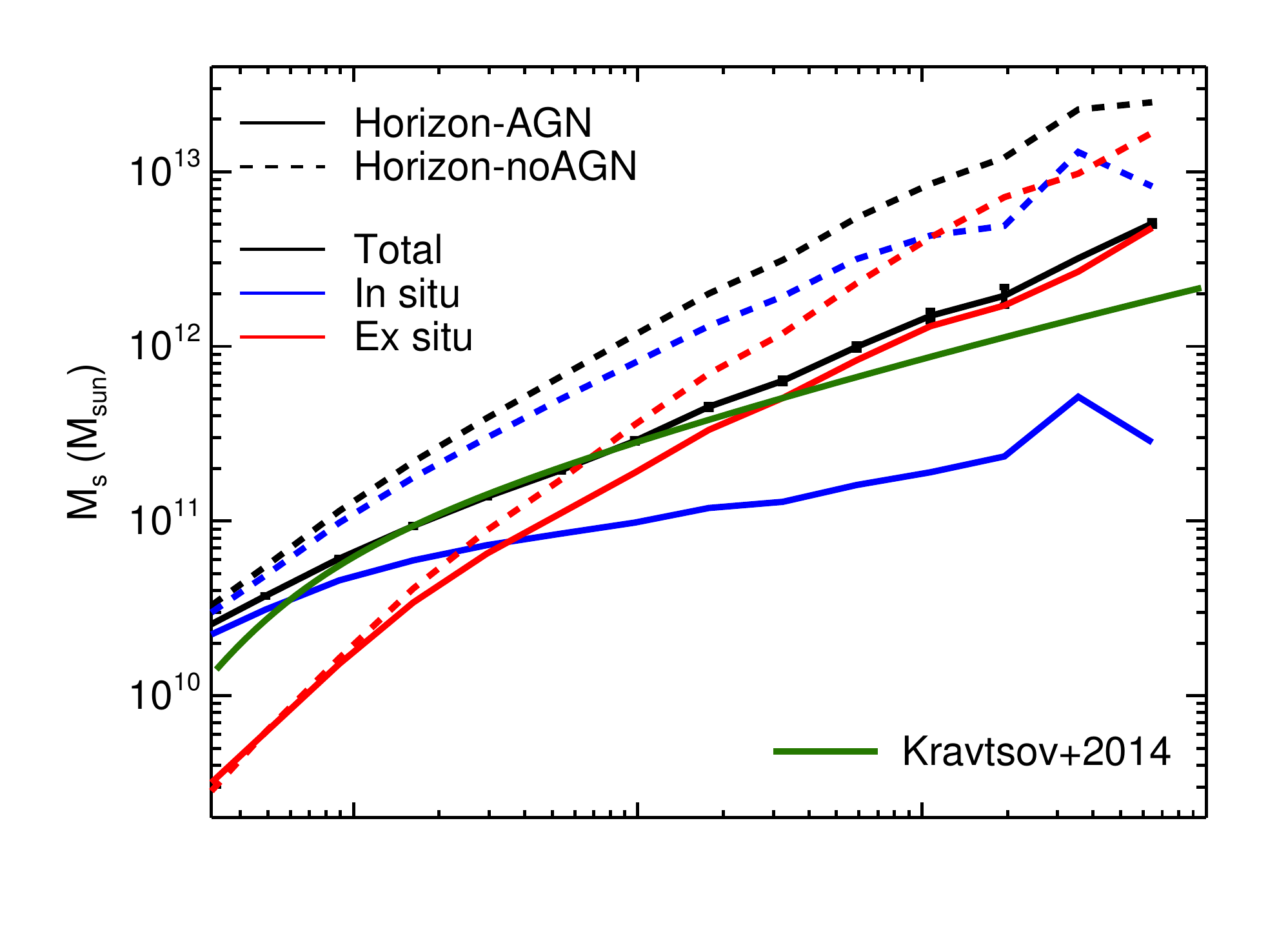}\vspace{-1.5cm}
\center \includegraphics[width=0.995\columnwidth]{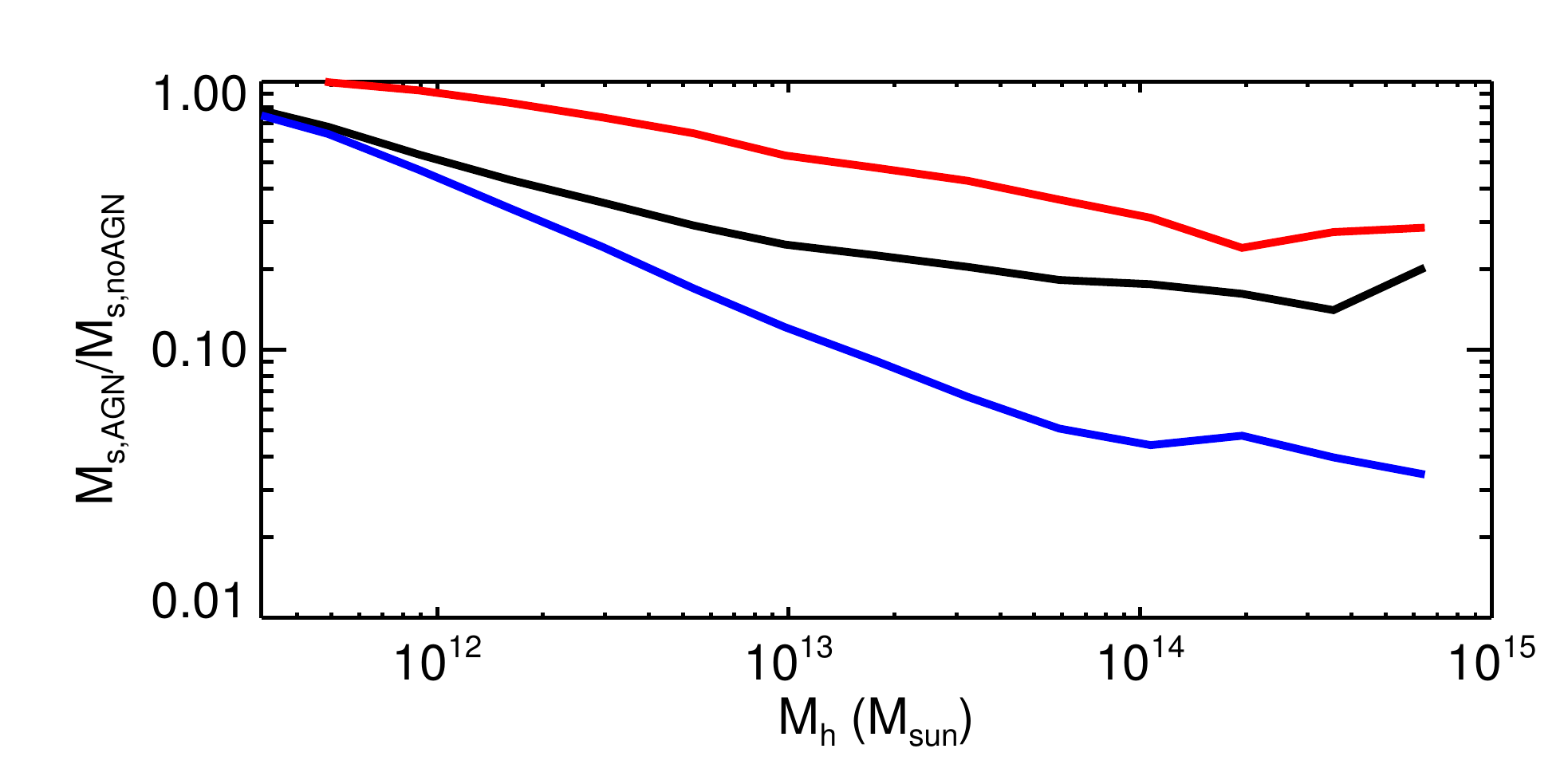}
 \caption{Top panel: mean stellar mass in the central galaxy as a function of halo mass (solid black) for \hagn\, and \hnoagn\, (dashed black) at $z=0$. The in situ and ex situ components are plotted in blue and red, respectively. The error bars (only plotted for the total mass component in \hagn\, for the sake of clarity) stand for the error around the mean. The green line corresponds to the observational data by~\citet{kravtsovetal14}. Bottom panel: relative mass of stars for the different stellar components of \hagn\, relative to \hnoagn.
  Both the in situ and accreted mass components are reduced by the AGN activity, though with a much stronger suppression for the former than for the latter.}
\label{fig:mbudget}
\end{figure}

Fig.~\ref{fig:mbudget} shows how much the in situ and ex situ matter  is affected respectively by AGN feedback in \hagn\, and \hnoagn\, as a function of  halo mass.
As outlined in Figs.~\ref{fig:finsitu} and~\ref{fig:finsituvsz}, in situ star formation is strongly affected by AGN activity.
This decrease of the in situ mass of stars can reach more than 90 \% above group-scale haloes $M_{\rm h}>10^{13}\,\rm M_\odot$ and this in situ mass decrement increases with mass.
In parallel, the fraction of stellar mass acquired by mergers is also reduced, mostly because lower-mass galaxies have their in situ mass also suppressed by AGN activity, which end up contributing to the accreted mass content of more massive galaxies.
Hence, AGN feedback significantly suppresses in situ star formation in galaxies above $M_{\rm h}>10^{13}\,\rm M_\odot$, and also decreases the amount of merger-acquired stars, though in smaller proportion.
We have also plotted the data from~\cite{kravtsovetal14} for comparison: \hagn\, compares favourably well with observations, while \hnoagn\, can be an order of magnitude above the observational predictions.
As it clearly appears from this last figure, to get the correct stellar-to-halo mass relation for massive galaxies, AGN feedback needs to suppress the in situ star formation in a large range of galaxy (or halo) mass.
Hence, AGN feedback must suppress the in situ mass in the central galaxy \emph{and} significantly reduce the ex situ mass, which is achieved by decreasing the in situ mass of satellites. 

\section{Morphometry and the origin of stars}
\label{section:morphotoinsitu}

\subsection{Kinematics}
\label{section:kintoinsitu}

\begin{figure}
\center \includegraphics[width=0.995\columnwidth]{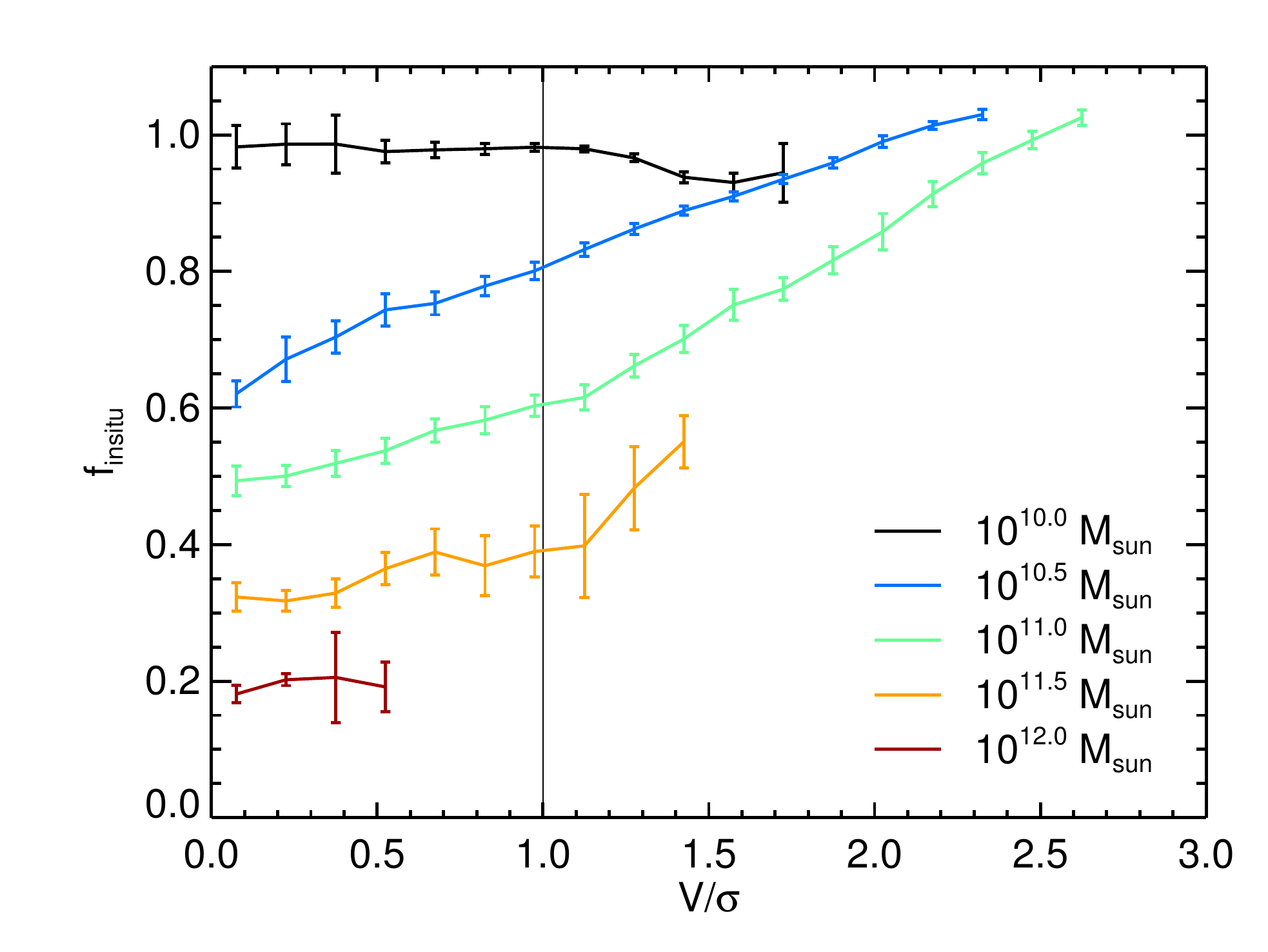}
\caption{Mean fraction of in situ formed stars $f_{\rm insitu}$ as a function of the $V/\sigma$ ratio in \hagn\, at $z=0$ for different galaxy masses ($\pm10\%$) as indicated in the panel with different line styles (increasing mass from top to bottom). Error bars indicate the error around the mean and the vertical line separates ellipticals from discs. At constant stellar mass, disc galaxies ($V/\sigma>1$) have larger in situ fractions than elliptical galaxies ($V/\sigma<1$): the morphological transformation of galaxies, besides being a mass effect, is also a cosmic accretion history effect.}
\label{fig:finsituvsvsig}
\end{figure}

\begin{figure}
\includegraphics[width=0.95\columnwidth]{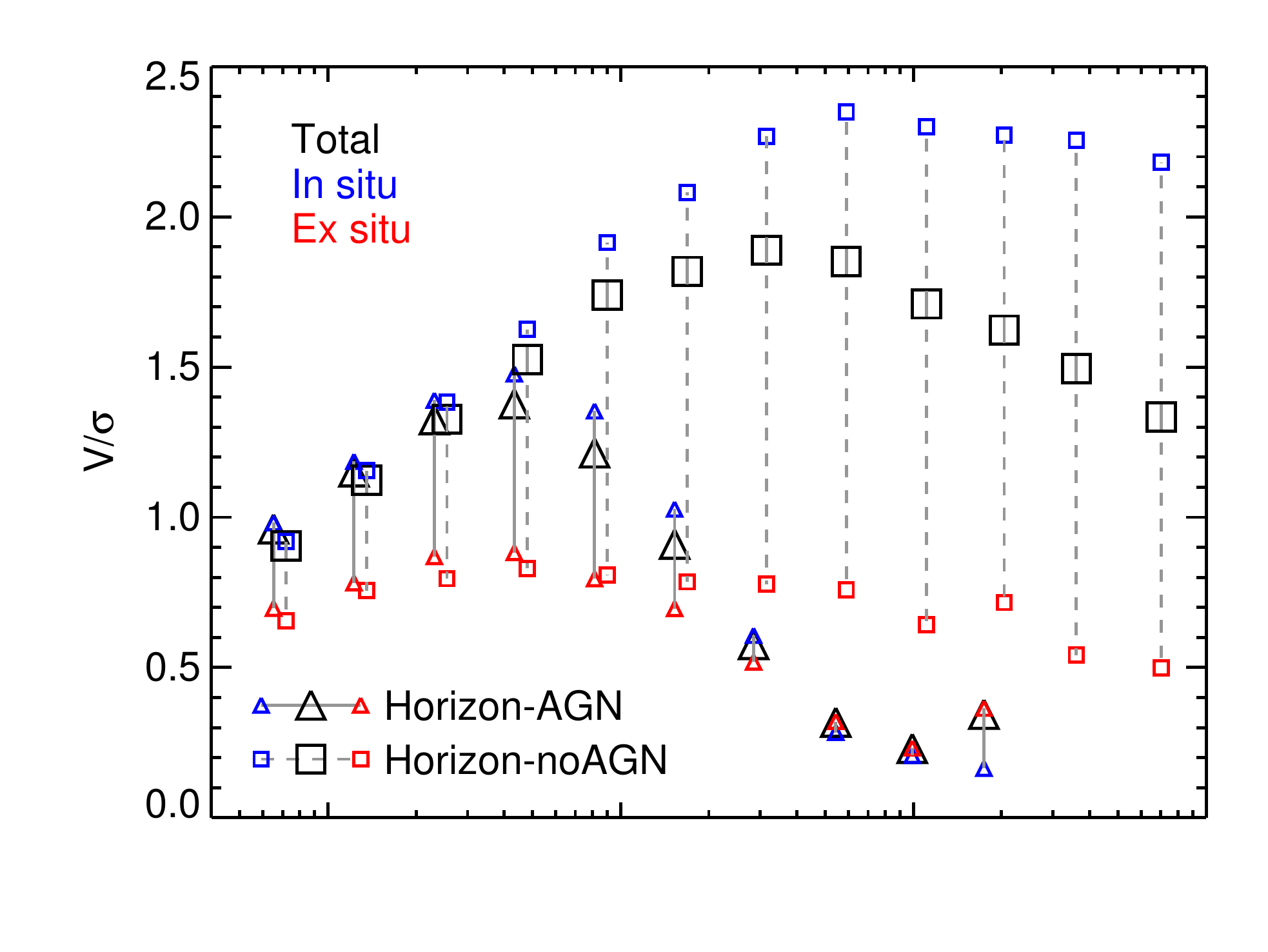}\vspace{-1.3cm}
\includegraphics[width=0.95\columnwidth]{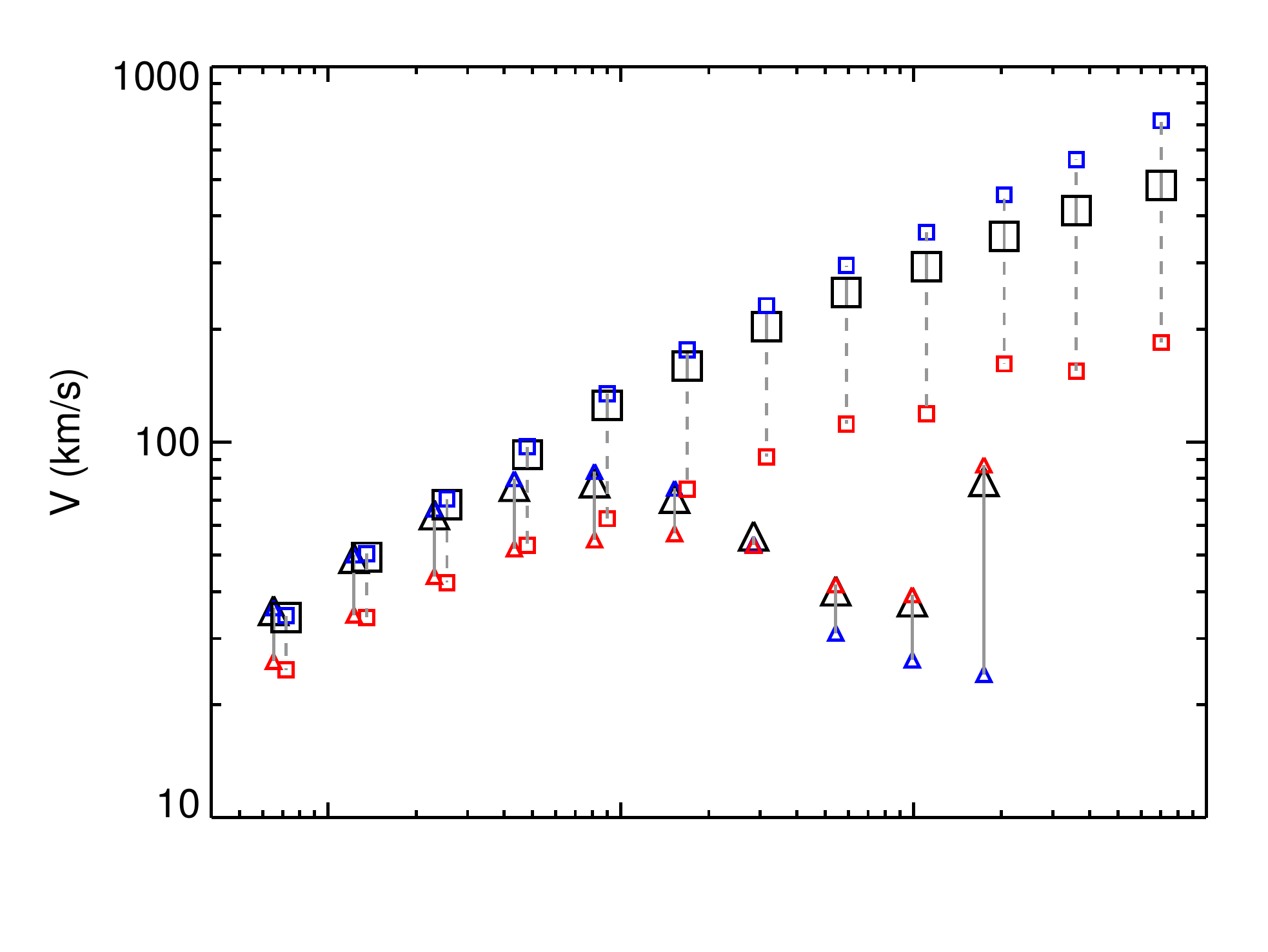}\vspace{-1.3cm}
\includegraphics[width=0.95\columnwidth]{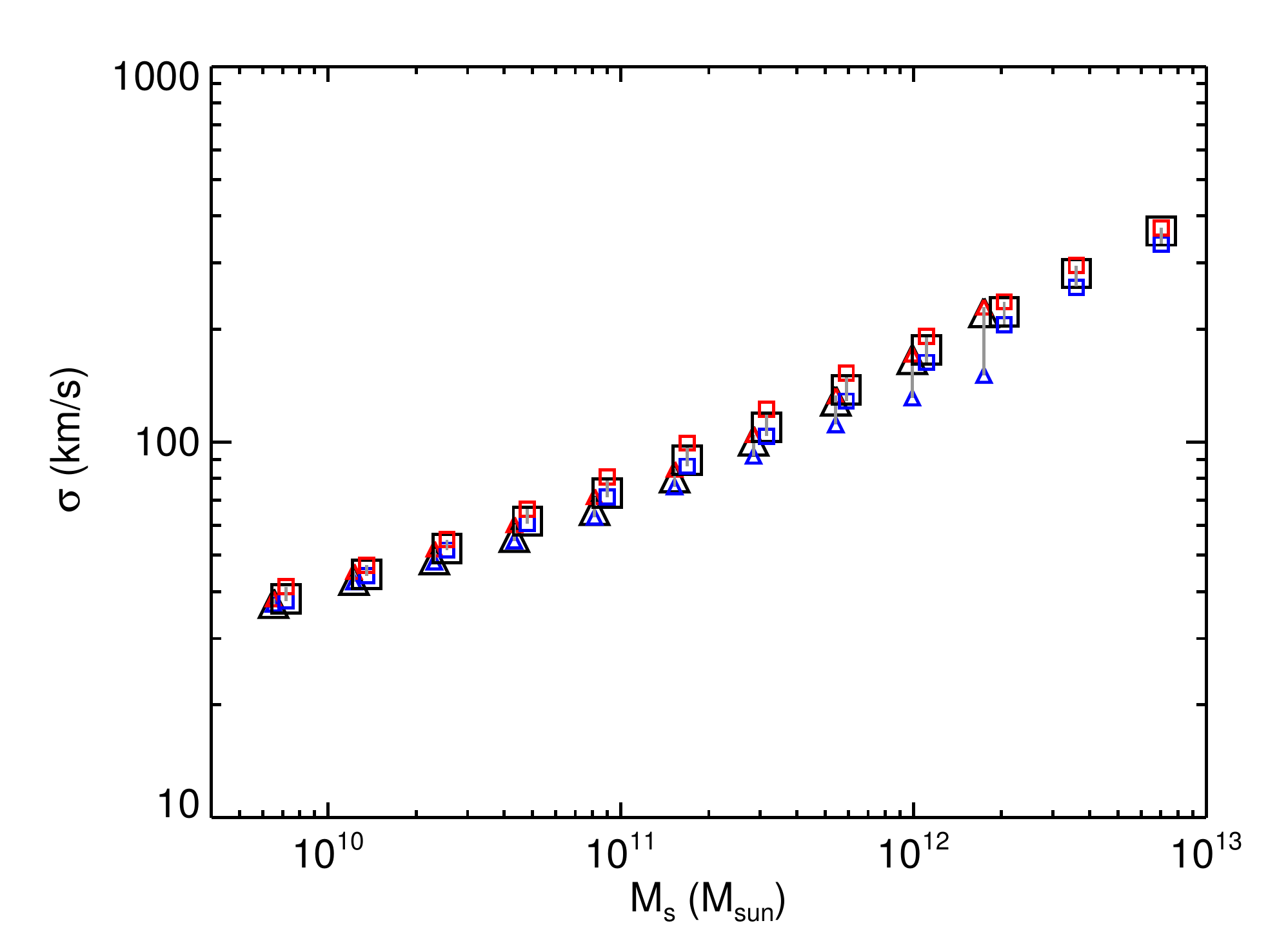}
 \caption{Top panel: mean ratio of $V/\sigma$ of stars as a function of stellar mass $M_{\rm s}$ for the total (black), in situ (blue), and ex situ (red) components in \hagn\, (triangles with solid grey lines) and in \hnoagn\, (squares with dashed grey lines) at $z=0$. For the sake of readability, we have multiplied the $x$-axis values of
  \hagn\, and \hnoagn\, by a factor of $0.95$ and $1.05$, respectively. In the middle and bottom panels, we plot the mean rotation velocity and the mean velocity dispersion as a function of stellar mass, 
  respectively. AGN feedback reduces $V/\sigma$ of both the in situ and ex situ component by suppressing the amount of rotation $V$ in the galaxy, while the velocity dispersion $\sigma$ keeps the same relation with stellar mass.}
\label{fig:vsig_insitu}
\end{figure}

\begin{figure}
\center \includegraphics[width=0.995\columnwidth]{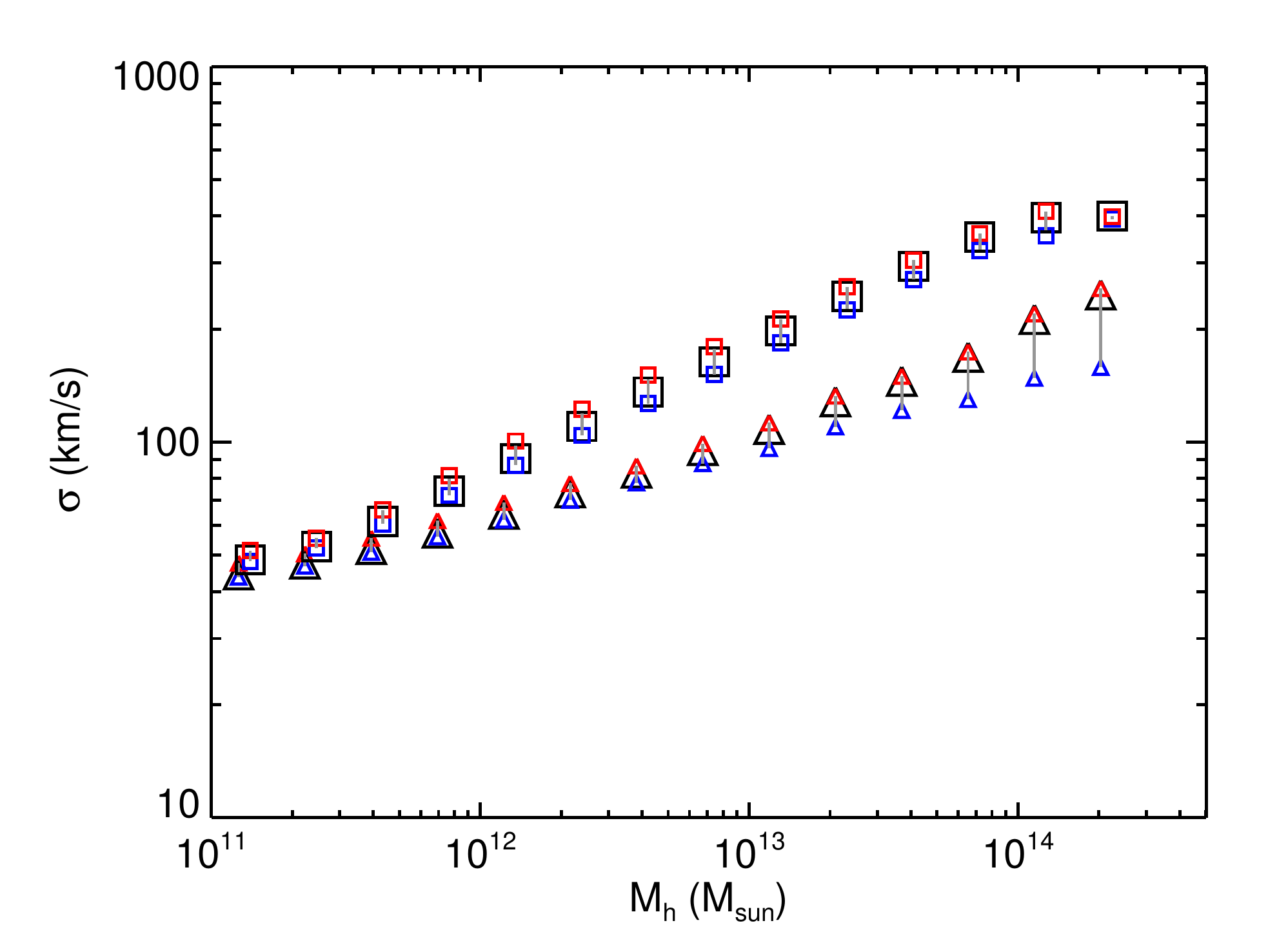}
 \caption{Same as the bottom panel of Fig.~\ref{fig:vsig_insitu} as a function of halo mass instead of stellar mass. 
 It is now apparent that, at constant halo mass, the velocity dispersion of stars is reduced by AGN feedback.}
\label{fig:vsig_insitu_mh}
\end{figure}

In Fig.~\ref{fig:finsituvsvsig}, we show the fraction of in situ formed stars as a function of $V/\sigma$, i.e. our galaxy morphology proxy, for different galaxy stellar masses in \hagn\, at $z=0$.
We witness that the more rotation-dominated the galaxy, the larger the fraction of in situ formed stars, except for the two extreme bins $M_{\rm s}\simeq 10^{10}\,\rm M_\odot$ and $M_{\rm s}\simeq 10^{12}\,\rm M_\odot$ where the distribution is flat ($f_{\rm in situ}\simeq 1$ and $\simeq0.2$, respectively).
Therefore, the presence of elliptical galaxies is not purely driven by \emph{mass} but is also driven by the nature of \emph{cosmic accretion}: at constant  stellar mass, their morphology also depends on the relative stellar mass content acquired through ex situ versus in situ components, i.e. by mergers or by gas accretion (leading to in situ star formation).
Also, for a given $V/\sigma$, the fraction of in situ formed stars depends on the galaxy stellar mass: more massive galaxies require a larger fraction of stellar mass acquired by mergers to form an elliptical.

\begin{figure}
\center \includegraphics[width=0.995\columnwidth]{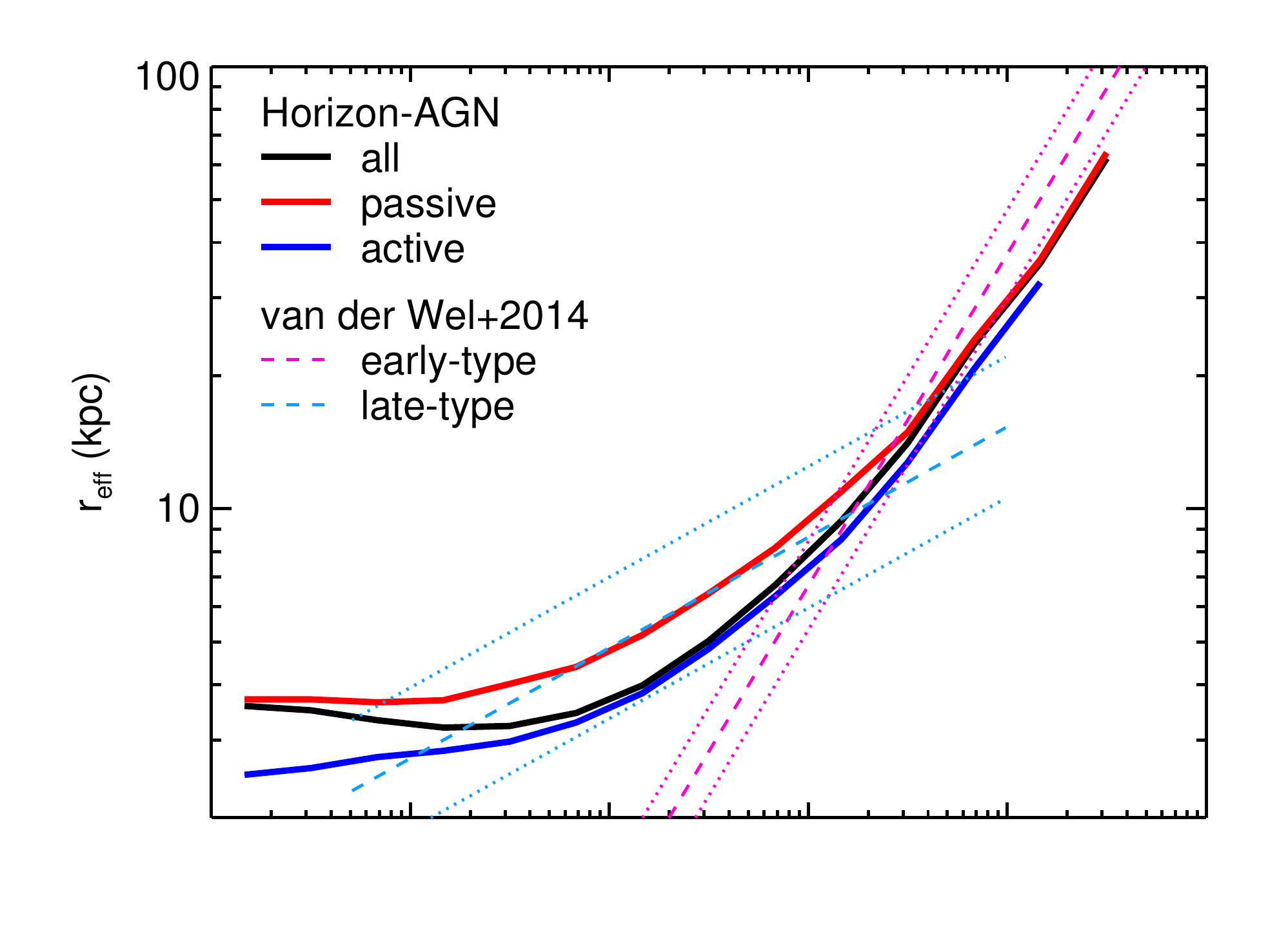}\vspace{-1.5cm}
\center \includegraphics[width=0.995\columnwidth]{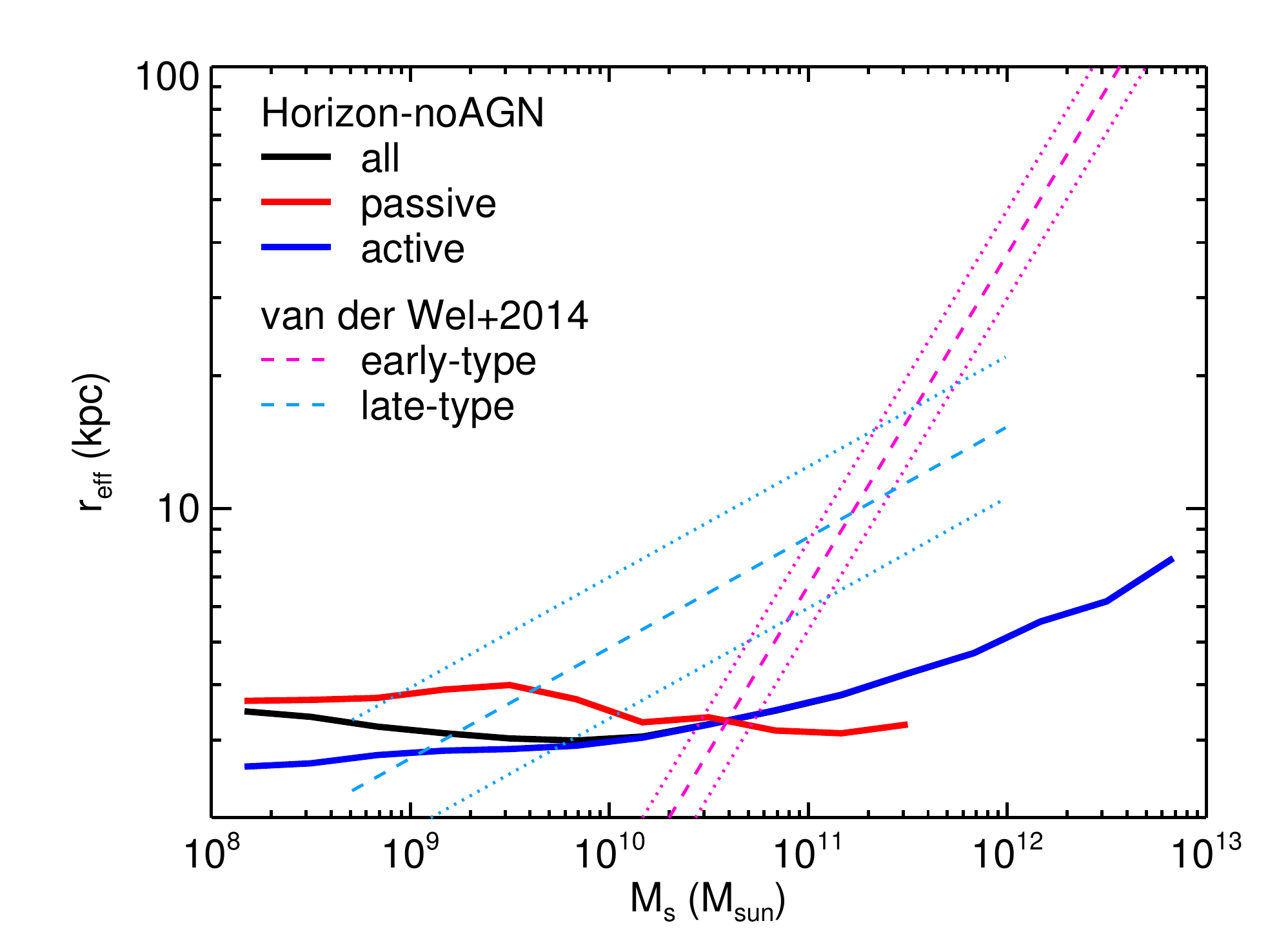}
 \caption{Size-mass relation for simulated galaxies in \hagn\ (solid lines in top panel) and \hnoagn\, (solid lines in bottom panel) at $z=0.25$ compared to observations from~\citet{vanderweletal14} for late-type galaxies (dashed cyan) and early-type galaxies (dashed magenta) and their corresponding scatter (in $\log r_{\rm eff}$) as dotted lines. We have plotted the result for our disc galaxies ($V/\sigma\ge1$ in blue) and elliptical galaxies ($V/\sigma<1$ in red).
AGN feedback allows for more extended galaxies, while without AGN feedback, massive galaxies are too compact compared to observations.
Discs and high-mass ellipticals in \hagn\, closely follow the observational relation.
However, low-mass ellipticals are not compact enough as their size approach the resolution limit (1 kpc).}
\label{fig:rvsmcomp}
\end{figure}

\begin{figure}
\center \includegraphics[width=0.995\columnwidth]{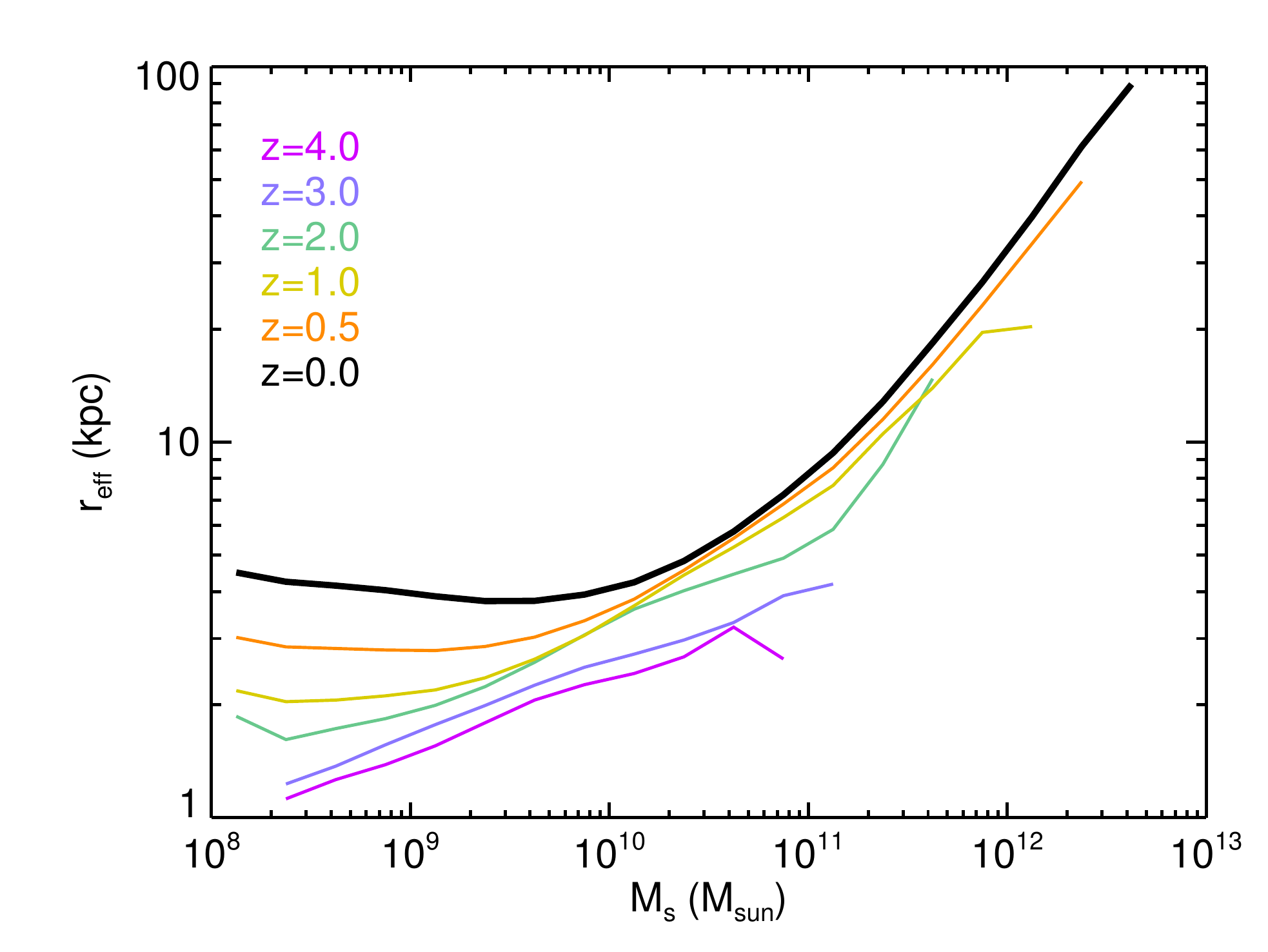}
 \caption{Size-mass relation for galaxies in \hagn\  at different redshifts indicated by the different colors. Galaxies are more compact at high redshift and the effect is seen for all galaxy masses.}
\label{fig:rvsmvariousz}
\end{figure}

\begin{figure}
\center \includegraphics[width=0.995\columnwidth]{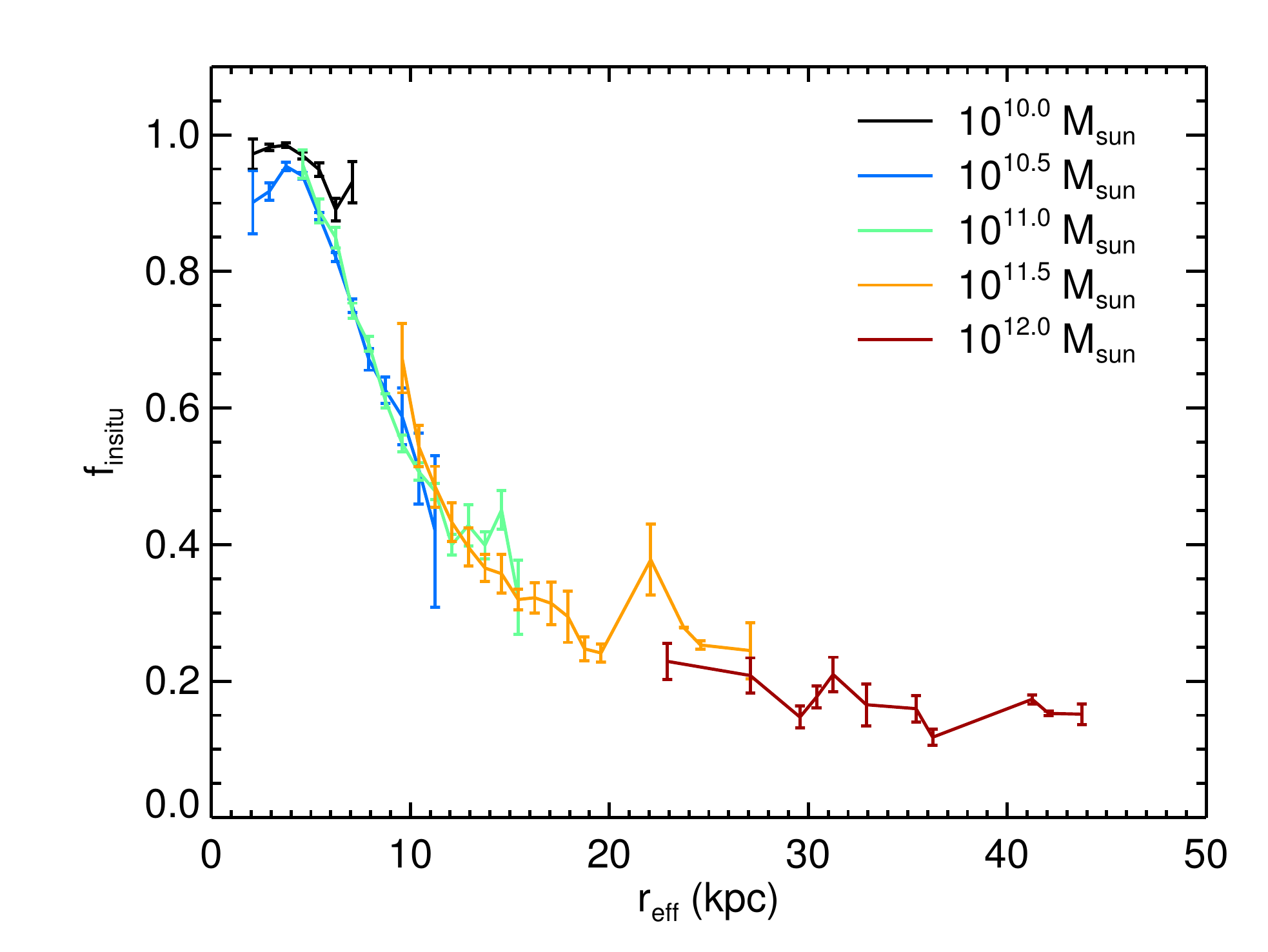}
\caption{Fraction of in situ formed stars $f_{\rm insitu}$ as a function of the the effective radius $r_{\rm eff}$ in \hagn\, at $z=0$ for different galaxy masses ($\pm10\%$) as indicated in the panel with different line styles (increasing mass from top to bottom). Error bars indicate the error around the mean. At constant stellar mass, compact galaxies (low $r_{\rm eff}$) have larger in situ fractions than extended galaxies}
\label{fig:finsituvsvreff}
\end{figure}

Fig.~\ref{fig:vsig_insitu} shows the ratio of rotation-to-dispersion of stars as a function of galaxy stellar mass at $z=0$ for the total, in situ and ex situ components of stars.
Note that the mass that is quoted in the plot for the in situ and ex situ components is that of the total stellar mass of the galaxy (not in situ and ex situ).
We see, as for Fig.~\ref{fig:vsig}, that galaxies are more likely ellipticals (low $V/\sigma$) with AGN feedback.
As expected, and in most cases, the ex situ component exhibits lower values of $V/\sigma$ compared to the in situ component: ex situ stars assembled in the galaxy by mergers are more dispersion-supported than stars that are formed in situ from centrifugally supported discs of gas.
The presence of AGN feedback, as it reduces the in situ fraction of stars, leads to lower values of $V/\sigma$ since the ex situ component dominates over the in situ component.
However, AGN feedback not only turns in situ-dominated galaxies into ex situ-dominated galaxies, it also changes the kinematics of the two components by decreasing the value of $V/\sigma$ of the ex situ and the in situ components.

As massive galaxies get their in situ star formation suppressed by AGN activity, the in situ component gets older than what it would have been in the absence of AGN feedback.
Therefore, as it gets older, the in situ component has more time to get perturbed by mergers (see Fig.~\ref{fig:minsituvsz}, where more massive galaxies, i.e. lower $f_{\rm in situ}$, have older in situ populations), instead of being continuously rejuvenated at low redshift.

The middle and bottom panels of Fig.~\ref{fig:vsig_insitu} replicate the top panel for the rotation velocity $V$ and velocity dispersion $\sigma$, respectively (instead of $V/\sigma$).
They show that AGN activity, through suppression of star formation, essentially leads to a suppression of rotation in the stellar kinematics of massive galaxies with more than an order of magnitude difference for massive galaxies.
Since the gas settles in rotation in galaxies (the gas turbulent velocities are of the order $10\,\rm km\,s^{-1}$ in disc galaxies), they lead to in situ formed stars with similar velocity patterns at formation, hence, dominated by centrifugal support.
Once AGN activity has suppressed in situ star formation, the rotating pattern of stars can be erased by mergers, as we can see it for massive galaxies.
Note that the rotation of the ex situ component is also significantly reduced by the presence of AGN feedback, though much less than that of the in situ component.
This effect can be attributed to transfers of angular momentum between the in situ and ex situ components (through non-axisymmetric densities) that enforce the components to have close rotational velocities.
Since galaxies are more gas-rich without AGN feedback compared to galaxies simulated with AGN feedback~\citep{duboisetal12}, mergers mostly consist of dissipative disc-disc mergers and hence the ex-situ component of stars is more likely to preserve a large rotation after the merger has occurred.
In contrast, the velocity dispersion of stars shows very little difference between \hagn\, and \hnoagn\, galaxies at constant stellar mass.
However, this mostly reflects the fact that the stellar velocity dispersion is also reduced by AGN feedback at the same time that the stellar mass is reduced.
This is shown in Fig.~\ref{fig:vsig_insitu_mh} with the dispersion of stars as a function of halo mass, and it is now clear that the dispersion of stars is also reduced at constant halo mass, i.e. for the same galaxy.
Therefore, if the velocity dispersion of a galaxy at the centre of a given halo is reduced by AGN feedback, it should correspond to a galaxy size increase.

\begin{figure*}
\includegraphics[width=0.75\columnwidth]{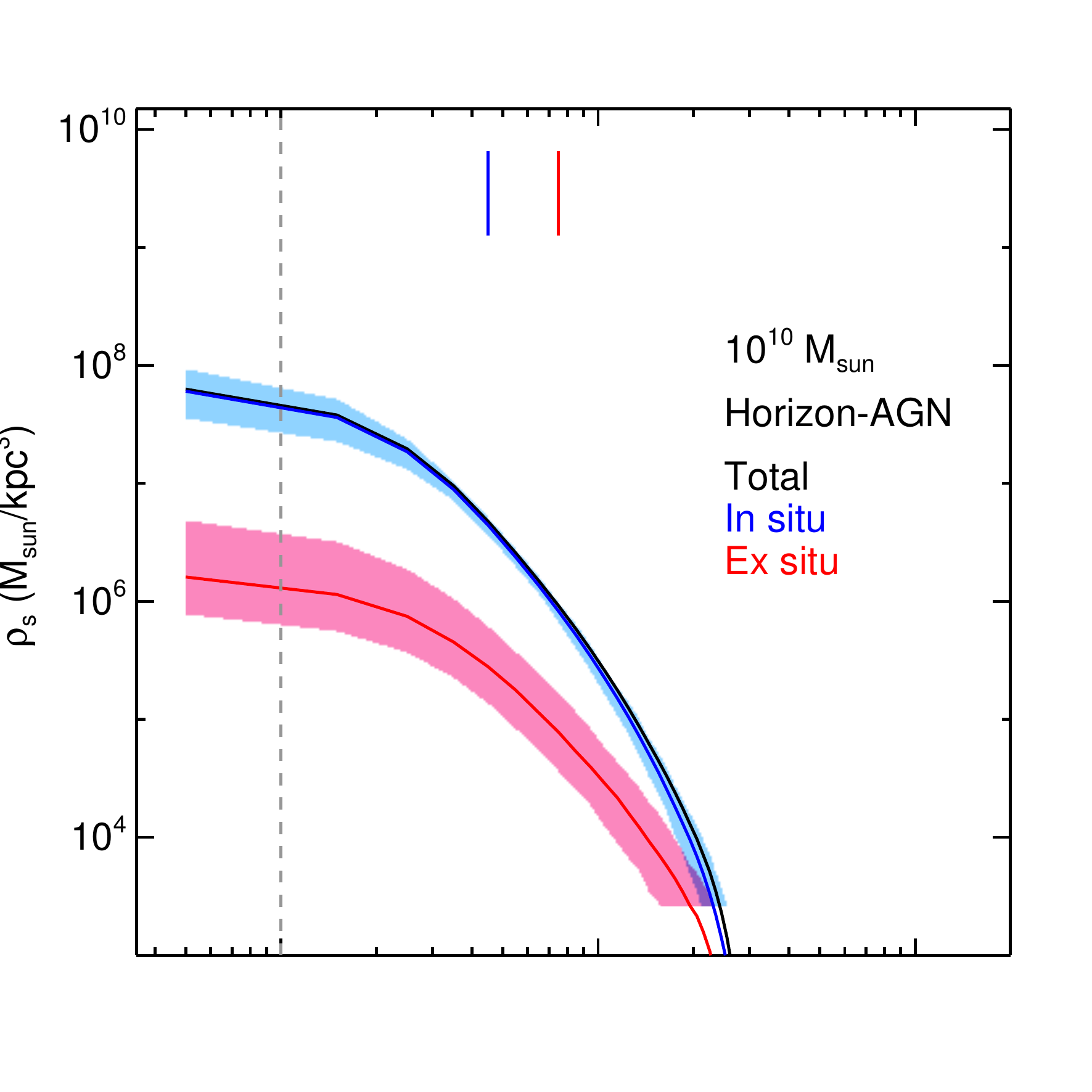}\hspace{-1.3cm}
\includegraphics[width=0.75\columnwidth]{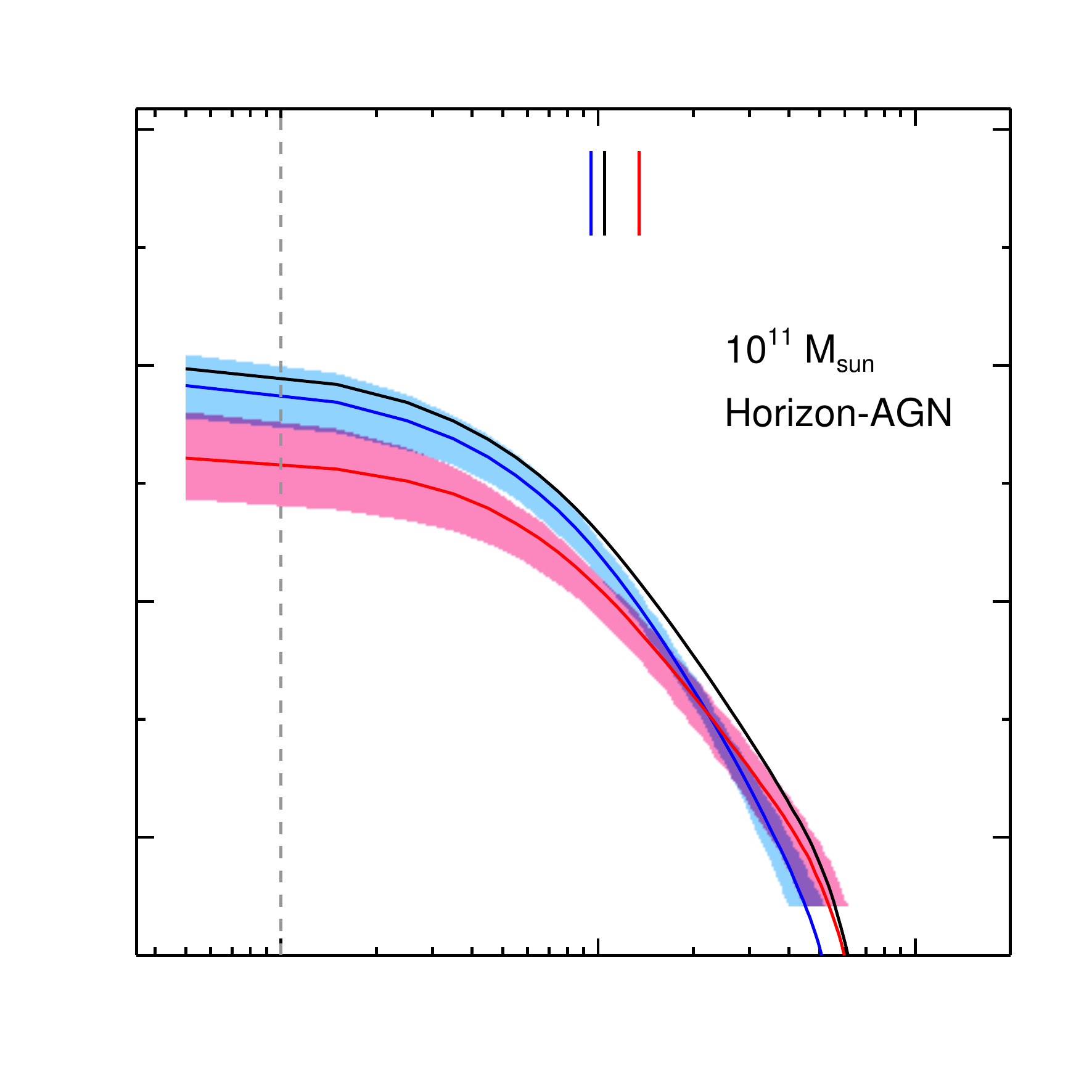}\hspace{-1.3cm}
\includegraphics[width=0.75\columnwidth]{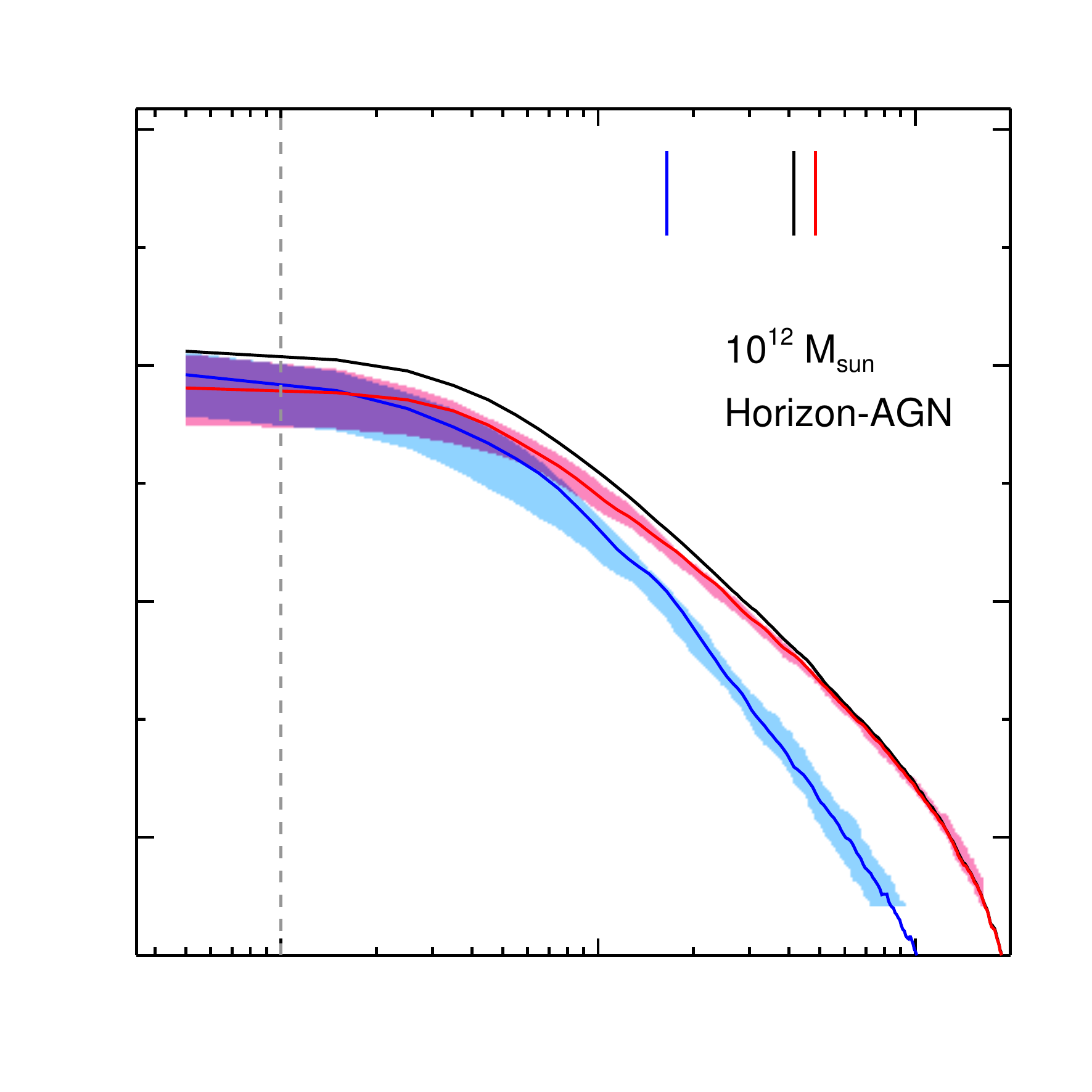}\vspace{-1.4cm}
\includegraphics[width=0.75\columnwidth]{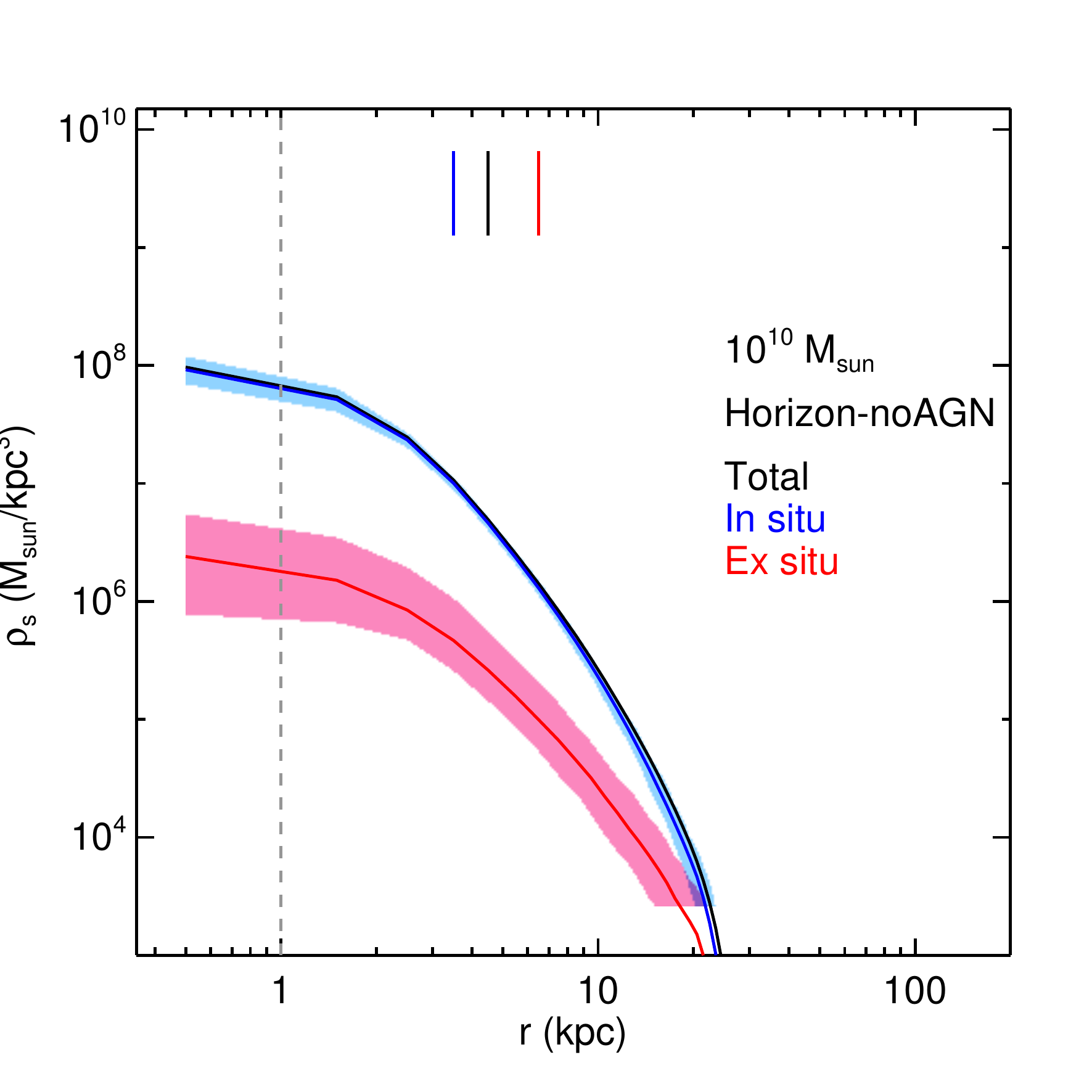}\hspace{-1.3cm}
\includegraphics[width=0.75\columnwidth]{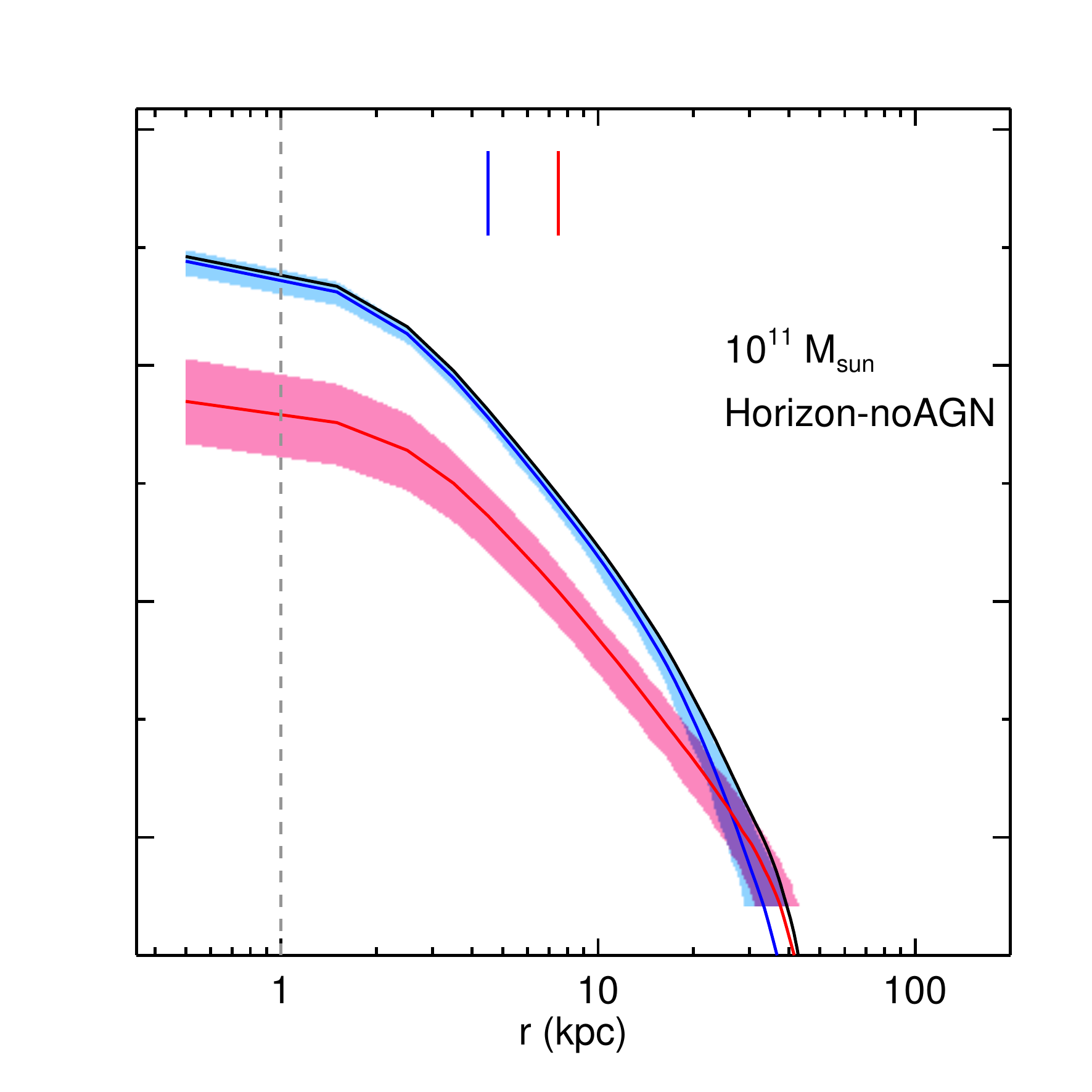}\hspace{-1.3cm}
\includegraphics[width=0.75\columnwidth]{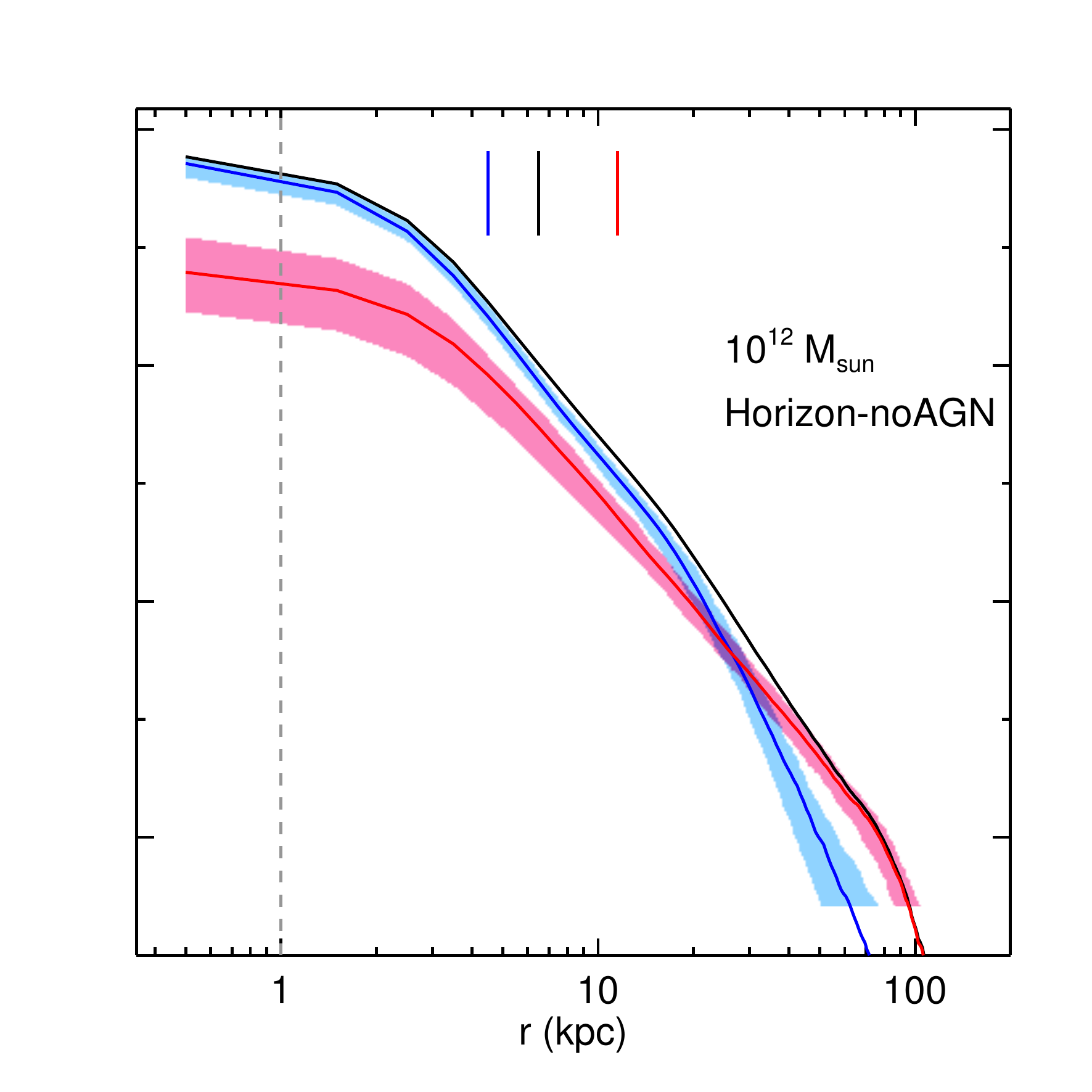}
 \caption{Stellar density distribution in spherical shells as a function of distance for different galaxy masses at $z=0$ as indicated in the different panels. {\it Top panels} are for \hagn\, and {\it bottom panels} for \hnoagn. The median of total stellar density distribution is in black, while the in situ and ex situ components are in blue and red respectively, with their 20 and 80 percentile interval as indicated by shaded areas. The three small vertical solid lines stand for the half-mass radius of the total (black), in situ (blue) and ex situ (red) components. The dashed vertical line indicates the resolution limit of the simulation.
 In \hnoagn, galaxies are always extremely cuspy in their centre where the density profile is dominated by the in situ component and up to large distance.
 On the opposite, in \hagn, intermediate and massive galaxies are more cored with a flatter stellar density in the centre, and the ex situ component is more significant in the central regions due to a strong decrease of the in situ component in the centre of galaxies.
 }
\label{fig:sdens}
\end{figure*}

\subsection{Sizes}
\label{section:size}

\begin{figure}
\center \includegraphics[width=0.995\columnwidth]{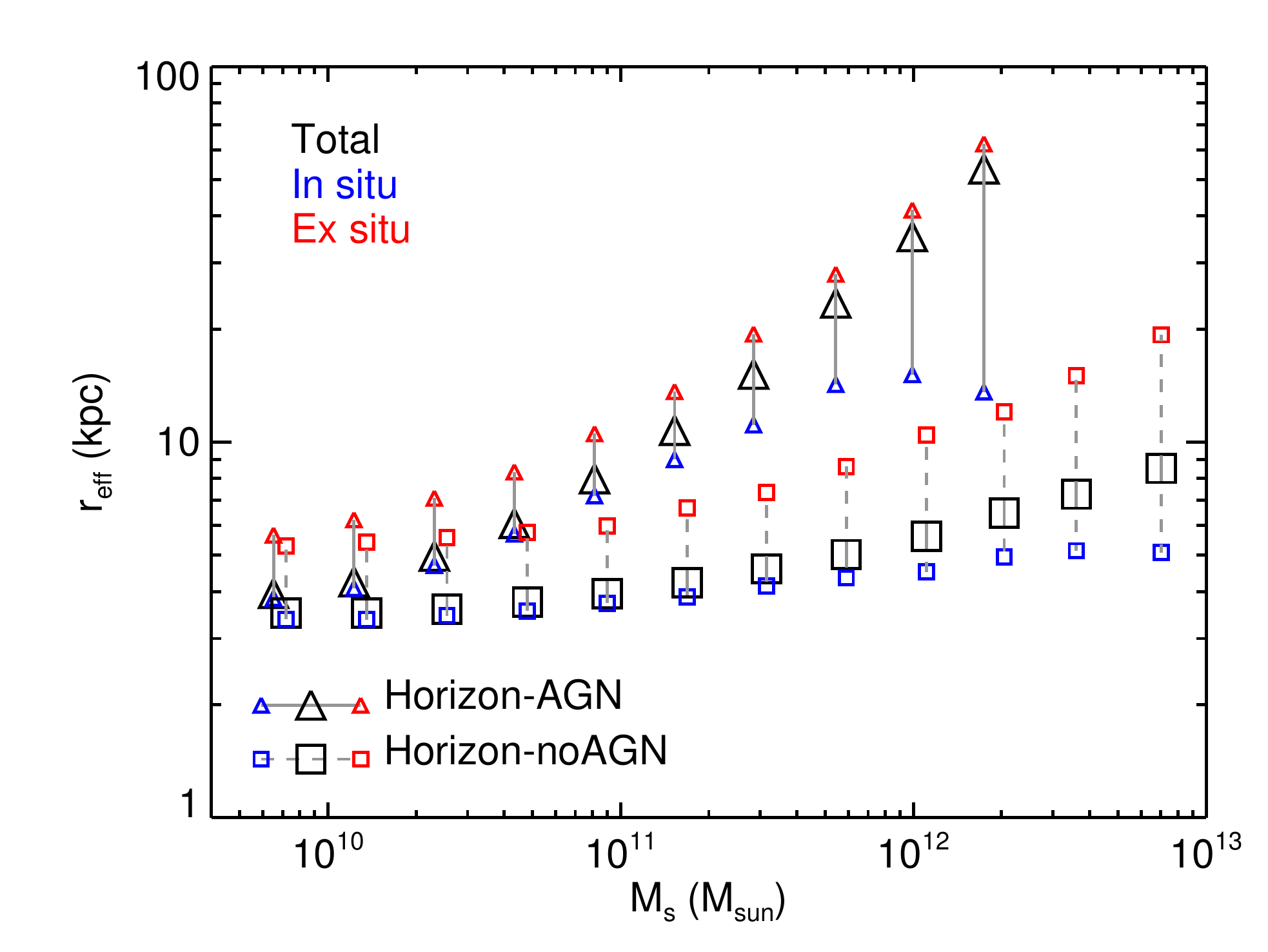}
 \caption{Effective radius ($r_{\rm eff}$) of stars as a function of stellar mass $M_{\rm s}$ for the total (black), in situ (blue), and ex situ (red) components in \hagn\, (triangles with solid grey lines) and in \hnoagn\, (squares with dashed grey lines) at $z=0$. For the sake of readability, we have multiplied the $x$-axis values of \hagn\, and \hnoagn\, by a factor of $0.95$ and $1.05$, respectively.}
\label{fig:reff_insitu}
\end{figure}

We now investigate how  galaxy sizes depend on the origin of stars.
We compute the effective radius of galaxies by first projecting their two-dimensional stellar density along each Cartesian axis of the simulation box $x$, $y$ and $z$. 
The two-dimensional half-mass radius is found for each projection and the  geometric mean of all projection is defined to be  the measured effective radius $r_{\rm eff}$ of the galaxy.

Fig.~\ref{fig:rvsmcomp} shows the effective radius of galaxies simulated in \hagn\, and \hnoagn\, at $z=0.25$ and a comparison to the observational data from~\cite{vanderweletal14}.
Without AGN feedback, the most massive galaxies end up being extremely compact with a radius up to an order of magnitude below the expected value at this redshift.
In contrast, AGN feedback allows for a fairly good match of the size-mass relation at high masses, with a power-law slope, $0.75$, of the order of the observed one, though somewhat shallower.
There is also a transition from a constant size for galaxies at low masses (though this flatness is probably a spurious effect from limited spatial resolution, see below), to intermediate mass galaxies (in the mass range $10^{10}<M_{\rm s}<10^{11}\,\rm M_\odot$) which increase their size with mass, and finally to the steeper slope observed at larger masses.

Separating the galaxies into discs or ellipticals leads to slight difference: discs appear more compact than ellipticals at all masses in \hagn\, (which is only true at low mass in \hnoagn).
At low-to-intermediate masses, observations suggest that ellipticals are more compact than discs~\citep{vanderweletal14}.
Discs are close to the empirical estimate at all masses, within a factor of 2 in size.
However, ellipticals are only a good match at the high-mass end, as they can be several times the empirical sizes at intermediate masses.
This effect could be attributed to limited spatial resolution, as the size of low-to-intermediate mass galaxies is only a few times the spatial resolution, and can get some spurious dynamical support.
Therefore, galaxy sizes are supposed to converge to $\sim \Delta x=1\,\rm kpc$ at the low-mass end, and it is well plausible that the low-mass galaxies will get more compact with increased spatial resolution.
Note than in our comparison to observations by~\cite{vanderweletal14} we separate discs from ellipticals based on kinematics, while they separate galaxies from passive and active (supposedly ellipticals and discs respectively in our case).
We have also tested to separate galaxy sizes by their sSFR, and the results are very similar. 

Fig.~\ref{fig:rvsmvariousz} shows the size-mass evolution with redshift in \hagn, and we see that galaxies are more compact at high redshift at all masses.
While above $M_{\rm s}\gtrsim5\times 10^{10}\,\rm M_\odot$, it can be explained by the increased content of ex situ acquired stars, at lower mass, galaxies have a very small fraction of mergers. 
Therefore, the size-mass increase with time at the low-mass end can be explained by the fact that the cosmological gas accretion at low redshift has a higher angular momentum than at high redshift~\citep{pichonetal11, kimmetal11}.
At high redshift, the break in slope is smaller, as galaxies are more gas rich, hence, mergers are more dissipative and lead to more compact galaxies. 
We refer readers to \cite{welkeretal16discs} for the details on how the gas content and the morphology of the progenitors influence the size evolution of galaxies in \hagn: rapid size evolution is preferentially driven by gas-poor and elliptical progenitors.

With Fig.~\ref{fig:finsituvsvreff}, we see that, at constant stellar mass, the lower the fraction of in situ formed stars in the galaxy, the larger the effective radius of the galaxy.
From this figure, it is also apparent that the $f_{\rm insitu}$-$r_{\rm eff}$ relation is mass-insensitive: for a given galaxy size, on average, the in situ fraction does not vary with galaxy mass.
Therefore, the size evolution of galaxies should  mostly be the product of the nature of the cosmic accretion history of the galaxy.
Since more massive galaxies have lower in situ fractions, as they quench star formation earlier, the trend with mass is also seen on this figure.

In order to have a better understanding of how the ex situ and in situ matter respectively contribute to the overall stellar mass distribution, and, thus, the size of the galaxy, we measure the stellar density profile of the two different populations of stars by tagging the stellar particles with their formation origin.
The one-dimensional average stellar density profiles of the total, in situ and ex situ identified stars for three different galaxy mass bins in \hagn\, and \hnoagn\, are plotted in Fig.~\ref{fig:sdens}.
In \hnoagn\, the core of the galaxy is always dominated by the in situ component of stars, even for the most massive objects~\citep[see also][]{naabetal09,oseretal12}, while, in \hagn\,, the contribution of the ex situ component is equivalent to the in situ component in the core of the stellar distribution for the most massive galaxies $M_{\rm s}\simeq 10^{12}\,\rm M_\odot$~\citep[see also][]{rodriguez-gomezetal16}.
At the same time, the stellar density profile of galaxies is also
flattened with AGN, as a response of DM haloes being more likely cored (see Peirani et al., in preparation, for more details).
It is clear that the in situ component of the stellar density at the centre (1 kpc) experiences most of the decrease due to AGN feedback with $-0.2$ dex for $M_{\rm s}=10^{10}\,\rm M_\odot$, $-1$ dex for $M_{\rm s}=10^{11}\,\rm M_\odot$, $-1.5$ dex for $M_{\rm s}=10^{12}\,\rm M_\odot$.
However, there is also a substantial decrease, due to AGN, of the ex situ component in the core (except for $M_{\rm s}=10^{10}\,\rm M_\odot$) with $-0.5$ dex for $M_{\rm s}=10^{11}\,\rm M_\odot$ and $-1$ dex for $M_{\rm s}=10^{12}\,\rm M_\odot$.
Correspondingly, the ex situ and the in situ stellar density are higher at larger radius for \hagn\, than for \hnoagn.
As a consequence, the in situ, the ex situ and the total components expand and produce larger half-mass radii.
The stellar density in the centre of galaxies remarkably depends little on mass when AGN feedback is present, while it can vary by several orders of magnitude in absence of BH activity.

\begin{figure*}
\includegraphics[width=2.05\columnwidth]{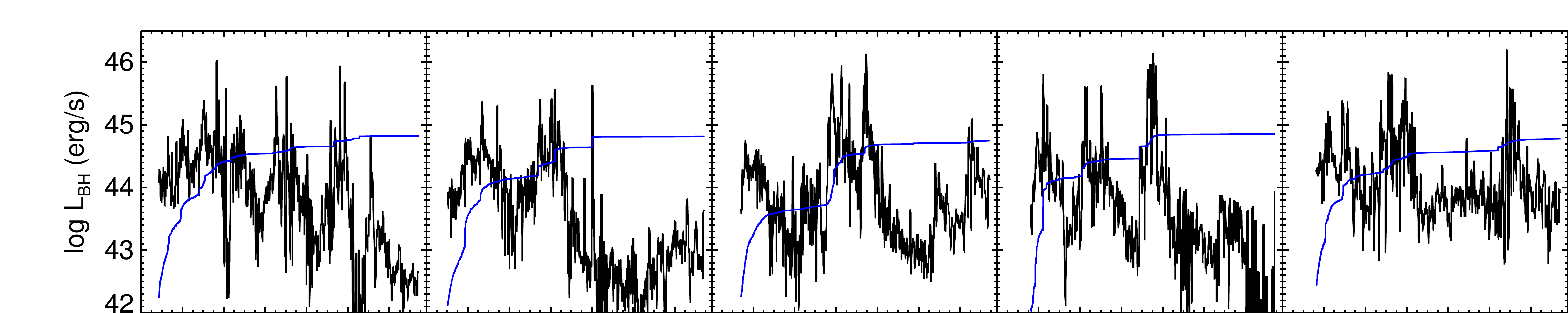}\vspace{-0.72cm}\\
\hspace{0.cm} \includegraphics[width=2.05\columnwidth]{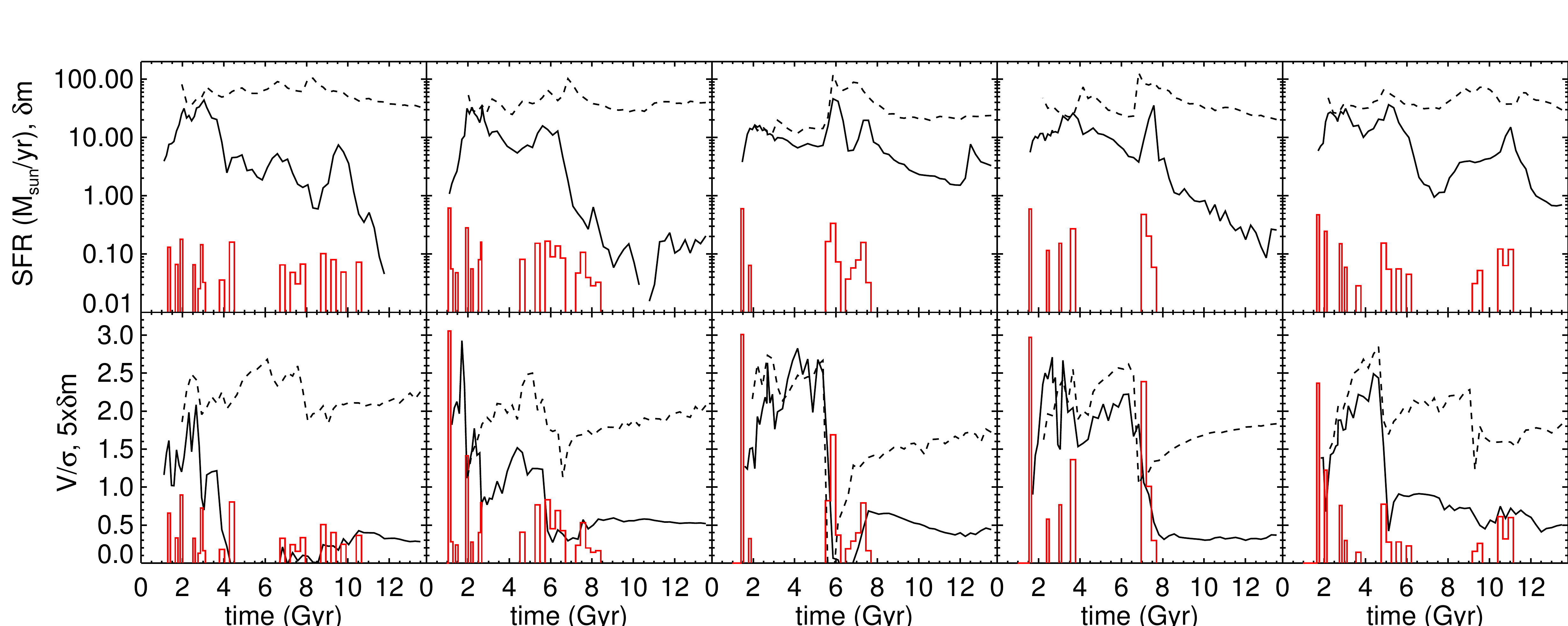}
 \caption{Five randomly selected galaxies of mass $M_{\rm s}\simeq 2\times 10^{11}\,\rm M_\odot$ at $z=0$ in \hagn\, (solid lines) with their matched galaxy in \hnoagn\, (dashed lines). Red histograms indicate significant mergers with their mass ratio $\delta m$ (only those with $\delta m>1/40$ are depicted) in the \hagn\, simulation (note that in third row those values are multiplied by 5). The first row shows the BH bolometric luminosity $L_{\rm BH}$ as a black line (smoothed over 50 Myr for readability) and 1 per cent of the Eddington luminosity as a blue line, the second row shows the SFR evolution, and the third row shows the $V/\sigma$ evolution. Mergers drive a rapid transformation of discs into ellipticals. This transformation is stabilised by the merger-induced burst of AGN feedback (\hagn) in a quasar mode (above 1 per cent of the Eddington luminosity) that durably quenches the SFR in the galaxy remnant, preventing the possibility to form back a disc.}
\label{fig:individuals}
\end{figure*}

Finally, the transition radius, which  separates the in situ-dominated (central parts) from the ex situ-dominated (outskirts), is decreasing with increasing galaxy mass~\citep[see also][]{rodriguez-gomezetal16}.
Without AGN feedback this transition radius is always larger than the half-mass radius of the galaxy since the in situ component always dominates the stellar mass budget.
In contrast, in \hagn, the transition radius becomes smaller than the half-mass radius  for the most massive galaxies ($M_{\rm s}=10^{12}\,\rm M_\odot$) and close to the resolution limit, which clearly shows that the ex situ component completely dominates the stellar mass budget at all distances from the centre of the galaxy.
At intermediate galactic mass ($M_{\rm s}=10^{11}\,\rm M_\odot$), the transition radius is also decreased by AGN feedback though it remains larger than the half-mass radius of the galaxy.

The contribution of, respectively, the in situ and ex situ components to the size-mass relation is now shown in a more compact form in Fig.~\ref{fig:reff_insitu}.
In both \hagn\, and \hnoagn, the ex situ component is more extended than the in situ component.
Without AGN feedback, galaxy sizes closely follows that of the in situ component, and, since the in situ size varies very little with mass in that case, this can account for the very weak evolution of the size-mass relation in \hnoagn.
However, with AGN feedback, the ex situ fraction of stars largely dominates at the high-mass end, which explains  the size-mass evolution observed in \hagn.
In addition to this transition from in situ to ex situ dominated galaxies, both the in situ and ex situ components are also larger with AGN feedback.
This effect is two-folds: there is less adiabatic contraction of stars due to the gas concentration (Peirani et al., in preparation), and in situ star formation halts earlier in the presence of AGN feedback, thus, in situ stars can freely expand due to mergers (ex situ component), which are drier and, thus, with fewer dissipation.
Finally, in Appendix~\ref{appendix}, we show the measurement of the effective radius of the total, in situ and ex situ components as a function of halo mass, and similar trends appear to those as a function of stellar shown in Fig.~\ref{fig:reff_insitu}: at a given halo mass, the central galaxy becomes more extended in all components due to the presence of AGN feedback.

\begin{figure}
\center{\includegraphics[width=0.90\columnwidth]{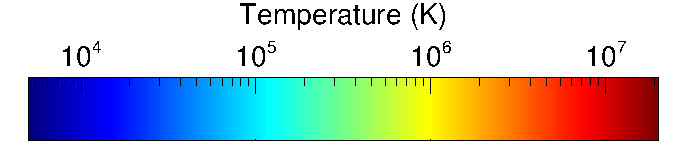}}\\
\includegraphics[width=0.48\columnwidth]{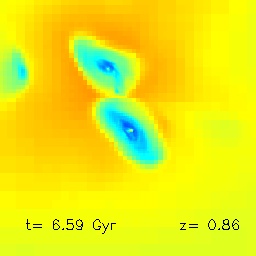}\hspace{0.1cm}
\includegraphics[width=0.48\columnwidth]{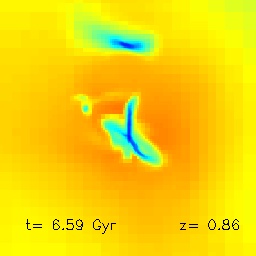}\\
\includegraphics[width=0.48\columnwidth]{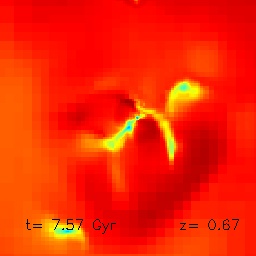}\hspace{0.1cm}
\includegraphics[width=0.48\columnwidth]{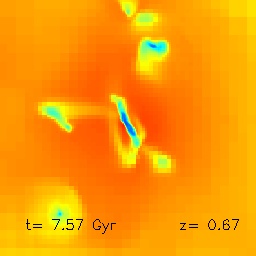}\\
\hspace{0.05cm}\includegraphics[width=0.48\columnwidth]{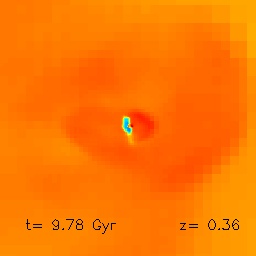}\hspace{0.1cm}
\includegraphics[width=0.48\columnwidth]{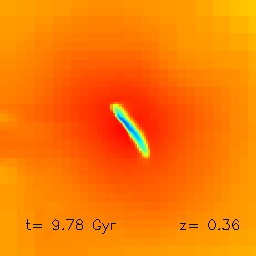}
 \caption{Illustration of the merger-enhanced AGN activity with mass-weighted temperature maps for the galaxy depicted in the fourth column of Fig.~\ref{fig:individuals} at around $t\sim7 \,\rm Gyr$ when the last major merger occurs. The image size is $250\,\rm kpc$ on the side, and the left and right columns are for the \hagn\, and the \hnoagn\, simulations, respectively, where galaxies are matched with their DM halo host. The burst of AGN activity increases the gas temperature around the merger remnant (second row), and leads to the disappearance of the disc of cold star-forming gas due to the quenching of the gas accretion onto the galaxy.}
\label{fig:individualsmaps}
\end{figure}

\section{Merger-induced morphological change}
\label{section:mergers}

In the previous sections we have seen that  only when AGN feedback is included, which reduces the in situ star formation to the benefit of the ex situ component in the overall stellar mass budget, then massive galaxies are able to become ellipticals.
Now, we illustrate the rapid morphological change of galaxies due to mergers by following a few randomly selected galaxies amongst massive ellipticals in \hagn, and we compare their evolution with its counterpart  in \hnoagn.

Fig.~\ref{fig:individuals} shows the evolution of the BH bolometric luminosity $L_{\rm BH}=\epsilon_{\rm r} \dot M_{\rm BH}c^2$ (where $\epsilon_{\rm r}=0.1$ is the radiative efficiency of the accretion disc, $\dot M_{\rm BH}$ the BH mass accretion rate, and $c$ the speed of light), the SFR and the ratio $V/\sigma$ as a function of time for 5 selected galaxies of mass $M_{\rm s}\simeq 2\times 10^{11}$ at $z=0$, and which end up as clear elliptical galaxy $V/\sigma<0.5$ in \hagn.
In the same figure, the SFR and $V/\sigma$ of the same galaxy (i.e. at the centre of the same DM halo) in \hnoagn\, are also drawn.
We have indicated the galaxy mergers in \hagn\, with their corresponding mass ratio, those with mass ratio above $1:40$.
In the top row, the value of 1 per cent of the Eddington luminosity is also shown, which value separates the radio mode (below) from the quasar mode (above) of AGN feedback implemented in the \hagn\ simulation.
At early times, most of the depicted galaxies in \hagn\, have levels of SFR close to that of \hnoagn, though they are lower in \hagn, but they significantly differ after a few Gyr of evolution due to the intensive early AGN activity.

Two effects induced by mergers appear clearly in this figure.
Galaxies experiencing a significant merger (or a succession of several mergers) first have a rapid increase in star formation, followed by a decrease.
One significant difference between simulated galaxies with and without AGN is that, in \hagn, after a merger the SFR decreases by several orders of magnitude, while, in \hnoagn, the SFR endures only a moderate decrease (roughly a factor of 2).
Thus, mergers induce a burst of SFR and a peak of AGN activity entering into a quasar mode that quenches the SFR of the galaxy during its quiescent evolution.
In the absence of AGN feedback, the enhancement of the SN activity during the merger-induced starburst is not sufficient to significantly reduce the SFR (see Fig.~\ref{fig:individuals} for the selected galaxies of \hnoagn).

\begin{figure}
\center \includegraphics[width=\columnwidth]{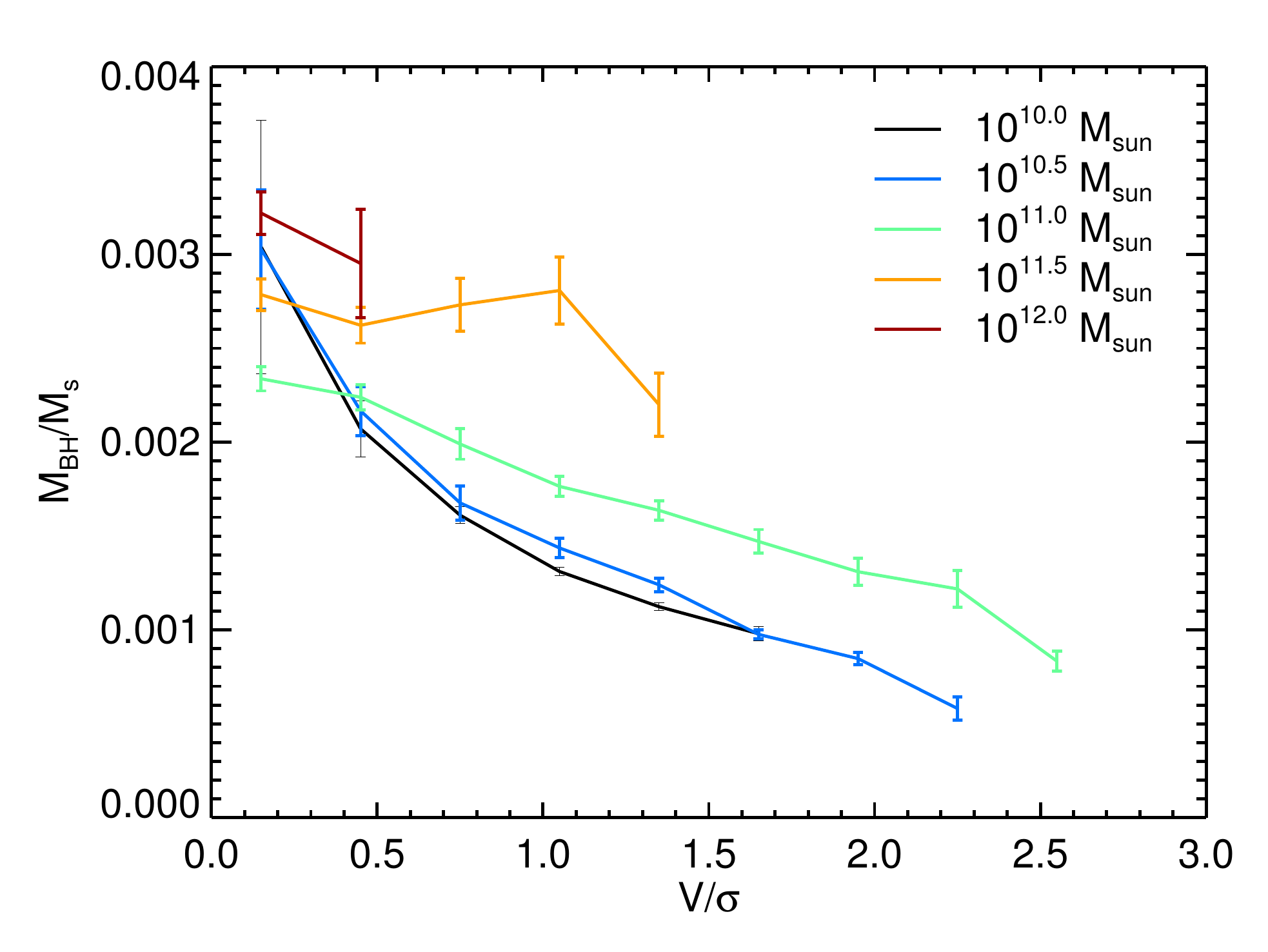}
 \caption{BH to galaxy mass ratio $M_{\rm BH}/M_{\rm s}$ as a function of $V/\sigma$ at $z=0$ for different galaxy mass bins as indicated on the panel ($\pm 10\%$). Elliptical galaxies (those with low $V/\sigma<1$) have overgrown BHs, while disc galaxies have undergrown BHs. Mergers induce a morphological transformation of disc into ellipticals and trigger an excess of BH growth (AGN activity).}
\label{fig:bhvsvsig}
\end{figure}

The second effect of mergers is on  galactic kinematics.
After each significant merger, the ratio $V/\sigma$ decreases, hereby, turning the disc into an elliptical, and this effect is seen in both \hagn\, and \hnoagn.
The fundamental difference between the two simulations that lead to the difference in galaxy morphology between the two runs is that the galaxy that experiences a merger is prevented by AGN feedback to efficiently replenish its cold gas content, and, hence, to rebuild its disc.
Consequently, the merger remnant and the progenitors (due to their past histories) contain less gas in \hagn\, than in  \hnoagn, and, thus, the remnant is more elliptical being the product of a less dissipative merger.
This is seen in Fig.~\ref{fig:individuals}, where the decrease in $V/\sigma$ after a merger is larger in \hagn\, than in \hnoagn, and where this ratio is less likely to be increasing after a merger (disc rebuilding, see~\citealp{welkeretal16discs}).
This effect, indeed, varies from one galaxy to another, depending on the gas richness of the merger, on the impact parameter of the merger, its mass ratio, etc.; however, it is very robust and is seen for a large range of galaxy mass, whatever the galaxy's final morphology.

In Fig.~\ref{fig:individualsmaps}, we show the evolution of the temperature around the peak of the merger-enhanced AGN activity depicted for the galaxy in the fourth column of Fig.~\ref{fig:individuals} at $t\sim 7\,\rm Gyr$.
The two main progenitors of the galaxy remnant have a disc component of cold gas (though less prominent in \hagn), which disappears during the merger and the post-merger phase in \hagn.
On the contrary, in \hnoagn, the cold disc of gas is still present during the merger and the post-merger phase, allowing for a continuous building of the stellar disc through in situ star formation.
The reason of the removal of the disc of gas in \hagn\, is the quasar AGN activity triggered during the merger of the two galaxies that raise the gas temperature in the circum-galactic gas (middle left panel of Fig.~\ref{fig:individualsmaps}), hereby, decreasing the gas accretion on to the galaxy, and preventing the rebuilding of the disc at later times (bottom left panel of Fig.~\ref{fig:individualsmaps}).
The late radio mode of AGN feedback is also mandatory to gently shock-heat the gas in order to
prevent the gas to experience the cooling catastrophe~\citep{duboisetal10,duboisetal11}, which can be seen in the temperature ripples (left and right sides of the central galaxy) of the bottom left panel of Fig.~\ref{fig:individualsmaps}.

Fig.~\ref{fig:bhvsvsig} shows the ratio of BH mass to galaxy stellar mass $M_{\rm BH}/M_{\rm s}$ as a function of $V/\sigma$ at $z=0$ for different galaxy mass bins.
To associate BHs with galaxies, we take the most massive BH within 10 per cent of the virial radius of the galaxy, or $3\,\rm kpc$ if this value is smaller (see~\citealp{volonterietal16} for the BH-to-galaxy-mass relation in \hagn). 
Galaxies with low $V/\sigma$ have relatively more massive BHs, while those with high $V/\sigma$ have relatively less massive BHs.
This behaviour is the signature of merger-enhanced AGN activity (thus, BH growth), which is seen at all galaxy mass.
Mergers trigger bursts of BH growth, and the AGN energy release decreases the cold gas content, which durably settles the morphological transformation into an elliptical induced by the merger.

In the absence of mergers, gas accretion can replenish the reservoir of cold gas in the galaxy, which would lead to the formation of a stellar disc component~\citep{welkeretal16discs}.
It is exactly the mechanism at work here: for a given stellar mass, we have seen that elliptical galaxies have a higher fraction of ex situ mass than disc galaxies, therefore, ellipticals endured more mergers in their past history than discs.
In light of this result, and the depicted behaviour of the SFR and stellar kinematics after a merger, we can safely conclude that mergers are the triggering mechanism of the galaxy morphological transformation (turning a disc into an elliptical).
That transformation is secured by the subsequent burst of AGN activity that is mandatory to quench the SFR during the long-term quiescent evolution of the galaxy.
Therefore, elliptical galaxies are the product of a cosmological accretion where mergers have been preferred over diffuse cosmic gas infall, triggering an AGN-regulated quenching and allowing to freeze the post-merger morphology of a galaxy.

\section{Conclusion}
\label{section:conclusion}

By comparing results from two otherwise identical state-of-the-art hydrodynamical cosmological simulations with and without AGN feedback (respectively named \hagn\, and \hnoagn), 
we have explored its impact on the cosmic evolution of the morphological mix of galaxies (Hubble sequence).
We have shown that AGN feedback is responsible for freezing the post-merger morphology of a galaxy, most likely elliptical, hence matching many observed geometric properties of galaxies. 
In brief our finding are as follows:
\begin{itemize}
\item The simulated population in \hagn\, is a fairly good match to the morphological diversity of galaxies, and the  diversity-mass relation is found to slightly evolve with redshift.
\item AGN feedback is essential in order to produce massive galaxies that resemble ellipticals. Without BH activity, massive galaxies are disc-like with kinematics  dominated by rotational support.
\item The role of AGN feedback in the morphological transformation is indirect: massive galaxies turn ellipticals with AGN because the ex situ component (accreted by mergers) dominates the stellar mass budget, as AGN suppress the in situ star formation.
\item Thanks to AGN feedback and the suppression of in situ star formation, the galaxy-halo mass relation is in much better agreement with observations.
\item \hagn\, establishes a clear relation between the morphology of galaxies and the origin of stars (in situ fraction) at constant stellar mass. Hence, morphology is demonstrated not to be purely driven by mass but also by the nature of cosmic accretion (in situ SF versus mergers).
\item When stars in galaxies are split into an in situ and an ex situ stellar components, the former is more rotation-dominated than the latter.  AGN feedback allows massive galaxies to get their  $V/\sigma$ closer to their ex situ component (less in situ star formation), and at the same time it reduces the $V/\sigma$ ratio for all individual components (both in situ and ex situ). This decrease of $V/\sigma$ is due to the suppression of rotation in massive galaxies by up to an order of magnitude, while the dispersion is also reduced although with lower significance.
\item The role of AGN feedback is also key to get more realistic size-mass relations. In absence of AGN feedback massive galaxies are too compact.
In situ fraction strongly correlates with the size of galaxies at constant stellar mass: this is again the  effect of  the nature of cosmic accretion.
\item As expected, the distribution of ex situ stars is more extended than that of  in situ stars. Hence AGN feedback, as it suppresses in situ star formation, also allows for less concentrated distribution of both in situ and ex situ stars.
\end{itemize}

 It should be stressed that the sub-grid physics of AGN feedback in \hagn\, was {\it not} calibrated to match any 
morphological properties (only calibrated to reproduce the BH-galaxy mass relation, see~\citealp{duboisetal12, volonterietal16}).
Hence, the good agreement to observed morphometric properties 
 presented in this paper provides independent consistency checks for the underlying galaxy 
 formation model. 
Massive galaxies in \hagn\, appear rounder, redder and less compact than in \hnoagn.
The simulated size-mass and fraction-of-ellipticals-mass relations 
 now match observations, in contrast to the case without AGN.
 The so-called main sequence of $\rm sSFR$ versus galaxy stellar mass,
 and its decrease at the high-mass end~\citep{kavirajetal16}, is now also better recovered with AGN feedback.
In the simulation without AGN feedback, these massive galaxies are rotationally supported disc galaxies, while AGN feedback allows for the formation of dispersion-dominated elliptical galaxies at the high-mass end.
Mergers by themselves cannot prevent subsequent cold gas infall. 
AGN feedback is critical to prevent in situ star formation 
produced in the subsequently infalling gas to rebuild 
 a new massive stellar disc.

In this paper, we have shown how the nature of the cosmic infall (in situ versus ex situ) is key to  galaxy morphology.
However, we have overlooked some important details of the cosmic infall that have to be addressed in future work.
First, we have not investigated how the gas is brought to the galaxy and how angular momentum is transferred to it, changing the extent of the galaxy's disc.
 Finally, parameters of the mergers such as its impact parameter, orbital angular momentum, mass ratio, gas content and distributions over time  -- and how these are encoded in the initial conditions -- should be  investigated in more details.

\section*{Acknowledgements}
This work was granted access to the HPC resources of CINES under the allocations 2013047012, 2014047012 and 2015047012 made by GENCI.
This research is part of the Spin(e) (ANR-13-BS05-0005, \url{http://cosmicorigin.org}) and Horizon-UK projects. 
This work has made use of the Horizon cluster on which the simulation was post-processed, hosted by the Institut d'Astrophysique de Paris.
Part of the analysis of the simulations was carried out on the DiRAC Facility jointly funded by BIS and STFC.
We warmly thank S.~Rouberol for  running it smoothly.
SP acknowledges support from the Japan Society for the Promotion of
Science (JSPS long-term invitation fellowship).
The research of JD is supported by Adrian Beecroft, The Oxford Martin School and STFC.
MV acknowledges funding from the ERC under the European Community's Seventh Framework Programme (FP7/2007-2013 Grant Agreement no. 614199, project `BLACK').
\vspace{-0.5cm}

\bibliographystyle{mn2e}
\bibliography{author}

\begin{thebibliography}{71}
\expandafter\ifx\csname natexlab\endcsname\relax\def\natexlab#1{#1}\fi

\bibitem[{{Aubert}, {Pichon} \& {Colombi}(2004){Aubert}, {Pichon}, \&
  {Colombi}}]{aubertetal04}
{Aubert} D., {Pichon} C., {Colombi} S., 2004, \mnras, 352, 376

\bibitem[{{Binney}(1977)}]{binney77}
{Binney} J., 1977, \apj, 215, 483

\bibitem[{{Birnboim} \& {Dekel}(2003)}]{birnboim&dekel03}
{Birnboim} Y., {Dekel} A., 2003, \mnras, 345, 349

\bibitem[{{Bournaud}, {Jog} \& {Combes}(2007){Bournaud}, {Jog}, \&
  {Combes}}]{bournaudetal07}
{Bournaud} F., {Jog} C.~J., {Combes} F., 2007, \aap, 476, 1179

\bibitem[{{Bruzual} \& {Charlot}(2003)}]{bruzual&charlot03}
{Bruzual} G., {Charlot} S., 2003, \mnras, 344, 1000

\bibitem[{{Bundy} {et~al}\mbox{.}(2009){Bundy}, {Fukugita}, {Ellis}, {Targett},
  {Belli}, \& {Kodama}}]{bundyetal09}
{Bundy} K., {Fukugita} M., {Ellis} R.~S., {Targett} T.~A., {Belli} S., {Kodama}
  T., 2009, \apj, 697, 1369

\bibitem[{{Bundy} {et~al}\mbox{.}(2010){Bundy}, {Scarlata}, {Carollo}, {Ellis},
  {Drory}, {Hopkins}, {Salvato}, {Leauthaud}, {Koekemoer}, {Murray}, {Ilbert},
  {Oesch}, {Ma}, {Capak}, {Pozzetti}, \& {Scoville}}]{bundyetal10}
{Bundy} K. {et~al.}, 2010, \apj, 719, 1969

\bibitem[{{Cimatti} {et~al}\mbox{.}(2008){Cimatti}, {Cassata}, {Pozzetti},
  {Kurk}, {Mignoli}, {Renzini}, {Daddi}, {Bolzonella}, {Brusa}, {Rodighiero},
  {Dickinson}, {Franceschini}, {Zamorani}, {Berta}, {Rosati}, \&
  {Halliday}}]{cimattietal08}
{Cimatti} A. {et~al.}, 2008, \aap, 482, 21

\bibitem[{{Conselice}(2006)}]{conselice06}
{Conselice} C.~J., 2006, \mnras, 373, 1389

\bibitem[{{Croton} {et~al}\mbox{.}(2006){Croton}, {Springel}, {White}, {De
  Lucia}, {Frenk}, {Gao}, {Jenkins}, {Kauffmann}, {Navarro}, \&
  {Yoshida}}]{crotonetal06}
{Croton} D.~J. {et~al.}, 2006, \mnras, 365, 11

\bibitem[{{Daddi} {et~al}\mbox{.}(2005){Daddi}, {Renzini}, {Pirzkal},
  {Cimatti}, {Malhotra}, {Stiavelli}, {Xu}, {Pasquali}, {Rhoads}, {Brusa}, {di
  Serego Alighieri}, {Ferguson}, {Koekemoer}, {Moustakas}, {Panagia}, \&
  {Windhorst}}]{daddietal05}
{Daddi} E. {et~al.}, 2005, \apj, 626, 680

\bibitem[{{Danovich} {et~al}\mbox{.}(2015){Danovich}, {Dekel}, {Hahn},
  {Ceverino}, \& {Primack}}]{danovichetal15}
{Danovich} M., {Dekel} A., {Hahn} O., {Ceverino} D., {Primack} J., 2015,
  \mnras, 449, 2087

\bibitem[{{Dav{\'e}}, {Oppenheimer} \& {Finlator}(2011){Dav{\'e}},
  {Oppenheimer}, \& {Finlator}}]{daveetal11}
{Dav{\'e}} R., {Oppenheimer} B.~D., {Finlator} K., 2011, \mnras, 415, 11

\bibitem[{{De Lucia} \& {Blaizot}(2007)}]{delucia&blaizot07}
{De Lucia} G., {Blaizot} J., 2007, \mnras, 375, 2

\bibitem[{{Dekel} \& {Birnboim}(2006)}]{dekel&birnboim06}
{Dekel} A., {Birnboim} Y., 2006, \mnras, 368, 2

\bibitem[{{Dekel} {et~al}\mbox{.}(2009){Dekel}, {Birnboim}, {Engel},
  {Freundlich}, {Goerdt}, {Mumcuoglu}, {Neistein}, {Pichon}, {Teyssier}, \&
  {Zinger}}]{dekeletal09}
{Dekel} A. {et~al.}, 2009, \nat, 457, 451

\bibitem[{{Dekel} \& {Silk}(1986)}]{dekel&silk86}
{Dekel} A., {Silk} J., 1986, \apj, 303, 39

\bibitem[{{Dubois} {et~al}\mbox{.}(2010){Dubois}, {Devriendt}, {Slyz}, \&
  {Teyssier}}]{duboisetal10}
{Dubois} Y., {Devriendt} J., {Slyz} A., {Teyssier} R., 2010, \mnras, 409, 985

\bibitem[{{Dubois} {et~al}\mbox{.}(2012){Dubois}, {Devriendt}, {Slyz}, \&
  {Teyssier}}]{duboisetal12}
{Dubois} Y., {Devriendt} J., {Slyz} A., {Teyssier} R., 2012, \mnras, 420, 2662

\bibitem[{{Dubois} {et~al}\mbox{.}(2011){Dubois}, {Devriendt}, {Teyssier}, \&
  {Slyz}}]{duboisetal11}
{Dubois} Y., {Devriendt} J., {Teyssier} R., {Slyz} A., 2011, \mnras, 417, 1853

\bibitem[{{Dubois} {et~al}\mbox{.}(2013){Dubois}, {Gavazzi}, {Peirani}, \&
  {Silk}}]{duboisetal13}
{Dubois} Y., {Gavazzi} R., {Peirani} S., {Silk} J., 2013, \mnras, 433, 3297

\bibitem[{{Dubois} {et~al}\mbox{.}(2014){Dubois}, {Pichon}, {Welker}, {Le
  Borgne}, {Devriendt}, \& {et al.}}]{duboisetal14}
{Dubois} Y., {Pichon} C., {Welker} C., {Le Borgne} D., {Devriendt} J., {et
  al.}, 2014, \mnras, 444, 1453

\bibitem[{{Dubois} {et~al}\mbox{.}(2015){Dubois}, {Volonteri}, {Silk},
  {Devriendt}, {Slyz}, \& {Teyssier}}]{duboisetal15snbh}
{Dubois} Y., {Volonteri} M., {Silk} J., {Devriendt} J., {Slyz} A., {Teyssier}
  R., 2015, \mnras, 452, 1502

\bibitem[{{Duc} {et~al}\mbox{.}(2015){Duc}, {Cuillandre}, {Karabal},
  {Cappellari}, {Alatalo}, {Blitz}, {Bournaud}, {Bureau}, {Crocker}, {Davies},
  {Davis}, {de Zeeuw}, {Emsellem}, {Khochfar}, {Krajnovi{\'c}}, {Kuntschner},
  {McDermid}, {Michel-Dansac}, {Morganti}, {Naab}, {Oosterloo}, {Paudel},
  {Sarzi}, {Scott}, {Serra}, {Weijmans}, \& {Young}}]{ducetal15}
{Duc} P.-A. {et~al.}, 2015, \mnras, 446, 120

\bibitem[{{Franx} {et~al}\mbox{.}(2008){Franx}, {van Dokkum}, {F{\"o}rster
  Schreiber}, {Wuyts}, {Labb{\'e}}, \& {Toft}}]{franxetal08}
{Franx} M., {van Dokkum} P.~G., {F{\"o}rster Schreiber} N.~M., {Wuyts} S.,
  {Labb{\'e}} I., {Toft} S., 2008, \apj, 688, 770

\bibitem[{{Haardt} \& {Madau}(1996)}]{haardt&madau96}
{Haardt} F., {Madau} P., 1996, \apj, 461, 20

\bibitem[{{Habouzit}, {Volonteri} \& {Dubois}(2016){Habouzit}, {Volonteri}, \&
  {Dubois}}]{habouzitetal16}
{Habouzit} M., {Volonteri} M., {Dubois} Y., 2016, ArXiv e-prints

\bibitem[{{Hilz}, {Naab} \& {Ostriker}(2013){Hilz}, {Naab}, \&
  {Ostriker}}]{hilzetal13}
{Hilz} M., {Naab} T., {Ostriker} J.~P., 2013, \mnras, 429, 2924

\bibitem[{{Ilbert} {et~al}\mbox{.}(2010){Ilbert}, {Salvato}, {Le Floc'h},
  {Aussel}, {Capak}, {McCracken}, {Mobasher}, {Kartaltepe}, {Scoville},
  {Sanders}, {Arnouts}, {Bundy}, {Cassata}, {Kneib}, {Koekemoer}, {Le
  F{\`e}vre}, {Lilly}, {Surace}, {Taniguchi}, {Tasca}, {Thompson}, {Tresse},
  {Zamojski}, {Zamorani}, \& {Zucca}}]{ilbertetal10}
{Ilbert} O. {et~al.}, 2010, \apj, 709, 644

\bibitem[{{Karim} {et~al}\mbox{.}(2011){Karim}, {Schinnerer},
  {Mart{\'{\i}}nez-Sansigre}, {Sargent}, {van der Wel}, {Rix}, {Ilbert},
  {Smol{\v c}i{\'c}}, {Carilli}, {Pannella}, {Koekemoer}, {Bell}, \&
  {Salvato}}]{karimetal11}
{Karim} A. {et~al.}, 2011, \apj, 730, 61

\bibitem[{{Kaviraj} {et~al}\mbox{.}(2013){Kaviraj}, {Cohen}, {Windhorst},
  {Silk}, {O'Connell}, {Dopita}, {Dekel}, {Hathi}, {Straughn}, \&
  {Rutkowski}}]{kavirajetal13}
{Kaviraj} S. {et~al.}, 2013, \mnras, 429, L40

\bibitem[{{Kaviraj} {et~al}\mbox{.}(2015){Kaviraj}, {Devriendt}, {Dubois},
  {Slyz}, {Welker}, {Pichon}, {Peirani}, \& {Le Borgne}}]{kavirajetal15}
{Kaviraj} S., {Devriendt} J., {Dubois} Y., {Slyz} A., {Welker} C., {Pichon} C.,
  {Peirani} S., {Le Borgne} D., 2015, \mnras, 452, 2845

\bibitem[{{Kaviraj} {et~al}\mbox{.}(2016){Kaviraj}, {Laigle}, {Kimm},
  {Devriendt}, {Dubois}, {Pichon}, {Slyz}, {Chisari}, \&
  {Peirani}}]{kavirajetal16}
{Kaviraj} S. {et~al.}, 2016, ArXiv e-prints

\bibitem[{{Kere{\v s}} {et~al}\mbox{.}(2005){Kere{\v s}}, {Katz}, {Weinberg},
  \& {Dav{\'e}}}]{keresetal05}
{Kere{\v s}} D., {Katz} N., {Weinberg} D.~H., {Dav{\'e}} R., 2005, \mnras, 363,
  2

\bibitem[{{Khandai} {et~al}\mbox{.}(2015){Khandai}, {Di Matteo}, {Croft},
  {Wilkins}, {Feng}, {Tucker}, {DeGraf}, \& {Liu}}]{khandaietal15}
{Khandai} N., {Di Matteo} T., {Croft} R., {Wilkins} S., {Feng} Y., {Tucker} E.,
  {DeGraf} C., {Liu} M.-S., 2015, \mnras, 450, 1349

\bibitem[{{Khochfar} \& {Silk}(2006)}]{khochfar&silk06}
{Khochfar} S., {Silk} J., 2006, \apjl, 648, L21

\bibitem[{{Kimm} {et~al}\mbox{.}(2011){Kimm}, {Devriendt}, {Slyz}, {Pichon},
  {Kassin}, \& {Dubois}}]{kimmetal11}
{Kimm} T., {Devriendt} J., {Slyz} A., {Pichon} C., {Kassin} S.~A., {Dubois} Y.,
  2011, ArXiv e-prints

\bibitem[{{Komatsu} {et~al}\mbox{.}(2011){Komatsu}, {Smith}, {Dunkley},
  {Bennett}, {Gold}, {Hinshaw}, {Jarosik}, {Larson}, {Nolta}, {Page},
  {Spergel}, {Halpern}, {Hill}, {Kogut}, {Limon}, {Meyer}, {Odegard}, {Tucker},
  {Weiland}, {Wollack}, \& {Wright}}]{komatsuetal11}
{Komatsu} E. {et~al.}, 2011, \apjs, 192, 18

\bibitem[{{Kravtsov}, {Vikhlinin} \& {Meshscheryakov}(2014){Kravtsov},
  {Vikhlinin}, \& {Meshscheryakov}}]{kravtsovetal14}
{Kravtsov} A., {Vikhlinin} A., {Meshscheryakov} A., 2014, ArXiv e-prints

\bibitem[{{Lackner} {et~al}\mbox{.}(2012){Lackner}, {Cen}, {Ostriker}, \&
  {Joung}}]{lackneretal12}
{Lackner} C.~N., {Cen} R., {Ostriker} J.~P., {Joung} M.~R., 2012, \mnras, 425,
  641

\bibitem[{{Lee} \& {Yi}(2013)}]{lee&yi13}
{Lee} J., {Yi} S.~K., 2013, \apj, 766, 38

\bibitem[{{Naab}, {Johansson} \& {Ostriker}(2009){Naab}, {Johansson}, \&
  {Ostriker}}]{naabetal09}
{Naab} T., {Johansson} P.~H., {Ostriker} J.~P., 2009, \apjl, 699, L178

\bibitem[{{Nipoti} {et~al}\mbox{.}(2009){Nipoti}, {Treu}, {Auger}, \&
  {Bolton}}]{nipotietal09}
{Nipoti} C., {Treu} T., {Auger} M.~W., {Bolton} A.~S., 2009, \apjl, 706, L86

\bibitem[{{Ocvirk}, {Pichon} \& {Teyssier}(2008){Ocvirk}, {Pichon}, \&
  {Teyssier}}]{ocvirketal08}
{Ocvirk} P., {Pichon} C., {Teyssier} R., 2008, \mnras, 390, 1326

\bibitem[{{Oppenheimer} {et~al}\mbox{.}(2010){Oppenheimer}, {Dav{\'e}},
  {Kere{\v s}}, {Fardal}, {Katz}, {Kollmeier}, \&
  {Weinberg}}]{oppenheimeretal10}
{Oppenheimer} B.~D., {Dav{\'e}} R., {Kere{\v s}} D., {Fardal} M., {Katz} N.,
  {Kollmeier} J.~A., {Weinberg} D.~H., 2010, \mnras, 406, 2325

\bibitem[{{Oser} {et~al}\mbox{.}(2012){Oser}, {Naab}, {Ostriker}, \&
  {Johansson}}]{oseretal12}
{Oser} L., {Naab} T., {Ostriker} J.~P., {Johansson} P.~H., 2012, \apj, 744, 63

\bibitem[{{Oser} {et~al}\mbox{.}(2010){Oser}, {Ostriker}, {Naab}, {Johansson},
  \& {Burkert}}]{oseretal10}
{Oser} L., {Ostriker} J.~P., {Naab} T., {Johansson} P.~H., {Burkert} A., 2010,
  \apj, 725, 2312

\bibitem[{{Pichon} {et~al}\mbox{.}(2011){Pichon}, {Pogosyan}, {Kimm}, {Slyz},
  {Devriendt}, \& {Dubois}}]{pichonetal11}
{Pichon} C., {Pogosyan} D., {Kimm} T., {Slyz} A., {Devriendt} J., {Dubois} Y.,
  2011, \mnras, 418, 2493

\bibitem[{{Power} {et~al}\mbox{.}(2003){Power}, {Navarro}, {Jenkins}, {Frenk},
  {White}, {Springel}, {Stadel}, \& {Quinn}}]{poweretal03}
{Power} C., {Navarro} J.~F., {Jenkins} A., {Frenk} C.~S., {White} S.~D.~M.,
  {Springel} V., {Stadel} J., {Quinn} T., 2003, \mnras, 338, 14

\bibitem[{{Rodriguez-Gomez} {et~al}\mbox{.}(2016){Rodriguez-Gomez},
  {Pillepich}, {Sales}, {Genel}, {Vogelsberger}, {Zhu}, {Wellons}, {Nelson},
  {Torrey}, {Springel}, {Ma}, \& {Hernquist}}]{rodriguez-gomezetal16}
{Rodriguez-Gomez} V. {et~al.}, 2016, \mnras, 458, 2371

\bibitem[{{Santini} {et~al}\mbox{.}(2014){Santini}, {Maiolino}, {Magnelli},
  {Lutz}, {Lamastra}, {Li Causi}, {Eales}, {Andreani}, {Berta}, {Buat},
  {Cooray}, {Cresci}, {Daddi}, {Farrah}, {Fontana}, {Franceschini}, {Genzel},
  {Granato}, {Grazian}, {Le Floc'h}, {Magdis}, {Magliocchetti}, {Mannucci},
  {Menci}, {Nordon}, {Oliver}, {Popesso}, {Pozzi}, {Riguccini}, {Rodighiero},
  {Rosario}, {Salvato}, {Scott}, {Silva}, {Tacconi}, {Viero}, {Wang}, {Wuyts},
  \& {Xu}}]{santinietal14}
{Santini} P. {et~al.}, 2014, \aap, 562, A30

\bibitem[{{Scannapieco} {et~al}\mbox{.}(2012){Scannapieco}, {Wadepuhl},
  {Parry}, {Navarro}, {Jenkins}, {Springel}, {Teyssier}, {Carlson}, {Couchman},
  {Crain}, {Dalla Vecchia}, {Frenk}, {Kobayashi}, {Monaco}, {Murante},
  {Okamoto}, {Quinn}, {Schaye}, {Stinson}, {Theuns}, {Wadsley}, {White}, \&
  {Woods}}]{scannapiecoetal12}
{Scannapieco} C. {et~al.}, 2012, \mnras, 423, 1726

\bibitem[{{Schaye} {et~al}\mbox{.}(2015){Schaye}, {Crain}, {Bower}, {Furlong},
  {Schaller}, {Theuns}, {Dalla Vecchia}, {Frenk}, {McCarthy}, {Helly},
  {Jenkins}, {Rosas-Guevara}, {White}, {Baes}, {Booth}, {Camps}, {Navarro},
  {Qu}, {Rahmati}, {Sawala}, {Thomas}, \& {Trayford}}]{schayeetal15}
{Schaye} J. {et~al.}, 2015, \mnras, 446, 521

\bibitem[{{Schweizer} \& {Seitzer}(1988)}]{schweizer&seitze88}
{Schweizer} F., {Seitzer} P., 1988, \apj, 328, 88

\bibitem[{{Shankar} {et~al}\mbox{.}(2013){Shankar}, {Marulli}, {Bernardi},
  {Mei}, {Meert}, \& {Vikram}}]{shankaretal13}
{Shankar} F., {Marulli} F., {Bernardi} M., {Mei} S., {Meert} A., {Vikram} V.,
  2013, \mnras, 428, 109

\bibitem[{{Silk} \& {Rees}(1998)}]{silk&rees98}
{Silk} J., {Rees} M.~J., 1998, \aap, 331, L1

\bibitem[{{Stewart} {et~al}\mbox{.}(2013){Stewart}, {Brooks}, {Bullock},
  {Maller}, {Diemand}, {Wadsley}, \& {Moustakas}}]{stewartetal13}
{Stewart} K.~R., {Brooks} A.~M., {Bullock} J.~S., {Maller} A.~H., {Diemand} J.,
  {Wadsley} J., {Moustakas} L.~A., 2013, \apj, 769, 74

\bibitem[{{Sutherland} \& {Dopita}(1993)}]{sutherland&dopita93}
{Sutherland} R.~S., {Dopita} M.~A., 1993, \apjs, 88, 253

\bibitem[{{Teyssier}(2002)}]{teyssier02}
{Teyssier} R., 2002, \aap, 385, 337

\bibitem[{{Toomre} \& {Toomre}(1972)}]{toomre&toomre72}
{Toomre} A., {Toomre} J., 1972, \apj, 178, 623

\bibitem[{{Trujillo} {et~al}\mbox{.}(2006){Trujillo}, {F{\"o}rster Schreiber},
  {Rudnick}, {Barden}, {Franx}, {Rix}, {Caldwell}, {McIntosh}, {Toft},
  {H{\"a}ussler}, {Zirm}, {van Dokkum}, {Labb{\'e}}, {Moorwood},
  {R{\"o}ttgering}, {van der Wel}, {van der Werf}, \& {van
  Starkenburg}}]{trujilloetal06}
{Trujillo} I. {et~al.}, 2006, \apj, 650, 18

\bibitem[{{Tweed} {et~al}\mbox{.}(2009){Tweed}, {Devriendt}, {Blaizot},
  {Colombi}, \& {Slyz}}]{tweedetal09}
{Tweed} D., {Devriendt} J., {Blaizot} J., {Colombi} S., {Slyz} A., 2009, \aap,
  506, 647

\bibitem[{{van der Wel} {et~al}\mbox{.}(2014){van der Wel}, {Franx}, {van
  Dokkum}, {Skelton}, {Momcheva}, {Whitaker}, {Brammer}, {Bell}, {Rix},
  {Wuyts}, {Ferguson}, {Holden}, {Barro}, {Koekemoer}, {Chang}, {McGrath},
  {H{\"a}ussler}, {Dekel}, {Behroozi}, {Fumagalli}, {Leja}, {Lundgren},
  {Maseda}, {Nelson}, {Wake}, {Patel}, {Labb{\'e}}, {Faber}, {Grogin}, \&
  {Kocevski}}]{vanderweletal14}
{van der Wel} A. {et~al.}, 2014, \apj, 788, 28

\bibitem[{{van der Wel} {et~al}\mbox{.}(2008){van der Wel}, {Holden}, {Zirm},
  {Franx}, {Rettura}, {Illingworth}, \& {Ford}}]{vanderweletal08}
{van der Wel} A., {Holden} B.~P., {Zirm} A.~W., {Franx} M., {Rettura} A.,
  {Illingworth} G.~D., {Ford} H.~C., 2008, \apj, 688, 48

\bibitem[{{van Dokkum}(2005)}]{vandokkum05}
{van Dokkum} P.~G., 2005, \aj, 130, 2647

\bibitem[{{van Dokkum} {et~al}\mbox{.}(2010){van Dokkum}, {Whitaker},
  {Brammer}, {Franx}, {Kriek}, {Labb{\'e}}, {Marchesini}, {Quadri}, {Bezanson},
  {Illingworth}, {Muzzin}, {Rudnick}, {Tal}, \& {Wake}}]{vandokkumetal10}
{van Dokkum} P.~G. {et~al.}, 2010, \apj, 709, 1018

\bibitem[{{Vogelsberger} {et~al}\mbox{.}(2014){Vogelsberger}, {Genel},
  {Springel}, {Torrey}, {Sijacki}, {Xu}, {Snyder}, {Bird}, {Nelson}, \&
  {Hernquist}}]{vogelsbergeretal14}
{Vogelsberger} M. {et~al.}, 2014, \nat, 509, 177

\bibitem[{{Volonteri} {et~al}\mbox{.}(2016){Volonteri}, {Dubois}, {Pichon}, \&
  {Devriendt}}]{volonterietal16}
{Volonteri} M., {Dubois} Y., {Pichon} C., {Devriendt} J., 2016, \mnras, 460,
  2979

\bibitem[{{Welker} {et~al}\mbox{.}(2015{\natexlab{a}}){Welker}, {Dubois},
  {Devriendt}, {Pichon}, {Kaviraj}, \& {Peirani}}]{welkeretal16discs}
{Welker} C., {Dubois} Y., {Devriendt} J., {Pichon} C., {Kaviraj} S., {Peirani}
  S., 2015{\natexlab{a}}, ArXiv e-prints

\bibitem[{{Welker} {et~al}\mbox{.}(2015{\natexlab{b}}){Welker}, {Dubois},
  {Pichon}, {Devriendt}, \& {Chisari}}]{welkeretal16plane}
{Welker} C., {Dubois} Y., {Pichon} C., {Devriendt} J., {Chisari} E.~N.,
  2015{\natexlab{b}}, ArXiv e-prints

\bibitem[{{Wellons} {et~al}\mbox{.}(2015){Wellons}, {Torrey}, {Ma},
  {Rodriguez-Gomez}, {Vogelsberger}, {Kriek}, {van Dokkum}, {Nelson}, {Genel},
  {Pillepich}, {Springel}, {Sijacki}, {Snyder}, {Nelson}, {Sales}, \&
  {Hernquist}}]{wellonsetal15}
{Wellons} S. {et~al.}, 2015, \mnras, 449, 361

\end{thebibliography}

\appendix

\section{Morphometric Properties of Galaxies at Fixed Halo Mass}
\label{appendix}

\begin{figure}
\center \includegraphics[width=0.995\columnwidth]{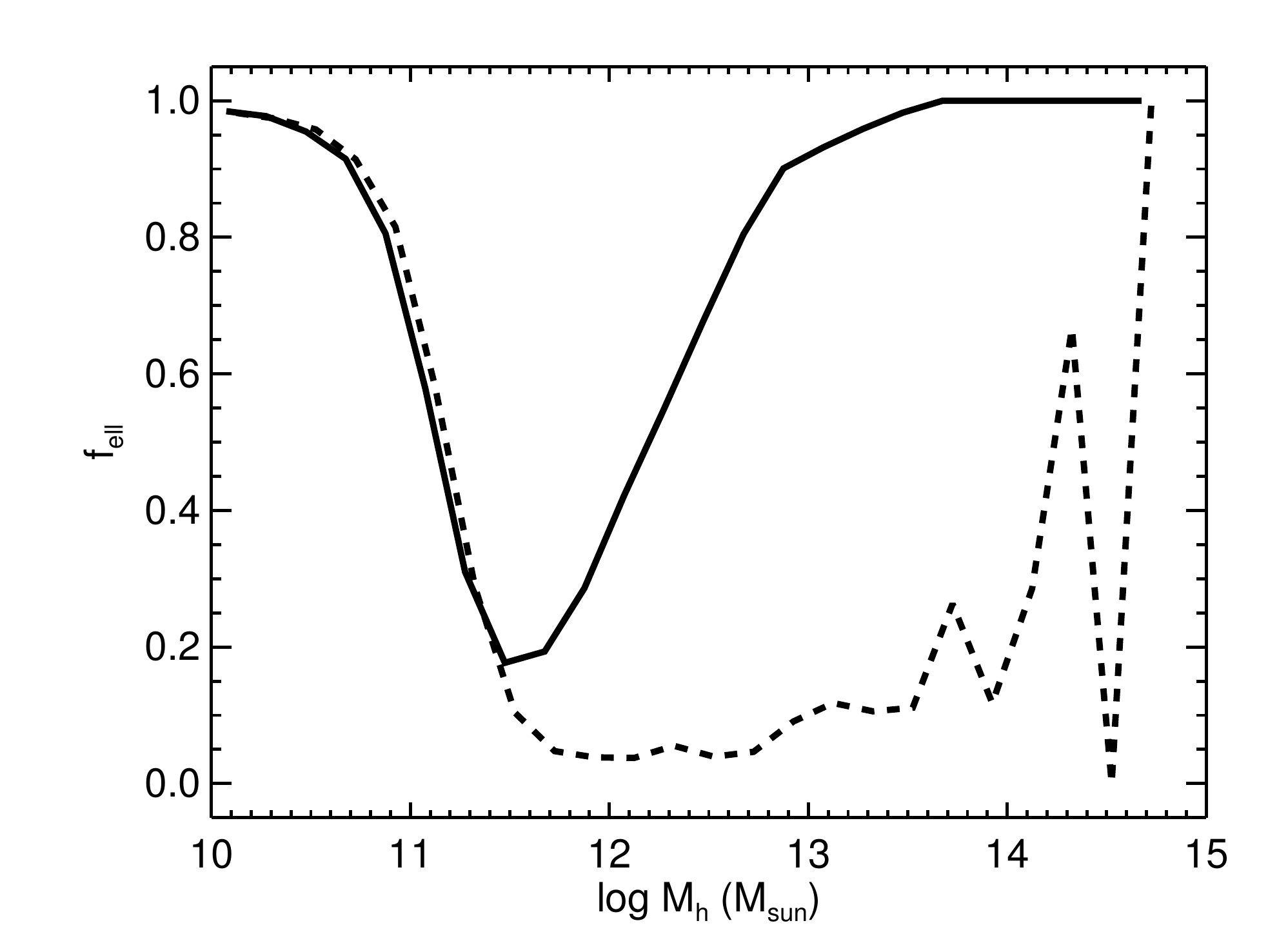}
 \caption{Fraction $f_{\rm ell}$ of elliptical galaxies (with $V/\sigma<1$) as a function of DM halo virial mass $M_{\rm h}$ at $z=0$ in the \hagn\, and \hnoagn\, simulations, respectively the black solid lines and the dashed lines. We see that the population of the most massive haloes $M_{\rm h}> 5\times 10^{12}\,\rm M_\odot$ is clearly dominated by elliptical galaxies in \hagn, whereas it is dominated by disc galaxies in \hnoagn.}
\label{fig:ellfractionmh}
\end{figure}

\begin{figure}
\center \includegraphics[width=0.995\columnwidth]{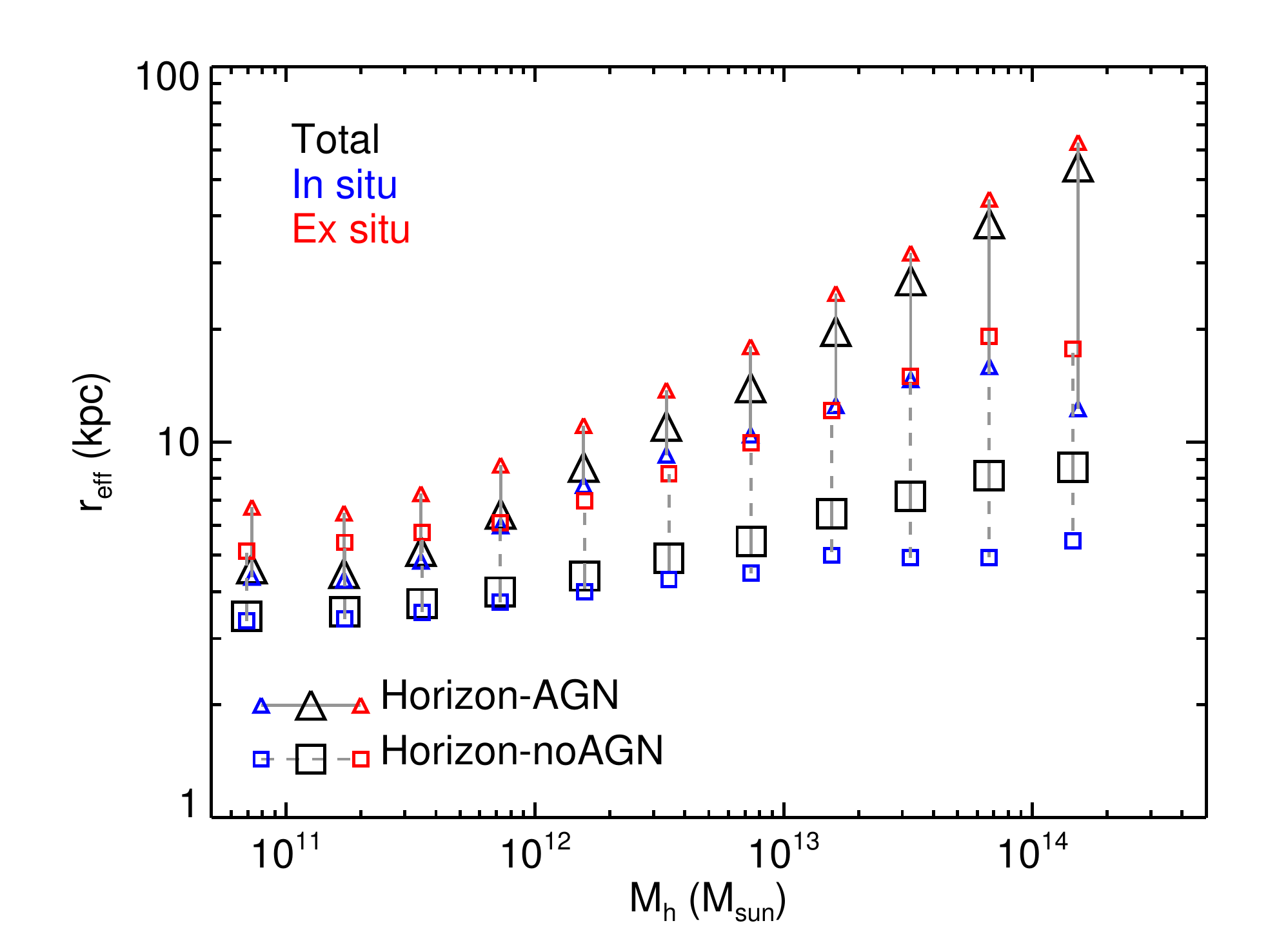}
 \caption{Effective radius $r_{\rm eff}$ of stars as a function of halo mass $M_{\rm h}$ for the total (black), in situ (blue), and ex situ (red) components in \hagn\, (triangles with solid grey lines) and in \hnoagn\, (squares with dashed grey lines) at $z=0$. Note that the sample of haloes corresponds to the sample of galaxies used to plot the same decomposed radius as a function of stellar mass in Fig.~\ref{fig:reff_insitu}.}
\label{fig:reff_insitumh}
\end{figure}

In this Appendix we show the morphological and size dependency of galaxies with their host halo mass instead of their galaxy stellar mass, in order to assess that the change in morphometric properties of galaxies at a given stellar mass is a real transformation of the galaxy and is not simply due to a change in galaxy stellar mass.

Fig.~\ref{fig:ellfractionmh} shows the fraction of ellipticals $f_{\rm ell}$ as a function of halo mass at $z=0$.
The trend with halo mass, that is seen in galaxy stellar mass in Fig.~\ref{fig:ellfraction}, is recovered.
In \hagn, there is clear transition from disc-dominated galaxies at intermediate halo mass (with a minimum of $f_{\rm ell}$ at $M_{\rm h}\simeq 4\times 10^{11}\,\rm M_\odot$) to elliptical galaxies above group-scale haloes $M_{\rm h}> 5\times 10^{12}\,\rm M_\odot$.
In \hnoagn, galaxies remain disc-like from intermediate haloes up to the cluster-scale haloes $M_{\rm h}\lesssim 10^{14}\,\rm M_\odot$ with strong variations in $f_{\rm ell}$ due to the limited sample above that halo mass (12 galaxy clusters).

Fig.~\ref{fig:reff_insitumh} shows the effective radius of galaxies, as well as the effective radius of the in situ and of the ex situ components, as a function of their host halo mass.
As for Fig.~\ref{fig:reff_insitu}, which shows those galaxy sizes as a function of the galaxy stellar mass, galaxies are more extended with AGN feedback for a given halo mass, and they show an increase in both the in situ and ex situ component.

Therefore, we can conclude that those morphometric changes induced by AGN feedback are not due to a naive rescaling in galaxy stellar mass, but correspond to real morphometric changes of individual galaxies.

\end{document}